\documentclass[final,5p]{elsarticle}

\usepackage{amssymb}
\usepackage{amsmath}
\usepackage{bm}
\usepackage{tikz}
\usetikzlibrary{positioning, shapes.geometric, arrows.meta}
\usepackage{braket}
\usepackage{enumitem}
\setlist[enumerate]{label=(\arabic*), itemsep=2pt, parsep=0pt, topsep=3pt, partopsep=0pt}
\usepackage[breaklinks]{hyperref}
\usepackage{microtype}
\usepackage{stfloats}
\usepackage{booktabs,tabularx,multirow}
\usepackage{graphicx}
\usepackage[version=4]{mhchem}
\graphicspath{{figure/}}
\journal{Computer Physics Communications}

\begin{document}

\begin{frontmatter}


\title{A Unified Heterogeneous Implementation of Numerical Atomic Orbitals-Based Real-Time TDDFT within the ABACUS Package}

\author[aff1]{Taoni Bao}
\author[aff1]{Yuanbo Li}
\author[aff1]{Zichao Deng}
\author[aff5]{Haotian Zhao}
\author[aff1]{Denghui Lu}
\author[aff2]{Yike Huang}
\author[aff3,aff4]{Chao Lian}
\author[aff5,aff6,aff7]{Lixin He}
\author[aff1]{Mohan Chen\corref{cor1}}

\affiliation[aff1]{organization={HEDPS, CAPT, School of Mechanics and Engineering Science and School of Physics, Peking University},
            city={Beijing},
            postcode={100871}, 
            country={P. R. China}}

\affiliation[aff5]{organization={Laboratory of Quantum Information, University of Science and Technology of China},
            city={Hefei},
            postcode={230026}, 
            country={P. R. China}}

\affiliation[aff2]{organization={Institute of Microelectronics, Chinese Academy of Sciences},
            city={Beijing},
            postcode={100029}, 
            country={P. R. China}}

\affiliation[aff3]{organization={Beijing National Laboratory for Condensed Matter Physics and Institute of Physics, Chinese Academy of Sciences},
            city={Beijing},
            postcode={100190}, 
            country={P. R. China}}

\affiliation[aff4]{organization={Songshan Lake Materials Laboratory},
            city={Dongguan},
            postcode={523808}, 
            country={P. R. China}}

\affiliation[aff6]{organization={Institute of Artificial Intelligence, Hefei Comprehensive National Science Center},
            city={Hefei},
            postcode={230088}, 
            country={P. R. China}}

\affiliation[aff7]{organization={Hefei National Laboratory, University of Science and Technology of China},
            city={Hefei},
            postcode={230088}, 
            country={P. R. China}}

\cortext[cor1]{Corresponding author. Email: mohanchen@pku.edu.cn}

%% Abstract
\begin{abstract}
We present a unified heterogeneous computing framework for real-time time-dependent density functional theory (RT-TDDFT) based on numerical atomic orbitals (NAOs), implemented in the ABACUS package. 
We introduce three co-designed abstraction layers, including unified data containers, unified linear algebra operators, and unified grid integration interfaces. 
These layers collectively accelerate the two most demanding parts of NAO-based RT-TDDFT: explicit real-time wavefunction propagation and real-space grid operations such as Hamiltonian construction and force evaluation under external fields. 
We validate the method by computing optical properties for systems ranging from finite molecules to periodic solids, showing excellent agreement with standard benchmarks. 
Performance evaluations on bulk silicon demonstrate that a single GPU can achieve substantial wall-clock speedup over a fully utilized dual-socket CPU node. 
Furthermore, distributed multi-GPU strong-scaling tests confirm high parallel efficiency over tens of GPUs. 
This work establishes a high-performance, portable platform for large-scale first-principles simulations of ultrafast electron dynamics.
\end{abstract}

\begin{graphicalabstract}
\centering
\includegraphics[width=0.9\textwidth]{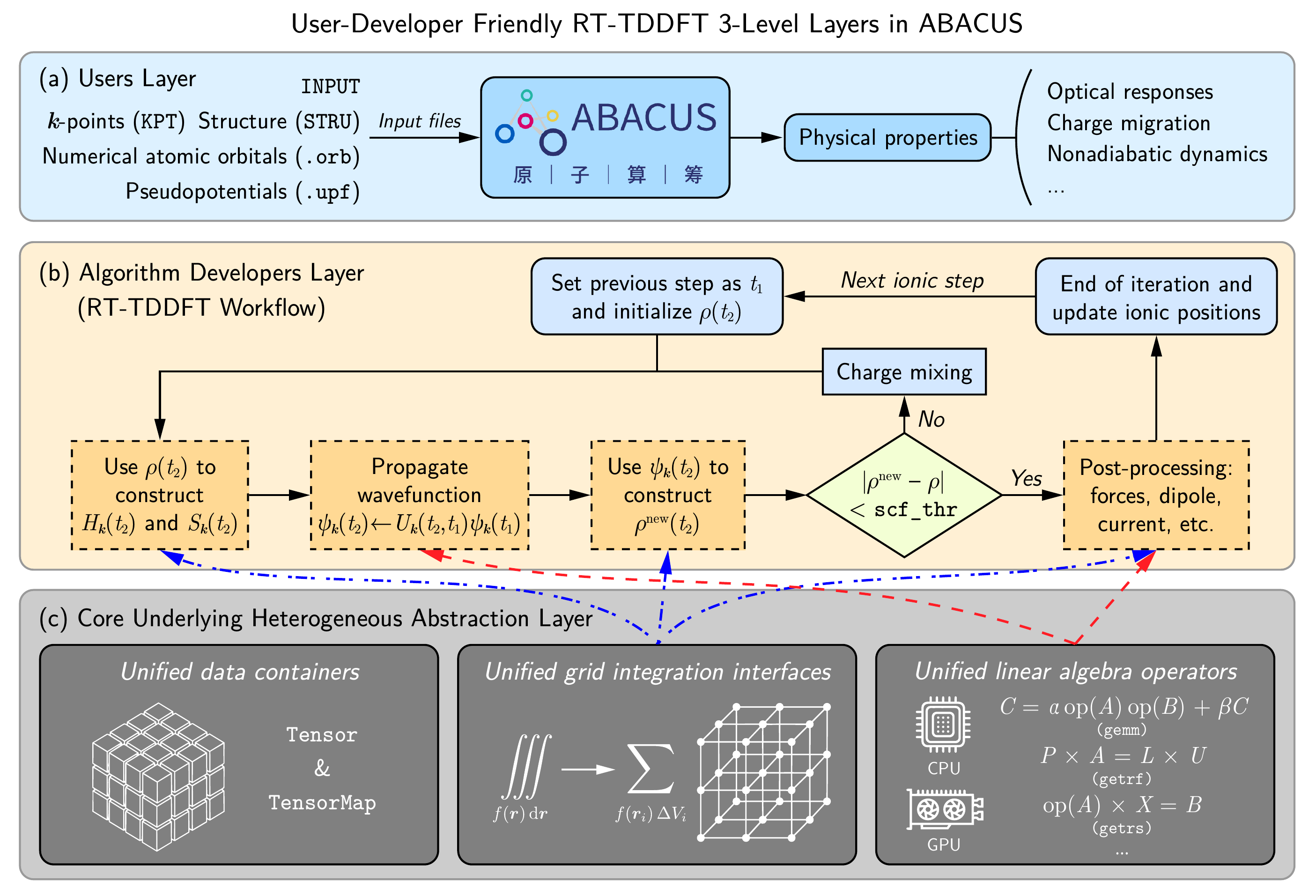}
\end{graphicalabstract}

\begin{highlights}
\item Unified heterogeneous framework for RT-TDDFT using numerical atomic orbitals
\item Hardware-agnostic abstraction decouples physical algorithms from device specifics
\item GPU-accelerated spherical integration eliminates velocity gauge bottlenecks
\item Achieves 12x speedup for key kernels and 76\% parallel efficiency on 40 GPUs
\end{highlights}

%% Keywords

\end{frontmatter}


\section{Introduction}

In recent years, real-time time-dependent density functional theory (RT-TDDFT) \cite{PhysRevLett.52.997,PhysRevB.54.4484,ullrich2012time} has become a cornerstone for simulating excited-state electronic dynamics and light-matter interactions beyond the ground-state description offered by conventional Kohn-Sham density functional theory (KSDFT) \cite{PhysRev.136.B864,PhysRev.140.A1133,RevModPhys.87.897}. By explicitly propagating the time-dependent Kohn-Sham orbitals under external field (such as ultrafast laser pulses), RT-TDDFT enables first-principles modeling of non-equilibrium phenomena \cite{jacs.3c08226} including photoabsorption \cite{adts.201800055,jctc.ct100289t,acs.jpcc.9b00296}, charge migration \cite{acs.jpclett.7b01652,Wang2016.ncomms11504,acs.jpclett.1c00772}, high-harmonic generation (HHG) \cite{PhysRevA.103.043117,PhysRevLett.118.087403,PhysRevB.99.184306}, and electron-phonon coupling \cite{PhysRevB.105.224311,PhysRevX.15.021053,j.physb.2025.417202}. Its ability to capture both linear and nonlinear optical responses without prior knowledge of excited-state eigenstates makes it particularly valuable for studying ultrafast processes on femtosecond to attosecond timescales \cite{RevModPhys.81.163}.

Although recent pioneering software developments have achieved impressive strong scaling on CPUs \cite{CHOI2024109349}, the ever-growing demand for computational efficiency in such time-resolved simulations has continuously driven the integration of graphics processing units (GPUs) into modern scientific computing workflows \cite{JPROC.2008.917757}. GPUs offer massive parallelism and high memory bandwidth, making them particularly well-suited for accelerating the compute-intensive linear algebra and grid-based operations prevalent in first-principles methods \cite{jcc.20829}. Several electronic structure packages have already embraced GPU acceleration, notably VASP \cite{PhysRevB.54.11169,jcc.23096}, CP2K \cite{jcp.5.0007045,acs.jcim.1c01538}, Quantum ESPRESSO \cite{Giannozzi_2017,jcp.5.0005082}, OpenMX \cite{OpenMX_CUDA_2014,JPSJ.94.124003}, PWmat \cite{JIA20139,JIA2013102}, INQ \cite{acs.jctc.1c00562} (a GPU-native TDDFT code based on a plane-wave and real-space dual representation), and QDD \cite{j.cpc.2023.108947} (Quantum Dissipative Dynamics, an RT-TDDFT code that employs a real-space grid), among many others. These efforts demonstrate that GPU offloading can yield considerable speedups, especially for large-scale or long-time simulations.

Nevertheless, the heterogeneous acceleration of first-principles software, especially for real-time methods, faces several challenges. First, most existing GPU-accelerated codes focus on ground-state KSDFT or linear-response TDDFT (LR-TDDFT), leaving real-time dynamics underexplored. This gap is especially pronounced for methods based on localized basis sets such as numerical atomic orbitals (NAOs) \cite{PhysRevB.40.3979,jcp.1.458452,PhysRevB.15.4716}. Second, developing and maintaining GPU code is inherently complex due to the need for explicit memory management, kernel optimization, and architecture-specific tuning, which hinders long-term software sustainability. Third, the rapid evolution and architectural diversity of modern high-performance computing platforms, including systems based on NVIDIA GPUs, AMD GPUs, and other emerging accelerators like Hygon DCUs \cite{hygon_dcu, CCF.Trans.HPC.s42514-024-00200-3} (Deep Computing Units), call for a portable programming model that avoids vendor-specific dependencies \cite{j.jpdc.2014.07.003}. A unified, hardware-agnostic framework for data representation and computation is thus essential to enable truly portable and maintainable heterogeneous scientific software.

To address these challenges, we build on the existing RT-TDDFT implementation in the open-source first-principles computational software package ABACUS (Atomic-orbital Based Ab-initio Computation at USTC) \cite{Chen_2010,j.commatsci.2015.07.004,jcp.5.0297563,ABACUS_code_repository}, which has been established and validated on CPU platforms and successfully applied to a variety of real-time simulations, including optical response, ultrafast charge dynamics, and coupled electron-ion dynamics \cite{acs.jpclett.1c00772,acs.jpclett.1c03442,PhysRevResearch.6.013123,acs.jctc.5c00111}. In this work, we further introduce a hardware-agnostic heterogeneous computing framework for ABACUS.
Central to this framework is the \texttt{Tensor} abstraction, a unified data container designed for high-dimensional arrays such as Hamiltonian and overlap matrices. By encapsulating memory management and linear algebra operations within this abstraction, we decouple the physical algorithms from hardware specifics, allowing a single codebase to execute efficiently across CPUs and GPUs. Based on this infrastructure, we present a fully heterogeneous implementation of RT-TDDFT using NAOs. A key technical contribution of this work is the implementation of a specialized GPU-accelerated spherical grid integration scheme. This scheme effectively overcomes the computational bottleneck arising from the position-dependent phase factors in the velocity gauge, making the formerly expensive gauge choice numerically efficient.

We rigorously validate the implementation across two primary regimes: pure electron dynamics under external fields and coupled electron-ion Ehrenfest dynamics. For the former, we compute optical absorption spectra and frequency-dependent dielectric functions for a hierarchy of systems: the finite anthracene molecule (C\textsubscript{14}H\textsubscript{10}) and (CdSe)\textsubscript{6} nanocluster, the one-dimensional hydrogen chain, the two-dimensional \textit{h}-BN monolayer, and the three-dimensional bulk silicon crystal. In all cases, our results exhibit excellent consistency with established benchmarks. For the latter, we demonstrate robust total energy conservation during both adiabatic (NaCl vibration) and highly nonadiabatic (torsional rotation of \protect\ce{H2C=NH2+}) processes. Performance evaluations on bulk silicon systems demonstrate that a single NVIDIA Ampere-architecture GPU achieves a wall-clock speedup of $3\times$--$4\times$ compared to a fully utilized dual-socket 56-core Intel Ice Lake CPU node. Notably, the specialized spherical grid integration kernel delivers a speedup of over $12\times$ compared to the optimized CPU baseline, effectively eliminating the algorithmic penalty of the velocity gauge. Furthermore, multi-GPU strong scaling tests show a parallel efficiency of approximately 76\% when scaling from 16 to 40 GPUs for systems containing up to 1728 atoms, confirming the code's capability for large-scale, long-timescale simulations.

The paper is organized as follows. In Sec.~\ref{sec:theory}, we outline the theoretical foundations of the RT-TDDFT implementation in ABACUS, detailing the real-time propagation schemes, the theoretical implications of different gauge choices (length, velocity, and hybrid), and Ehrenfest dynamics for ion-electron coupling within the NAO framework. Sec.~\ref{sec:methodology} describes the methodology and software architecture, introducing the unified \texttt{Tensor} abstraction, portable linear algebra interfaces, and the specific acceleration strategies employed for both uniform and spherical grid integrations. In Sec.~\ref{sec:results}, we present a comprehensive validation of our approach. We first demonstrate the physical accuracy of the implementation by assessing both fixed-ion optical properties and energy-conserving Ehrenfest dynamics, and then provide a detailed performance analysis, including system-size scaling and multi-GPU strong scaling benchmarks. Finally, we summarize our findings and discuss future directions in Sec.~\ref{sec:Conclusion}. Unless otherwise specified, all descriptions regarding the theoretical implementation and numerical benchmarks presented in this work are based on ABACUS version v3.9.0.24.

\section{Theory and Computational Approach}\label{sec:theory}

\subsection{Real-Time Propagation}

\begin{figure*}[t]
    \centering
    \includegraphics[width=0.9\textwidth]{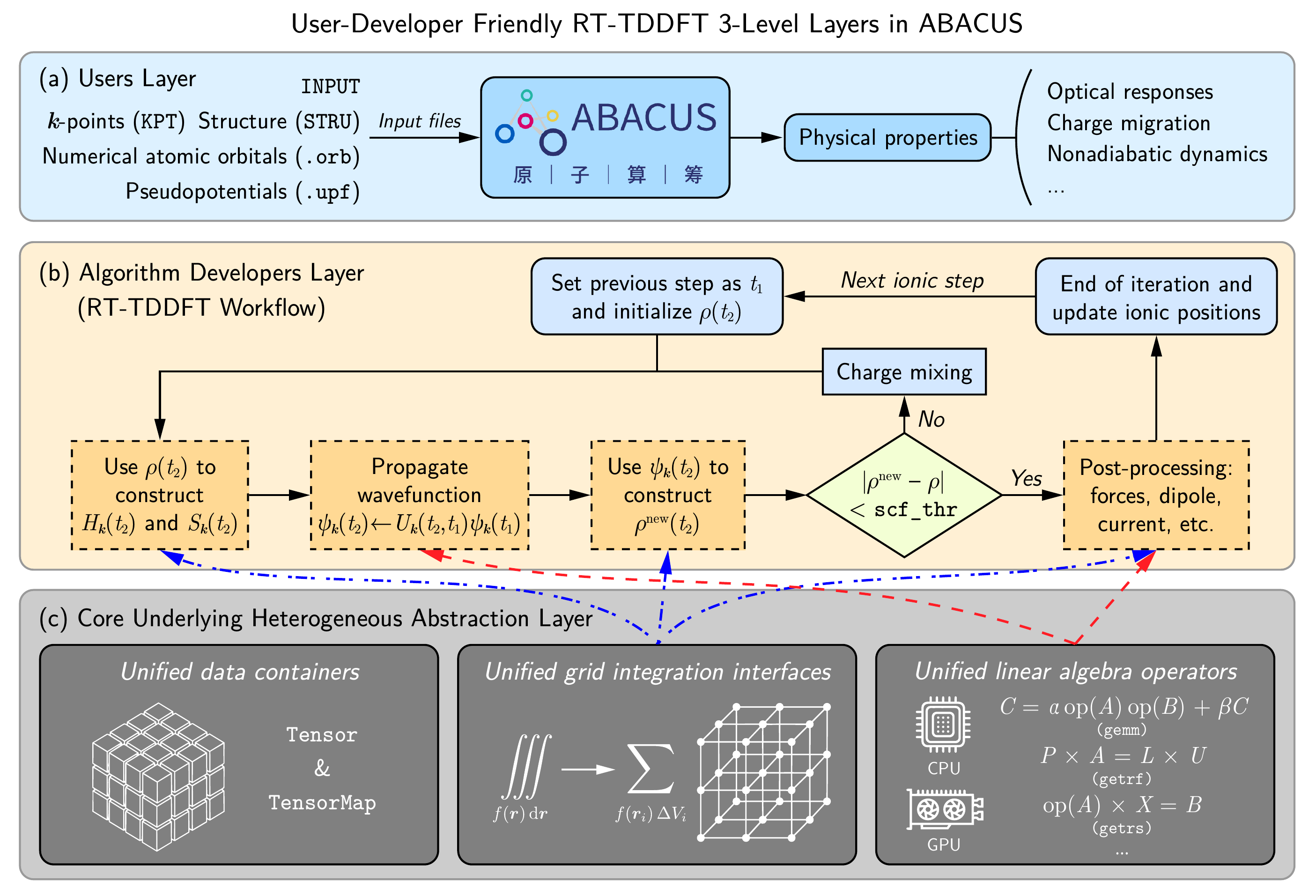}
    \caption{Schematic architecture of the heterogeneous RT-TDDFT implementation in ABACUS, organized into three logical layers: (a) The \textit{User Layer} handles standard inputs (e.g., structure, numerical atomic orbitals, pseudopotentials) and outputs physical properties such as optical responses. (b) The \textit{Algorithm Developer Layer} illustrates the main RT-TDDFT workflow, including the self-consistent time-evolution loop, Hamiltonian construction, wavefunction propagation, and Ehrenfest dynamics updates. (c) The \textit{Core Underlying Heterogeneous Abstraction Layer} underpins the simulation with unified data containers, grid integration interfaces, and portable linear algebra operators, bridging the physics algorithms with diverse hardware backends.}
    \label{fig:TDDFT_flowchart}
\end{figure*}

The overall RT-TDDFT algorithm in ABACUS is summarized in Fig.~\ref{fig:TDDFT_flowchart}. We first describe the real-time propagation scheme used in our ABACUS implementation. The real-space numerical atomic orbital $\phi_{\mu}(\bm{r})$ is decomposed into radial and angular (real spherical harmonics) \cite{Chen_2010,j.commatsci.2015.07.004,SIESTA_2002}:
\begin{equation}
    \phi_{\mu}(\bm{r}) = f_{Il\zeta}(r)\mathrm{Y}_{lm}(\hat{\bm{r}}),
\end{equation}
where $\mu = \{I, l, m, \zeta\}$ denotes orbital parameters (atom index $I$, angular momentum $l$, magnetic number $m$, and the multiplicity of the radial functions $\zeta$ for a given $l$). Under periodic boundary conditions, Kohn-Sham eigenfunctions are expanded in Bloch-like \cite{Bloch1929} NAO basis as
\begin{equation}
    \psi_{n\bm{k}}(\bm{r}) = \sum_{\bm{R}} \sum_\mu c_{n\mu,\bm{k}} \mathrm{e}^{\mathrm{i}\bm{k}\cdot\bm{R}} \phi_\mu(\bm{r} - \bm{\tau}_I - \bm{R}),
\end{equation}
where $\bm{R}$ labels lattice vectors, and $\bm{\tau}_I$ is the atomic position in the unit cell $\bm{R}$. This formulation preserves locality while enabling $\bm{k}$-point sampling.

The time evolution of electronic states follows the time-dependent Kohn-Sham (TDKS) equation \cite{PhysRevLett.52.997,PhysRevLett.82.3863}:
\begin{equation}
    H\psi_{n\bm{k}}(\bm{r},t)=\mathrm{i}\frac{\partial}{\partial t}\psi_{n\bm{k}}(\bm{r},t),
\end{equation}
where $H$ is the TDKS Hamiltonian, $\psi_{n\bm{k}}$ represents the time-dependent wavefunctions with band index $n$ and crystal momentum $\bm{k}$, while $\bm{r}$ and $t$ denote the spatial coordinates and time, respectively.

Projecting the TDKS equation onto the NAO basis yields the following matrix equation
\begin{equation}\label{eq:TDKS_matrix}
    S_{\bm{k}}^{-1}H_{\bm{k}} C_{\bm{k}}(t) = \mathrm{i} \frac{\partial}{\partial t} C_{\bm{k}}(t),
\end{equation}
where
\begin{equation}
    \begin{aligned}
        H_{\mu\nu}(\bm{k}) &= \sum_{\bm{R}} \mathrm{e}^{\mathrm{i}\bm{k}\cdot\bm{R}} \Braket{\phi_\mu(\bm{r}-\bm{\tau}_I)|H|\phi_\nu(\bm{r}-\bm{\tau}_J-\bm{R})} \\
        &\equiv \sum_{\bm{R}} \mathrm{e}^{\mathrm{i}\bm{k}\cdot\bm{R}} H_{\mu\nu}(\bm{R})
    \end{aligned}
\end{equation}
is the matrix element of $\bm{k}$-space Hamiltonian $H_{\bm{k}}$, and
\begin{equation}
    \begin{aligned}
        S_{\mu\nu}(\bm{k}) &= \sum_{\bm{R}} \mathrm{e}^{\mathrm{i}\bm{k}\cdot\bm{R}} \Braket{\phi_\mu(\bm{r}-\bm{\tau}_I)|\phi_\nu(\bm{r}-\bm{\tau}_J-\bm{R})} \\
        &\equiv \sum_{\bm{R}} \mathrm{e}^{\mathrm{i}\bm{k}\cdot\bm{R}} S_{\mu\nu}(\bm{R})
    \end{aligned}
\end{equation}
is the matrix element of $\bm{k}$-space overlap matrix $S_{\bm{k}}$. Here, the indices $\mu$ and $\nu$ are associated with atoms $I$ and $J$ at positions $\bm{\tau}_I$ and $\bm{\tau}_J$ within their respective unit cells, and $\bm{R}$ is the lattice vector connecting the unit cell of $\mu$ to that of $\nu$. $C_{\bm{k}}(t)$ is the column vector of coefficients $\{c_{n\mu,\bm{k}}(t)\}$.

The non-orthogonality of atomic orbitals necessitates explicit inclusion of the overlap matrix $S$ in the evolution equations. Here, the composite operator $S^{-1}H$ effectively assumes the role of a conventional Hamiltonian matrix. For a general time-dependent Hamiltonian $H_{\bm{k}}(t)$, the time-evolution operator \cite{PhysRev.75.486} in the NAO basis is formally expressed as
\begin{equation}
    U_{\bm{k}}(t_2,t_1) = \mathsf{T}\exp\!\left[-\mathrm{i}\int_{t_1}^{t_2} S_{\bm{k}}^{-1}H_{\bm{k}}(t')\,\mathrm{d}t'\right],
\end{equation}
where $\mathsf{T}$ denotes the time-ordering operator, and $U_{\bm{k}}(t_2,t_1)$ describes the propagator from the initial time $t_1$ to the final time $t_2$. For sufficiently small time steps, this operator simplifies to an exponential form, so the wavefunction at $t_2$ is then propagated from $t_1$ as
\begin{equation}\label{eq:TDKS_propagation}
    \begin{aligned}
        \psi_{n\bm{k}}(\bm{r},t_2) 
        &= U_{\bm{k}}(t_2,t_1) \psi_{n\bm{k}}(\bm{r},t_1) \\
        &\approx \exp\!\left[-\mathrm{i} S_{\bm{k}}^{-1} H_{\bm{k}}(t') \Delta t\right] \psi_{n\bm{k}}(\bm{r},t_1),
    \end{aligned}
\end{equation}
where $t' = (t_1 + t_2)/2$ represents the midpoint time, and $\Delta t = t_2 - t_1$ is the step size.

While $S_{\bm{k}}=S_{\bm{k}}(t')$ can be treated within the Born-Oppenheimer approximation \cite{Born_Oppenheimer_1927}, the electronic Hamiltonian $H_{\bm{k}}(t)$ undergoes rapid changes. Relevant studies demonstrate near-linear Hamiltonian evolution within $\Delta t \leqslant 0.5\ \mathrm{fs}$, making the midpoint approximation
\begin{equation}
    H_{\bm{k}}(t') \approx \frac{1}{2}\left[H_{\bm{k}}(t_1) + H_{\bm{k}}(t_2)\right]
\end{equation}
numerically applicable \cite{PhysRevB.87.205117,PhysRevLett.114.063004}. To maintain implementation consistency, the midpoint approximation is also applied to $S_{\bm{k}}$; however, practical tests indicate that the specific choice of the time point yields minor differences in the results.

In terms of numerical implementation of the propagator, the exponential operator cannot be computed exactly. ABACUS provides three approximation schemes for the time-evolution operator $U_{\bm{k}}$. The first one is the Crank-Nicolson method \cite{Crank_Nicolson_1947}:
\begin{equation}
    U_{\bm{k}} \approx \frac{S_{\bm{k}}-\mathrm{i}  H_{\bm{k}}(t') \Delta t / 2}{S_{\bm{k}}+\mathrm{i}  H_{\bm{k}}(t') \Delta t / 2}.
\end{equation}
The second one is the 4th-order Taylor expansion \cite{jcp.1.1774980}:
\begin{align}
    U_{\bm{k}} &\approx I + \hat{A} + \frac{1}{2}\hat{A}^2 + \frac{1}{6}\hat{A}^3 + \frac{1}{24}\hat{A}^4, \\
    \hat{A} &= -\mathrm{i} S_{\bm{k}}^{-1} H_{\bm{k}}(t')\Delta t.
\end{align}
The last one is the Enforced Time-Reversal Symmetry (ETRS) \cite{jcp.1.1774980} method:
\begin{equation}
    U_{\bm{k}} \approx \exp\!\left[-\mathrm{i} S_{\bm{k}}^{-1} H_{\bm{k}}(t+\Delta t)\frac{\Delta t}{2}\right] \exp\!\left[-\mathrm{i} S_{\bm{k}}^{-1} H_{\bm{k}}(t)\frac{\Delta t}{2}\right],
\end{equation}
in which the exponential operator is computed using the 4th-order Taylor expansion. Although these three schemes construct the evolution operator differently, their global time-stepping accuracy is universally restricted to $\mathcal{O}(\Delta t^2)$ due to either the midpoint approximation or the operator splitting error. Among them, the Crank-Nicolson method is chosen as the default propagator in ABACUS, as its mathematically symmetric form strictly preserves the unitarity of time evolution, thereby guaranteeing superior charge conservation and numerical stability for long-time simulations.

It is worth noting that the midpoint approximation introduces an implicit coupling in the propagation. Since the midpoint Hamiltonian $H_{\bm{k}}(t')$ requires the knowledge of $H_{\bm{k}}(t_2)$, the propagator $U_{\bm{k}}$ implicitly depends on the future electronic state. Specifically, the Hamiltonian $H_{\bm{k}}(t_2)$ is a functional of the electron density $\rho(\bm{r},t_2)$, which in turn is uniquely determined by the target wavefunctions $\psi_{n\bm{k}}(\bm{r},t_2)$ at the end of the time step. Consequently, Eq.~\eqref{eq:TDKS_propagation} forms a self-consistent loop where the unknown wavefunctions $\psi_{n\bm{k}}(\bm{r},t_2)$ on the left-hand side are required to construct the propagator on the right-hand side. To resolve this implicit dependence and ensure physical convergence, a self-consistent procedure must be performed at each individual time step.

\subsection{Selection of Gauges}

In the description of light-matter interactions within the RT-TDDFT formalism, the external interaction is introduced as a classical electromagnetic field. As is well known in electrodynamics, the electromagnetic potentials possess a gauge freedom: different choices of scalar and vector potentials that correspond to the same physical fields lead to equivalent physical predictions, provided the theory is implemented exactly. However, in practical numerical simulations, the choice of gauge can have a profound impact on both accuracy and feasibility.

\subsubsection{Length Gauge}\label{sec:Length_Gauge}

The most straightforward choice is the length gauge \cite{PhysRevB.54.4484}, where the interaction with a time-dependent electric field $\bm{E}(t)$ is introduced via the dipole coupling term $\bm{E}(t) \cdot \bm{r}$ in the Hamiltonian:
\begin{equation}
    H_{\text{len}}(t) = H_0 + \bm{E}(t) \cdot \bm{r}.
\end{equation}
Here, $H_0$ denotes the field-free Kohn-Sham Hamiltonian. In this gauge, the time-dependent dipole moment $\bm{P}(t)$ is directly accessible and serves as the primary response quantity for finite systems:
\begin{equation}
    \bm{P}(t) = \int \rho(\bm{r}, t)\bm{r} \, \mathrm{d}\bm{r},
\end{equation}
The optical absorption spectrum is then obtained from the Fourier transform of $\bm{P}(t)$. However, the operator $\bm{r}$ is not compatible with periodic boundary conditions. In fact, it breaks translational symmetry because $\bm{r}$ is not a periodic function. Although one may attempt to enforce periodicity by introducing artificial opposite electric fields at the cell boundaries, such constructions are non-physical and generally fail for extended solids. Consequently, the length gauge is restricted to finite or surface-slab systems, where the dipole moment is well-defined.

\subsubsection{Velocity Gauge}\label{sec:Velocity_Gauge}

For periodic systems, the velocity gauge \cite{PhysRevB.62.7998} is traditionally adopted. Here, the electric field is expressed through a time-dependent vector potential $\bm{A}(t)$, related by $\bm{E}(t) = -\partial_t \bm{A}(t)$. The Hamiltonian becomes
\begin{equation}
    H_{\text{vel}}(t) = \frac{1}{2} \left[ -\mathrm{i}\nabla + \bm{A}(t) \right]^2 + V_{\text{H}} + V_{\text{xc}} + V_{\text{ps}},
\end{equation}
where $V_{\text{H}}$, $V_{\text{xc}}$ and $V_{\text{ps}}$ are the Hartree, exchange-correlation and pseudopotential terms, respectively. This formulation preserves the periodicity of the crystal lattice. In this gauge, the natural response quantity is the current density $\bm{J}(t)$, given by
\begin{align}
    \bm{J}(t) &= -\frac{1}{\Omega N_k} \sum_{n\bm{k}} f_{n\bm{k}} \operatorname{Re} \Braket{\psi_{n\bm{k}}(t) | \bm{\pi} | \psi_{n\bm{k}}(t)}, \label{eq:current_density}\\
    \bm{\pi} &= -\mathrm{i}\nabla + \bm{A}(t) + \mathrm{i}\left[\widetilde{V}_{\text{NL}}, \bm{r}\right],
\end{align}
where $\Omega$ is the unit-cell volume, $N_k$ the number of $\bm{k}$-points, and $\widetilde{V}_{\text{NL}}$ the nonlocal pseudopotential. The dielectric function is then constructed from the Fourier-transformed current.

However, implementing the velocity gauge in the NAO basis introduces a critical computational challenge. Under the gauge transformation, the nonlocal pseudopotential operator acquires a position-dependent phase factor \cite{PhysRevLett.87.087402}:
\begin{align}
    \widetilde{V}_{\text{NL}} &= \mathrm{e}^{-\mathrm{i}\bm{A}(t)\cdot\bm{r}} V_{\text{NL}} \mathrm{e}^{\mathrm{i}\bm{A}(t)\cdot\bm{r}}, \\
    V_{\text{NL}} &= \sum_{Ilm\zeta} D_{Il\zeta} \Ket{\beta_{Ilm\zeta}} \Bra{\beta_{Ilm\zeta}},
\end{align}
which modifies the matrix elements between NAOs and nonlocal pseudopotential projector functions $\beta_{Ilm\zeta}$ as
\begin{equation}
    \begin{aligned}
        \Braket{\phi_\mu | V_{\text{NL}} | \phi_\nu} \to
        \sum_{Ilm\zeta} D_{Il\zeta} 
        &\Braket{\phi_\mu | \mathrm{e}^{-\mathrm{i}\bm{A}(t)\cdot\bm{r}} | \beta_{Ilm\zeta}} \\
        &\Braket{\beta_{Ilm\zeta} | \mathrm{e}^{\mathrm{i}\bm{A}(t)\cdot\bm{r}} | \phi_\nu}.
    \end{aligned}
\end{equation}
Crucially, the overlap integral $\Braket{\phi_\mu | \mathrm{e}^{-\mathrm{i}\bm{A}(t)\cdot\bm{r}} | \beta_{Ilm\zeta}}$ no longer possesses the two-center structure that enables efficient analytic evaluation. Instead, it must be computed numerically on the real-space integration grid, which is significantly more expensive.

In ABACUS, this term constitutes one of the computational bottlenecks in velocity-gauge RT-TDDFT simulations with NAOs. To address this, our heterogeneous framework fully offloads the evaluation of these projector overlaps to the GPU.

\subsubsection{Hybrid Gauge}\label{sec:Hybrid_Gauge}

Despite its formal suitability for periodic systems, the velocity gauge suffers from a critical flaw in NAO-based codes: the basis functions $\phi_\mu(\bm{r})$ are centered on atoms and do not carry the global phase factor $\mathrm{e}^{-\mathrm{i}\bm{A}(t)\cdot\bm{r}}$ required by the exact gauge transformation. This omission leads to computational errors in the current, particularly manifesting as an unphysical low-frequency divergence in the dielectric function, as we are about to demonstrate in Sec.~\ref{sec:bulk_Si}.

To resolve this issue while retaining compatibility with localized bases, Zhao and He recently proposed a hybrid gauge approach that incorporates a local phase factor relative to each atomic center $\bm{\tau}_\mu$ \cite{acs.jctc.5c00111}:
\begin{equation}
    \phi_\mu(\bm{r}) \to \mathrm{e}^{-\mathrm{i}\bm{A}(t)\cdot(\bm{r} - \bm{\tau}_\mu)} \phi_\mu(\bm{r}).
\end{equation}
This yields a Hamiltonian matrix element that combines features of both gauges:
\begin{equation}
    H_{\mu\nu}^{\text{hyb}}(t) = \mathrm{e}^{-\mathrm{i}\bm{A}(t)\cdot(\bm{\tau}_\mu - \bm{\tau}_\nu)} \Braket{\phi_\mu | H_0 + \bm{E}(t) \cdot (\bm{r} - \bm{\tau}_\nu) | \phi_\nu},
\end{equation}
where $H_0$ is the field-free Hamiltonian. The displacement $\bm{r} - \bm{\tau}_\nu$ ensures the integral remains bounded and periodic, while the phase factor restores gauge consistency. An additional advantage of the hybrid gauge is that the nonlocal pseudopotential terms can still be evaluated efficiently through two-center integrals, without requiring grid-based integration. However, this comes at the cost of more complex expressions, which can make the implementation of advanced post-processing features (such as the derivation and coding of atomic forces) more challenging.

Our GPU implementation is also compatible with this hybrid gauge scheme, empowering users to select the most appropriate gauge based on their specific research needs.

\subsection{Ehrenfest Dynamics}

In RT-TDDFT simulations, the ionic degrees of freedom are typically treated classically while electrons evolve quantum-mechanically. To model electron-ion dynamics, we adopt the Ehrenfest approach \cite{Ehrenfest1927}, in which ions move on a mean-field potential energy surface determined by the instantaneous electronic state. This necessitates the accurate and efficient evaluation of atomic forces at every time step.

Unlike plane-wave methods where basis functions are independent of atomic coordinates, NAOs move with their host atoms. This explicit time-dependence of the basis functions introduces a ``moving spatial gauge'' during the time evolution. As demonstrated by Ojanperä \textit{et al.} \cite{jcp.1.3700800}, neglecting the basis function velocities leads to a violation of the norm of the electronic states and causes significant fluctuations in total energy conservation during nonadiabatic processes. To rigorously preserve these conservation laws, an additional matrix $P$ must be incorporated into the TDKS equation. Specifically, the Hamiltonian $H_{\bm{k}}$ in the evolution equation Eq.~\eqref{eq:TDKS_matrix} is effectively replaced by $H_{\bm{k}} + P_{\bm{k}}$:
\begin{equation}\label{eq:TDKS_Ehrenfest}
    S_{\bm{k}}^{-1}(H_{\bm{k}} + P_{\bm{k}}) C_{\bm{k}}(t) = \mathrm{i} \frac{\partial}{\partial t} C_{\bm{k}}(t),
\end{equation} 
where the matrix elements of $P_{\bm{k}}$ depend on the instantaneous atomic velocities $\dot{\bm{\tau}}_K$:
\begin{equation}
    P_{\mu\nu}(\bm{k}) = -\mathrm{i} \sum_{K} \dot{\bm{\tau}}_K \cdot \bm{D}_{K,\mu\nu}(\bm{k}).
\end{equation}
Here, $K$ is the atom index, and $\bm{D}_{K,\mu\nu}(\bm{k})$ represents the derivative of the basis functions with respect to the atomic coordinates:
\begin{equation}
    \bm{D}_{K,\mu\nu}(\bm{k}) = \sum_{\bm{R}} \mathrm{e}^{\mathrm{i}\bm{k}\cdot\bm{R}} \Braket{\phi_{\mu 0} | \frac{\partial \phi_{\nu\bm{R}}}{\partial \bm{\tau}_K}}.
\end{equation}
For notational simplicity in the expressions, we define $\phi_{\mu 0} \equiv \phi_\mu(\bm{r}-\bm{\tau}_I)$ and $\phi_{\nu\bm{R}} \equiv \phi_\nu(\bm{r}-\bm{\tau}_J-\bm{R})$, where $\bm{\tau}_I$ and $\bm{\tau}_J$ are the positions of the atoms on which the orbitals $\mu$ and $\nu$ are centered, respectively, and $\bm{R}$ is the lattice vector.

Although this $P$ matrix directly modifies the electronic propagation, its physical origin is fundamentally tied to the ionic kinematics in Ehrenfest dynamics, making its evaluation a crucial bridge between the electronic and ionic subsystems.

Furthermore, the moving basis introduces additional terms to the atomic forces beyond the standard Hellmann-Feynman \cite{Hellmann1937,PhysRev.56.340} contribution. The ion-electron interaction is described by norm-conserving pseudopotentials \cite{PhysRevLett.43.1494}, which are typically decomposed into a local potential and a separable fully non-local Kleinman-Bylander projection part \cite{PhysRevLett.48.1425}. The total energy-conserving force $\bm{F}^{\mathrm{EC}}_I$ acting on atom $I$ can be decomposed into four physically distinct components:
\begin{equation}
    \bm{F}^{\mathrm{EC}}_I = \bm{F}^{\mathrm{HF}}_I + \bm{F}^{\mathrm{Pulay}}_I + \bm{F}^{\mathrm{ortho}}_I + \bm{F}^{\mathrm{Ewald}}_I,
\end{equation}
where $\bm{F}^{\mathrm{HF}}_I$ is the Hellmann-Feynman force arising from explicit dependence of the Hamiltonian (specifically the local and non-local pseudopotentials) on ionic positions, $\bm{F}^{\mathrm{Pulay}}_I$ accounts for the implicit dependence through the motion of the basis orbitals, $\bm{F}^{\mathrm{ortho}}_I$ stems from the nonorthogonality of the NAO basis (often interpreted as a generalized Pulay correction), and $\bm{F}^{\mathrm{Ewald}}_I$ is the classical electrostatic force between ions, computed via the Ewald summation techniques \cite{Ewald1921}.

It should be noted that the following expressions for these force components are formulated within the length gauge. When employing other gauges, such as the velocity or hybrid gauges, additional terms arise in the force expression due to the explicit time-dependence of the gauge transformation and its effect on the quantum operators. A rigorous derivation of these gauge-dependent corrections is beyond the scope of the present work.

The explicit expressions for each term are as follows. First, the Hellmann-Feynman force is
\begin{equation}
    \bm{F}^{\mathrm{HF}}_{I} = - \sum_{\bm{R}} \sum_{\mu\nu} \rho_{\nu\mu}(\bm{R}) \Braket{\phi_{\mu 0} | \frac{\partial H}{\partial \bm{\tau}_{I}} | \phi_{\nu\bm{R}}},
\end{equation}
where $\rho_{\nu\mu}(\bm{R})$ represents the real-space density matrix element. It is constructed from the orbital coefficients $c_{n\mu,\bm{k}}$ via the Fourier transform
\begin{equation}
    \begin{aligned}
        \rho_{\mu\nu}(\bm{R}) &= \sum_{\bm{k}} \left[\frac{1}{N_k} \sum_{n} f_{n\bm{k}} c_{n\mu,\bm{k}} c_{n\nu,\bm{k}}^{*} \right] \mathrm{e}^{\mathrm{i}\bm{k}\cdot\bm{R}} \\
        &\equiv \sum_{\bm{k}} \rho_{\mu\nu}(\bm{k}) \mathrm{e}^{\mathrm{i}\bm{k}\cdot\bm{R}},
    \end{aligned}
\end{equation}
with $f_{n\bm{k}}$ denoting the occupation numbers and $N_k$ the number of $\bm{k}$-points used in the Brillouin zone sampling. The derivative of the Hamiltonian includes contributions from both local and nonlocal potentials. The local part is evaluated efficiently in reciprocal space \cite{PhysRevB.83.195102} using FFTs \cite{FFT1965}, while the nonlocal part is computed via two-center integrals over the projector functions and atomic orbitals.
Second, the Pulay force term is \cite{Pulay1969}
\begin{equation}
    \begin{aligned}
        \bm{F}^{\mathrm{Pulay}}_{I} &= - \sum_{\bm{R}} \sum_{\mu\nu} \rho_{\nu\mu}(\bm{R}) \\
        &\quad \left[ \Braket{\frac{\partial \phi_{\mu 0}}{\partial \bm{\tau}_{I}} | H | \phi_{\nu\bm{R}}} + \Braket{\phi_{\mu 0} | H | \frac{\partial \phi_{\nu\bm{R}}}{\partial \bm{\tau}_{I}}} \right],
    \end{aligned}
\end{equation}
which arises because the basis functions themselves depend on atomic positions. The derivatives of the NAOs are computed analytically for the angular part (real spherical harmonics) and numerically for the radial part.
Third, the nonorthogonality force term is \cite{SIESTA_2002,PhysRevB.51.1456}
\begin{equation}
    \bm{F}^{\mathrm{ortho}}_{I} = - \sum_{\bm{R}} \sum_{\mu\nu} E_{\mu\nu}(\bm{R}) \frac{\partial S_{\nu\mu}(\bm{R})}{\partial \bm{\tau}_{I}},
\end{equation}
where $E_{\mu\nu}(\bm{R})$ is the energy-density matrix (EDM) in real space, and $S_{\nu\mu}(\bm{R})$ is the overlap matrix. This term captures the response of the electronic energy to changes in the nonorthogonal basis metric.
Last, the Ewald force term is \cite{Perram1988}
\begin{equation}
    \bm{F}^{\mathrm{Ewald}}_{I} = - \frac{\partial E^{\mathrm{Ewald}}}{\partial \bm{\tau}_{I}},
\end{equation}
which is computed using standard Ewald summation techniques and is independent of the electronic information.

A critical distinction in RT-TDDFT, compared to conventional KSDFT, lies in the evaluation of the energy-density matrix $E_{\mu\nu}$. In ground-state calculations, the EDM element at a given $\bm{k}$-point is constructed from the occupied Kohn-Sham eigenstates as \cite{SIESTA_2002}
\begin{equation}
    E_{\mu\nu}(\bm{k}) = \frac{1}{N_k} \sum_{n} f_{n\bm{k}} \varepsilon_{n\bm{k}} c_{n\mu,\bm{k}} c_{n\nu,\bm{k}}^{*},
\end{equation}
where $\varepsilon_{n\bm{k}}$ are the eigenvalues of the Hamiltonian. Then, the real-space EDM is obtained via a Fourier transform:
\begin{equation}
    E_{\mu\nu}(\bm{R}) = \sum_{\bm{k}} E_{\mu\nu}(\bm{k}) \mathrm{e}^{\mathrm{i}\bm{k}\cdot\bm{R}}.
\end{equation}
However, in RT-TDDFT, the time-dependent wavefunctions $\psi_{n\bm{k}}(t)$ are generally not eigenstates of the instantaneous Hamiltonian $H_{\bm{k}}(t)$. Consequently, the above spectral decomposition is no longer valid.
Instead, the EDM must be computed directly from the fundamental definition (in matrix form) \cite{SIESTA_2002}:
\begin{equation}\label{eq:EDM_definition}
    E_{\bm{k}}(t) = \frac{1}{2} \left( S_{\bm{k}}^{-1} H_{\bm{k}} \rho_{\bm{k}} + \rho_{\bm{k}} H_{\bm{k}} S_{\bm{k}}^{-1} \right).
\end{equation}
This expression ensures that the force remains consistent with the variational principle even for non-stationary states. As demonstrated by Ojanperä \textit{et al.} \cite{jcp.1.3700800}, relying on the ground-state approximation, known as the incomplete basis set corrected (IBSC) force, for nonadiabatic processes causes unsatisfactory total energy drifts. Therefore, evaluating the exact EDM via the equation above is essential for numerical stability. Because this computation involves explicit matrix multiplications and inversions rather than a simple sum over precomputed eigenvalues (which are obtained via the KSDFT SCF procedure), it constitutes a computational hotspot in RT-TDDFT workflows.
In the heterogeneous implementation, this EDM construction pipeline, along with all other FFT-based and grid-based force components (e.g., the local part of the Hellmann-Feynman and Pulay forces), has been fully GPU-accelerated.

It is also worth noting that a strictly rigorous derivation of Ehrenfest dynamics from the quantum-classical action principle yields an additional momentum-conserving force term $\bm{F}^{\mathrm{MC}}_I$ associated with the moving spatial gauge \cite{jcp.1.3700800}. Analogous to the energy-density matrix $E_{\mu\nu}(\bm{R})$ used for the nonorthogonality force, we define a velocity-density matrix (VDM) to capture the nonlocal velocity coupling. In $\bm{k}$-space, the VDM is constructed as
\begin{equation}
    \Upsilon_{\bm{k}}(t) = S_{\bm{k}}^{-1} \left( \sum_K \dot{\bm{\tau}}_K \cdot \bm{D}_K(\bm{k}) \right) \rho_{\bm{k}} = \mathrm{i} S_{\bm{k}}^{-1} P_{\bm{k}} \rho_{\bm{k}},
\end{equation}
which is then Fourier transformed to real space:
\begin{equation}
    \Upsilon_{\mu\nu}(\bm{R}) = \sum_{\bm{k}} \Upsilon_{\mu\nu}(\bm{k}) \mathrm{e}^{\mathrm{i}\bm{k}\cdot\bm{R}}.
\end{equation}
By defining the velocity-gradient matrix element
\begin{equation}
    \bm{W}_{I, \mu\nu}(\bm{R}) = \dot{\bm{\tau}}_I \cdot \Braket{\frac{\partial \phi_{\mu 0}}{\partial \bm{\tau}_I} | \frac{\partial \phi_{\nu\bm{R}}}{\partial \bm{\tau}_I}},
\end{equation}
and reusing the spatial derivative matrix
\begin{equation}
    \bm{D}_{I, \mu\nu}(\bm{R}) = \Braket{\phi_{\mu 0} | \frac{\partial \phi_{\nu\bm{R}}}{\partial \bm{\tau}_I}},
\end{equation}
the momentum-conserving force acting on atom $I$ can be expressed as \cite{jcp.1.3700800,fhi_aims_tddft_tutorial}
\begin{equation}
    \begin{aligned}
        \bm{F}^{\mathrm{MC}}_I &= \mathrm{i} \sum_{\bm{R}} \sum_{\mu\nu} \Big\{ \rho_{\nu\mu}(\bm{R}) \left[ \bm{W}_{I, \nu\mu}^{*}(\bm{R}) - \bm{W}_{I, \mu\nu}(\bm{R}) \right] \\
        &\quad + \bm{D}_{I, \mu\nu}(\bm{R}) \Upsilon_{\mu\nu}^{*}(\bm{R}) - \bm{D}_{I, \mu\nu}^{*}(\bm{R}) \Upsilon_{\mu\nu}(\bm{R}) \Big\}.
    \end{aligned}
\end{equation}
Then the total force is given by
\begin{equation}
    \bm{F}_I = \bm{F}^{\mathrm{EC}}_I + \bm{F}^{\mathrm{MC}}_I.
\end{equation}

While $\bm{F}^{\mathrm{MC}}_I$ is formally required to guarantee macroscopic momentum balance, evaluating it entails calculating the velocity-density matrix $\Upsilon_{\mu\nu}(\bm{R})$ and executing a series of computationally intensive matrix multiplications. Consequently, its impact is often deemed marginal compared to the substantial computational cost, and it is frequently omitted in practical large-scale Ehrenfest MD implementations. Even Ojanperä \textit{et al.} \cite{jcp.1.3700800}, who explicitly derived these velocity-dependent correction terms, opted to use solely the energy-conserving force and deferred the implementation of the momentum-conserving terms to future work. Therefore, this specific momentum-conserving term is currently not implemented in ABACUS. Our benchmarks in Sec.~\ref{sec:Physical_Validation_Ehrenfest} indicate that the present implementation of the EC force already provides highly satisfactory total energy conservation for standard adiabatic and nonadiabatic simulations. The inclusion of the momentum-conserving term is being considered for future development to further extend the limits of long-term simulation stability.

Once the forces are computed, ionic trajectories are propagated using classical equations of motion. In the Ehrenfest scheme, the nuclear acceleration at time $t$ is given by
\begin{equation}
    M_I \ddot{\bm{\tau}}_I(t) = \bm{F}_I[\rho(\bm{r}, t)],
\end{equation}
where $M_I$ is the mass of the atom $I$, and the force depends parametrically on the time-dependent electron density. The Newton's equation of motion is then integrated using the standard velocity Verlet algorithm \cite{PhysRev.159.98,jcp.1.442716} in molecular dynamics simulations, which has been implemented in ABACUS.
This coupling allows for energy exchange between electronic and ionic subsystems, capturing phenomena such as laser-induced structural deformation or nonthermal melting \cite{PhysRevB.94.184310}. The integration of ionic motion is decoupled from the electronic propagation loop but synchronized at each time step, as illustrated in Fig.~\ref{fig:TDDFT_flowchart}. While the Ehrenfest method assumes a single mean-field trajectory and thus cannot describe branching or decoherence like the surface hopping method \cite{jcp.1.1675788,jcp.1.459170}, it remains a robust and computationally feasible choice for simulating ultrafast coherent dynamics in moderately nonadiabatic regimes.

An additional notice is that the implementation of the moving spatial gauge is currently an experimental feature and has not yet been merged into the main release as of version v3.9.0.24.

\section{Methodology and Implementation}\label{sec:methodology}

\subsection{Unified Data Containers}

A core component of our heterogeneous computing framework is the \texttt{Tensor} data container, a unified abstraction for multi-dimensional arrays that underpins all heterogeneous operations of RT-TDDFT implementation in ABACUS. \texttt{Tensor} serves as the general-purpose container for matrices and high-dimensional arrays such as Hamiltonians, overlap matrices and density matrices, as illustrated in the bottom panel of Fig.~\ref{fig:TDDFT_flowchart}(c).

Designed from first principles to support both CPU and accelerator-based execution without source-level specialization, \texttt{Tensor} encapsulates memory management, data layout, and device affinity within a single, consistent interface. This enables developers to write hardware-agnostic numerical kernels that operate identically across diverse platforms including NVIDIA GPUs, AMD GPUs, and Hygon DCUs, while retaining full control over precision and shape semantics.

The \texttt{Tensor} class is not a template but a runtime-polymorphic container parameterized by explicit enumerations for data type (e.g., single/double precision real or complex) and device type (CPU or GPU/DCU). Upon instantiation, it automatically allocates memory through a unified allocator system that dispatches to the appropriate backend (e.g., host \texttt{malloc} for CPUs, CUDA or HIP allocators for accelerators). 

Memory management follows the Resource Acquisition Is Initialization (RAII) principle: resources are acquired during construction and automatically released upon destruction, eliminating the need for manual deallocation or explicit host-device transfers. Internally, each \texttt{Tensor} holds a \texttt{TensorBuffer} object that encapsulates a backend-specific allocator. The buffer owns the allocated memory and, in its destructor, invokes the corresponding \texttt{free()} method of the allocator, ensuring correct deallocation on the appropriate device. The allocator itself is also owned by the buffer and deleted upon destruction, preventing resource leakage. This design guarantees exception-safe cleanup and abstracts away low-level pointer handling, significantly reducing the risk of memory leaks, dangling pointers, and platform-specific bugs common in traditional scientific codes that rely on raw pointer arithmetic.

Beyond basic storage, \texttt{Tensor} supports essential operations such as reshaping, slicing, deep copying, and zero initialization, all implemented in a device-transparent manner. For instance, a reshape operation validates the logical consistency of element counts and updates metadata without moving data, whereas a slice operation instantiates a new \texttt{Tensor} object and explicitly copies the data from the specified sub-region. These features facilitate flexible data manipulation required in software workflows, such as extracting sub-blocks of Hamiltonian or overlap matrices, or reorganizing coefficient arrays for batched linear algebra.

To bridge the gap between the unified container system and external data structures, particularly during the initialization and finalization phases of heterogeneous computation, we introduce \texttt{TensorMap}. Unlike \texttt{Tensor}, which owns its memory, \texttt{TensorMap} provides a non-owning ``view'' over an externally allocated buffer, mapping it into the same semantic framework with specified shape, data type, and device context. This allows seamless integration with legacy modules or third-party libraries that manage their own memory, enabling incremental adoption of the unified infrastructure without invasive refactoring. In practice, \texttt{TensorMap} is used at the entry and exit points of the heterogeneous pipeline to wrap raw pointers (e.g., from wavefunction initial guesses or output buffers) into tensor-like objects that can participate in standardized operations like synchronization or arithmetic.

The design of \texttt{Tensor} also anticipates future developments in computational science, particularly the convergence of electronic structure theory and artificial intelligence \cite{jcp.5.0297563}. Its interface and internal structure are compatible with modern tensor-centric programming paradigms, preserving the possibility of integrating advanced features such as automatic differentiation, capabilities that are increasingly central to next-generation scientific software. By decoupling algorithmic logic from hardware-specific memory mechanics, the \texttt{Tensor} abstraction enforces a clean separation of concerns: physicists and method developers focus on equations and numerical schemes, while the underlying runtime handles platform adaptation. This design directly supports the ``unified'' philosophy of our work: one codebase, multiple architectures, without sacrificing performance or expressiveness. Moreover, the architecture is extensible; future support for additional accelerators can be incorporated by adding new allocator and kernel backends, without modifying the high-level simulation logic.

\subsection{Unified Linear Algebra Operators}\label{sec:Linear_Algebra_Operators}

Building upon the foundation of the \texttt{Tensor} abstraction, our heterogeneous framework also provides a unified interface for linear algebra operations. This layer serves as the computational backbone for critical tasks in RT-TDDFT, particularly the construction and application of time-evolution operators including matrix multiplications and inversions, while ensuring consistent behavior across CPUs, GPUs, DCUs, and other accelerator architectures.

The design centers on a family of polymorphic operator templates that encapsulate standard dense linear algebra routines, including \texttt{axpy} (constant times a vector plus a vector), \texttt{gemm} (general matrix-matrix multiplication), \texttt{getrf} (LU factorization of a general matrix), and \texttt{getrs} (solving a linear system using a previously computed LU factorization), among others. Each operator is parameterized by data type (real or complex, single or double precision) and device context (CPU or accelerator), enabling compile-time dispatch to the appropriate backend: BLAS \cite{BLAS_2002} and LAPACK \cite{LAPACK_1999} on CPU, and cuBLAS and cuSOLVER or other architecture-dependent solvers on GPU, DCU or other accelerators (see Fig.~\ref{fig:TDDFT_flowchart}(c)). Crucially, the interface semantics, including input/output conventions, matrix storage layout (column-major), and error reporting, are preserved across platforms, guaranteeing numerical equivalence and simplifying validation.

In practice, these operators are invoked transparently within high-level physics routines. For instance, in the wavefunction propagation kernel \texttt{evolve\_k}, solving a linear system of the form 
\begin{equation}\label{eq:Crank_Nicolson}
    \left(S_{\bm{k}} + \frac{\mathrm{i}\Delta t}{2} H_{\bm{k}}\right) \psi_{n\bm{k}}(t+\Delta t) = \left(S_{\bm{k}} - \frac{\mathrm{i}\Delta t}{2} H_{\bm{k}}\right) \psi_{n\bm{k}}(t)
\end{equation}
is required for the Crank-Nicolson propagator, where $S_{\bm{k}}$ and $H_{\bm{k}}$ are the overlap and Hamiltonian matrices, respectively. The implementation directly solves this linear system for the wavefunctions $\psi_{n\bm{k}}(t+\Delta t)$. Specifically, it constructs the coefficient matrix $A_{\bm{k}} = S_{\bm{k}} + \mathrm{i}H_{\bm{k}}\Delta t/2$, factorizes it via \texttt{getrf}, and then performs \texttt{getrs} to obtain the updated states. This process is carried out using \texttt{Tensor}-backed data containers, ensuring that the surrounding physics logic remains entirely platform-independent, as the same operator interface is utilized regardless of the execution target.

Beyond dense linear solvers, ABACUS also supports high-level tensor operations such as the Einstein summation convention for arbitrary tensor contractions. While not required in the current RT-TDDFT implementation, such capabilities are essential for more advanced many-body methods, such as \textit{GW} approximation \cite{PhysRev.139.A796,PhysRevB.34.5390} for quasiparticle corrections.

Importantly, the linear algebra layer integrates seamlessly with the \texttt{Tensor} ecosystem. All operands are \texttt{Tensor} objects, enabling automatic device placement and RAII-based memory lifecycle. This tight coupling allows complex sequences, such as matrix assembly, factorization, inversion, and wavefunction update, to be expressed as a coherent pipeline of tensor operations, with hardware-specific details confined to the operator implementations.

By unifying linear algebra under a single, expressive interface, our design achieves three core objectives. First, it decouples physical algorithms from hardware-specific programming models. Second, it ensures numerical consistency across execution environments. Third, it empowers domain scientists to develop and validate methods without deep expertise in GPU programming. Such an abstraction is vital for ensuring that scientific software remains sustainable, portable, and maintainable amid the rapid rise of heterogeneous computing and AI for science.

\begin{figure*}[t]
    \centering
    \includegraphics[width=0.96\textwidth]{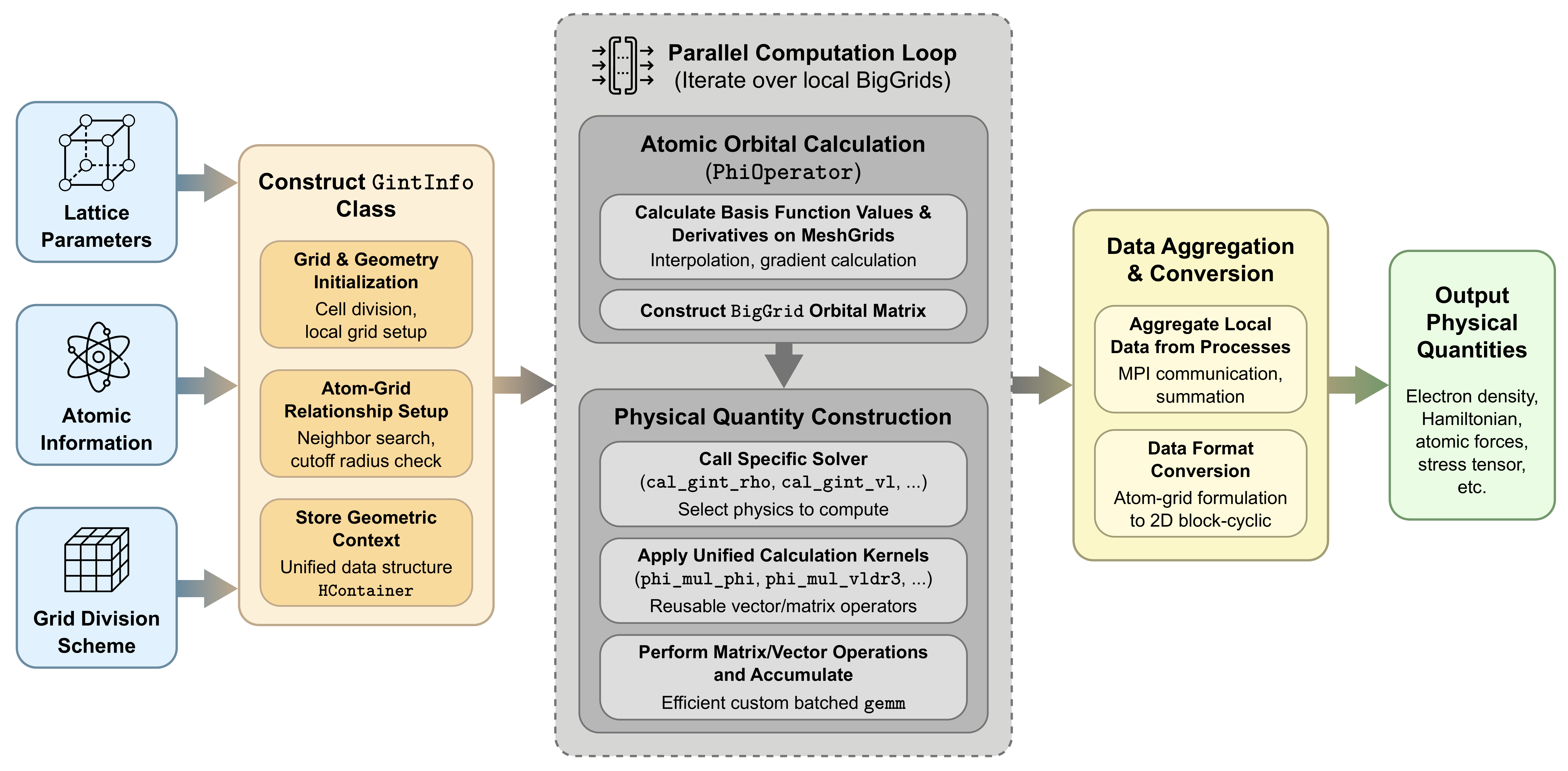}
    \caption{Data flow and architectural diagram of the \texttt{Gint} module. The workflow begins with the initialization of the \texttt{GintInfo} class (green block), which manages geometry, grid division, and atom-grid neighbor relationships. The core computation iterates over local grid blocks (\texttt{BigGrid}), utilizing the \texttt{PhiOperator} to evaluate atomic orbitals and unified calculation kernels to construct physical quantities such as density and Hamiltonian matrices. Finally, data is aggregated via MPI and converted to the target storage format.}
    \label{fig:Gint_flowchart}
\end{figure*}

\subsection{Unified Grid Integration Interfaces}

\subsubsection{Uniform Grid Integration for Usual Quantities}\label{sec:Uniform_Grid_Integration}

The evaluation of usual physical quantities defined on real-space grids, such as the charge density $\rho(\bm{r})$, the local potential $V^{\text{loc}}_{\mu\nu}$, and certain components of the atomic forces (e.g., the Pulay force term), constitutes one of the most computationally expensive parts of NAO-based DFT. To address this, we have developed a modular grid integration module (\texttt{Gint}) within ABACUS. This module provides a unified interface that decouples the geometric complexity of grid-atom associations from the numerical kernels used to compute specific physical quantities, represented as the unified grid integration interfaces in Fig.~\ref{fig:TDDFT_flowchart}(c).

A general overview of the \texttt{Gint} module is illustrated in Fig.~\ref{fig:Gint_flowchart}.
The core architecture of the \texttt{Gint} module relies on a two-level grid decomposition strategy, managed by the \texttt{GintInfo} class (green block in Fig.~\ref{fig:Gint_flowchart}). The continuous real space is discretized into a fine mesh (\texttt{MeshGrid}) for accurate numerical integration, which is further grouped into coarser blocks (\texttt{BigGrid}) during the initialization phase shown in Fig.~\ref{fig:Gint_flowchart}. Each \texttt{BigGrid} maintains a list of neighboring atoms that have non-zero orbital overlap with the grid points contained within it. This design maximizes memory locality and allows for efficient screening of sparse matrix elements. The computation of orbital values and their derivatives is encapsulated in the \texttt{PhiOperator} class, which serves as a unified kernel for various integration tasks.

For these usual quantities, the integration is performed over a uniform real-space mesh. As detailed in our recent work along with other ABACUS developers \cite{ABACUS_GPU}, we have implemented a fully heterogeneous acceleration scheme for these terms. The \texttt{Gint} module supports the calculation of several key quantities, which can all be expressed as summations over grid points $g$.

Firstly, the charge density. The real-space electron density $\rho(\bm{r})$ at a grid point $\bm{r}_g$ is constructed from the density matrix $\rho_{\mu\nu}(\bm{R})$ and the atomic orbitals:
\begin{equation}
    \begin{aligned}
        \rho(\bm{r}_g) &= \sum_{\bm{R},\bm{R}'} \sum_{\mu\nu} \rho_{\mu\nu}(\bm{R}) \phi_{\mu}(\bm{r}_g-\bm{\tau}_{I}-\bm{R}') \\
        &\hspace{4em}\times \phi_{\nu}(\bm{r}_g-\bm{\tau}_{J}-\bm{R}-\bm{R}').
    \end{aligned}
\end{equation}
Here, $\rho_{\mu\nu}(\bm{R})$ is the density matrix element associated with orbitals $\mu$ and $\nu$ separated by lattice vector $\bm{R}$.

Secondly, the Hamiltonian matrix elements. The matrix elements of the local potential $V^{\text{loc}}(\bm{r})$ (including local pseudopotentials, Hartree, and exchange-correlation potentials) are computed as:
\begin{equation}
    H_{\mu\nu}^{\text{loc}}(\bm{R}) = \sum_{g} \phi_\mu(\bm{r}_g-\bm{\tau}_I) V^{\text{loc}}(\bm{r}_g) \phi_\nu(\bm{r}_g-\bm{\tau}_J-\bm{R}) \Delta V,
\end{equation}
where $\Delta V$ represents the elementary volume (integration weight) of the uniform mesh.

Thirdly, the local part of the Pulay force. For the Pulay force on atom $I$, the local component is given by:
\begin{equation}
    \begin{aligned}
        &\bm{F}_{I}^{\text{loc, Pulay}} \\ 
        =\ &- \sum_{\bm{R}} \sum_{\mu\nu} \rho_{\nu\mu}(\bm{R}) \\
        &\sum_{g} \bigg[ 
        \frac{\partial \phi_{\mu}(\bm{r}_g-\bm{\tau}_I)}{\partial \bm{\tau}_I} V^{\text{loc}}(\bm{r}_g) \phi_{\nu}(\bm{r}_g-\bm{\tau}_J-\bm{R}) \\
        &\hspace{1em}+ \phi_{\mu}(\bm{r}_g-\bm{\tau}_I) V^{\text{loc}}(\bm{r}_g) \frac{\partial \phi_{\nu}(\bm{r}_g-\bm{\tau}_J-\bm{R})}{\partial \bm{\tau}_I} 
        \bigg] \Delta V.
    \end{aligned}
\end{equation}
This calculation requires the evaluation of orbital gradients $\partial \phi / \partial \bm{\tau}$ on the grid, which are handled by the \texttt{PhiOperator} kernel.

By leveraging the \texttt{BigGrid} structure, we transform the summation over individual grid points into efficient batched matrix operations. The sparse atomic-orbital vectors (and their gradients) on the grid are packed into dense matrix tiles, allowing the use of highly optimized custom variable-size batched \texttt{GEMM} kernels on GPUs to perform the matrix operations and accumulation depicted in Fig.~\ref{fig:Gint_flowchart}. In the current implementation, this is combined with multi-stream execution over \texttt{BigGrid} batches. This approach has been shown to achieve considerable speedups for large-scale systems, including twisted bilayer graphene with over 10000 atoms \cite{ABACUS_GPU}.

\subsubsection{Spherical Grid Integration for Nonlocal Potentials}\label{sec:Spherical_Grid_Integration}

In the velocity gauge formulation of RT-TDDFT, the time-dependent vector potential $\bm{A}(t)$ introduces a spatially modulated phase factor into the nonlocal pseudopotential matrix elements:
\begin{equation}
    \Braket{\phi_\mu | \mathrm{e}^{-\mathrm{i}\bm{A}(t)\cdot\bm{r}} | \beta_{Ilm\zeta}},
\end{equation}
which must be evaluated numerically at every time step, even if the atomic positions are fixed. Unlike the static two-center integrals $\Braket{\phi_\mu | \beta_{Ilm\zeta}}$ computed analytically, these phase-modulated overlaps cannot be handled by the uniform real-space grid infrastructure \texttt{Gint} used for charge density or local potential. The extreme localization of the projector functions $\beta_{Ilm\zeta}$ around atomic cores, combined with the rapid spatial oscillations of the phase factor $\mathrm{e}^{-\mathrm{i}\bm{A}(t)\cdot\bm{r}}$, renders integration over the global uniform mesh inaccurate unless an impractically high grid cutoff is employed.

To address this, we have developed a dedicated spherical-grid integration framework independent of the \texttt{Gint} module. This framework leverages atom-centered quadrature grids that respect the radial symmetry of the projectors: a 110-point Lebedev-Laikov grid \cite{lebedev1999quadrature} for angular integration and a 140-point Gauss-Legendre grid for the radial direction.
The CPU implementation computes these integrals via a straightforward nested loop over neighboring atoms and quadrature. Although we have performed multiple rounds of optimization on the CPU code (e.g., OpenMP parallelization), it is still difficult to apply advanced techniques such as memory coalescing or batched processing, and consequently remains a performance bottleneck in the long-time simulations of velocity-gauge RT-TDDFT.

To overcome this limitation, we implemented an optimized GPU-accelerated path centered around the kernel \texttt{snap\_psibeta\_atom\_batch\_gpu}. We adopt an atom-level batching scheme: instead of processing one neighboring atom at a time, the GPU kernel launches a single grid that simultaneously handles all neighboring orbitals of a central atom hosting the projectors $\beta$. This eliminates redundant kernel launches and dramatically improves occupancy.
Moreover, the integration loop structure is reorganized for GPU efficiency: each thread processes a unique angular quadrature point and accumulates over all radial points, enabling full vectorization across the 110-point Lebedev-Laikov grid.
Crucially, the reduction over quadrature points is performed using warp-level primitives (\texttt{\_\_shfl\_down\_sync}) \cite{harris2007optimizing} followed by a shared-memory cross-warp reduction, completely avoiding atomic operations. The CPU version, even with OpenMP, relies on sequential accumulation per orbital-projector pair and cannot exploit such fine-grained parallel reductions.
As a result, the GPU implementation of the spherical grid integration delivers significant speedup over the legacy CPU code, which is demonstrated in Sec.~\ref{sec:System_Size_Scaling_Test}.

\begin{figure*}[!htbp]
    \centering
    \includegraphics[width=0.96\textwidth]{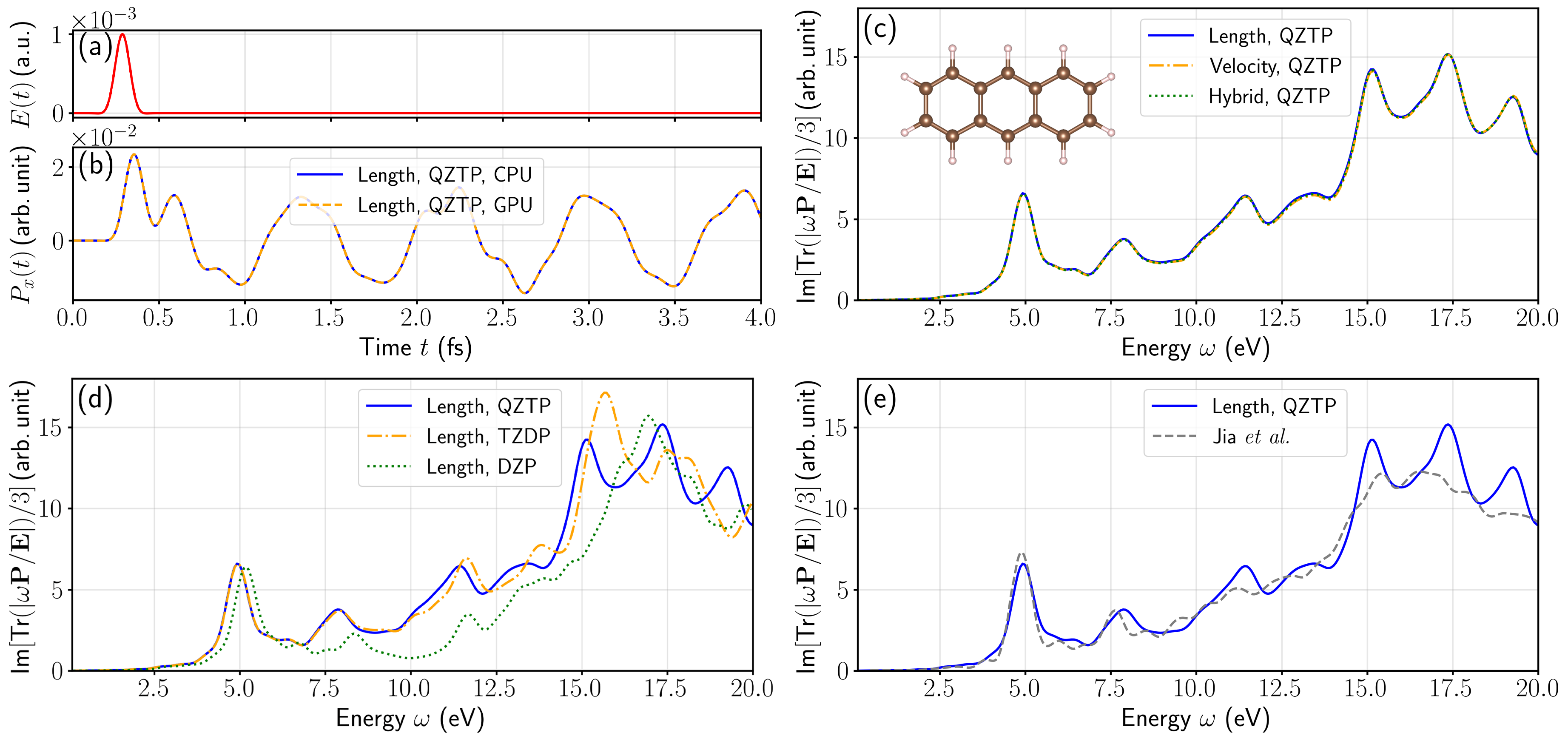}
    \caption{Optical properties of the anthracene molecule. (a) Time-dependent external electric field profile. (b) Time evolution of the induced dipole moment, demonstrating consistency between CPU and GPU calculations. (c) Absorption spectra calculated using length, velocity, and hybrid gauges with the QZTP basis. (d) Basis set convergence test. (e) Comparison with DGDFT benchmark results \cite{acs.jctc.8b00580}.}
    \label{fig:abs_C14H10}
\end{figure*}

\section{Results and Discussions}\label{sec:results}

\subsection{Physical Validation: Optical Properties with Fixed Ions}\label{sec:Physical_Validation_Fixed_Ions}

To rigorously validate the accuracy, numerical stability, and versatility of our GPU-accelerated RT-TDDFT implementation in ABACUS, we first performed a comprehensive suite of benchmark calculations for optical absorption spectra under fixed atomic positions. We selected five representative systems spanning different dimensionalities and chemical bonding characters: the anthracene molecule (C\textsubscript{14}H\textsubscript{10}), a bare (CdSe)\textsubscript{6} nanocluster, a dimerized hydrogen chain, a monolayer of hexagonal boron nitride (\textit{h}-BN), and bulk silicon. These systems cover zero-, one-, two-, and three-dimensional periodicities, serving as a robust test set for real-time electron dynamics simulations.

\subsubsection{Computational Details}

All numerical simulations were carried out using the Crank-Nicolson time propagator. The TDKS equations were propagated under a uniform external electric field with a Gaussian wave packet profile:
\begin{equation}
    E(t) = A \cos\left[2\pi f (t - t_0)\right] \exp\!\left[-\frac{(t - t_0)^2}{2\sigma^2}\right],
\end{equation}
where the amplitude was set to $A = 1.0\times 10^{-3}\ \mathrm{a.u.}$, the carrier frequency $f = 2.0\ \mathrm{fs}^{-1}$, the pulse width $\sigma = 0.05\ \mathrm{fs}$, and the center time $t_0 = 0.2904\ \mathrm{fs}$.

We utilized the SG15 Optimized Norm-Conserving Vanderbilt Pseudopotentials (ONCVPSP) \cite{PhysRevB.88.085117,j.cpc.2015.05.011}. The numerical atomic orbitals were custom-generated using the Systematically Improvable Atomic Basis (SIAB) method \cite{Chen_2010,Chen_2011,PhysRevB.103.235131}. A uniform energy cutoff of $E_{\mathrm{cut}} = 100\ \mathrm{Ry}$ was employed for the plane-wave basis during the reference calculation for orbital generation. Consistent with the SIAB generation scheme, the same $100\ \mathrm{Ry}$ cutoff (\texttt{ecutwfc}) was applied in all subsequent calculations to ensure optimal compatibility between the localized orbitals and the grid mesh. The exchange-correlation functionals were evaluated via the Libxc library \cite{j.softx.2017.11.002}.

For finite (non-periodic) systems, optical absorption spectra were computed from the dipole response in the length gauge. The complex polarizability tensor was obtained via Fourier transform of the induced dipole moment $\bm{P}(t)$, and the absorption was evaluated as
\begin{equation}
    \operatorname{Re}\sigma(\omega) \propto \frac{\omega}{3} \operatorname{Im} \operatorname{Tr} \left[ \frac{\bm{P}(\omega)}{\bm{E}(\omega)} \right],
\end{equation}
where the results were averaged over $x, y, z$ directions to account for anisotropy. For periodic systems, we utilized both the velocity gauge and the hybrid gauge approaches, and computed the current density $\bm{J}(t)$. The complex optical conductivity is then defined as $\sigma(\omega) = \bm{J}(\omega)/\bm{E}(\omega)$, and the dielectric function follows from the standard relation (under atomic units):
\begin{equation}
    \varepsilon(\omega) = 1 + \mathrm{i}\frac{4\pi\sigma(\omega)}{\omega}.
\end{equation}
In all cases, the resulting time-domain signals were Fourier-transformed with appropriate exponential windowing to obtain smooth spectral features.

\begin{figure}[!htbp]
    \centering
\includegraphics[width=0.48\textwidth]{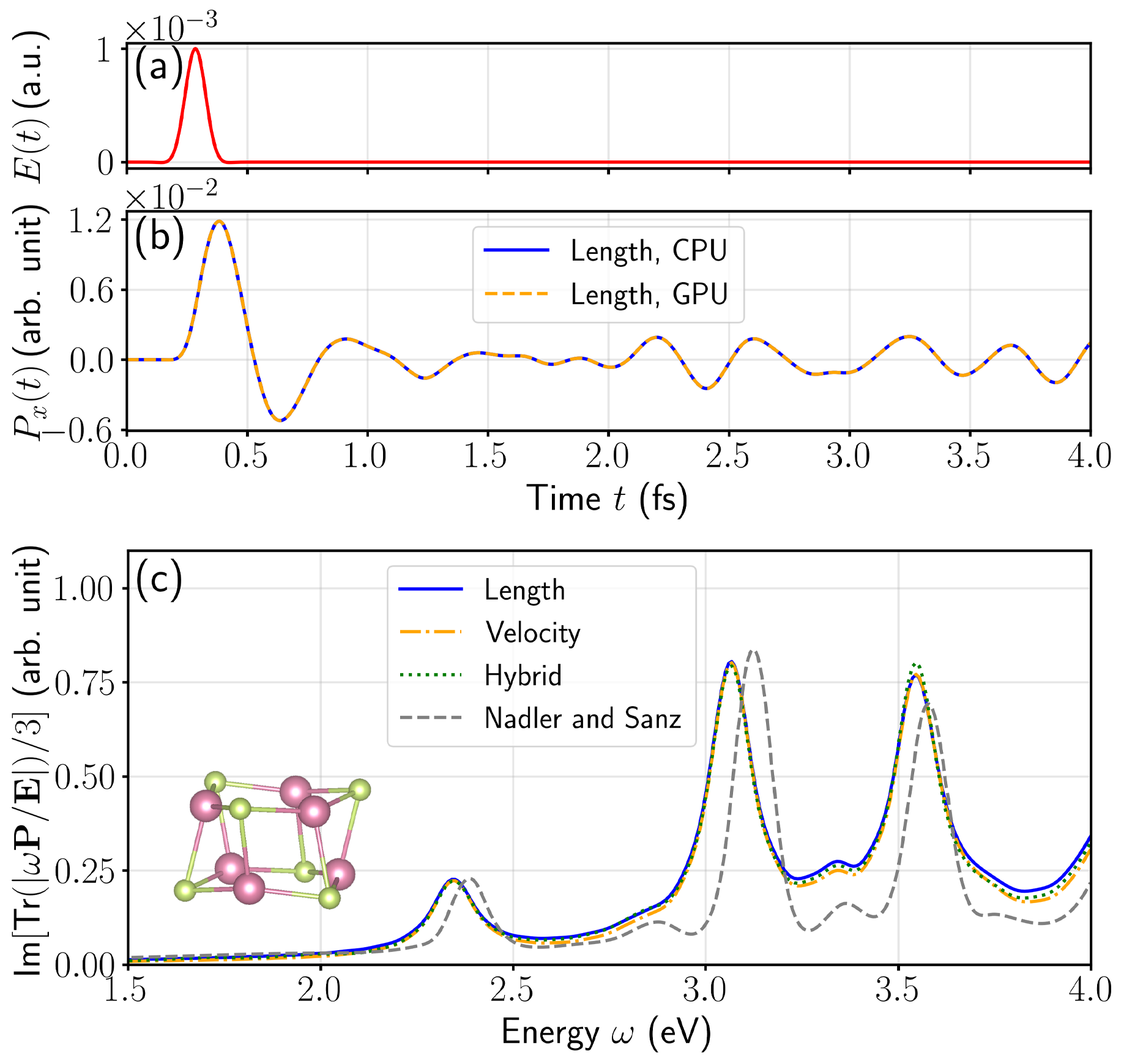}
    \caption{Optical properties of the bare (CdSe)\textsubscript{6} cluster. (a) Time-dependent external electric field profile. (b) Time evolution of the induced dipole moment, demonstrating consistency between CPU and GPU calculations. (c) Absorption spectra obtained using length, velocity, and hybrid gauges, compared with CP2K benchmark results \cite{Nadler2013}.}
    \label{fig:abs_CdSe6}
\end{figure}

Unless otherwise noted, the results presented correspond to calculations accelerated by GPUs. The consistency between CPU and GPU implementations has been strictly verified for all systems (see subplots (b) in Figs.~\ref{fig:abs_C14H10}--\ref{fig:abs_Si}), with results overlapping to numerical precision.

\subsubsection{Anthracene Molecule}\label{sec:Anthracene}

As a benchmark for organic semiconductors, we investigated the anthracene molecule (C\textsubscript{14}H\textsubscript{10}) using the Perdew-Burke-Ernzerhof (PBE) functional \cite{PhysRevLett.77.3865,PhysRevLett.78.1396}. The molecule was placed in a supercell with a $20\ \textrm{Å}$ vacuum layer to isolate periodic images. The time step was set to $\Delta t = 0.00242\ \mathrm{fs}$ with a total simulation duration of $2500$ steps.

We assessed the influence of basis set size using three distinct levels of precision, denoted as DZP, TZDP, and QZTP. Specifically, the basis configurations were: DZP (C: $2s2p1d$, H: $2s1p$), TZDP (C: $3s3p2d$, H: $3s2p$), and QZTP (C: $4s4p3d2f$, H: $4s3p$), all with a cutoff radius of $10\ \mathrm{a.u.}$. Notably, the QZTP basis for carbon includes additional $f$-orbitals to capture higher-order polarization effects.

Fig.~\ref{fig:abs_C14H10}(c) demonstrates excellent gauge invariance, with length, velocity, and hybrid gauge results for the QZTP basis virtually superimposing. We further note that, for isolated systems, evaluating the optical spectrum from the dipole polarization in the velocity gauge avoids the low-frequency divergence that is commonly encountered when the spectrum is extracted from the current \cite{acs.jctc.5c00111}.

The basis set convergence analysis in Fig.~\ref{fig:abs_C14H10}(d) reveals that while the low-energy spectrum ($<5\ \mathrm{eV}$) is basically converged at the DZP level, high-energy excitations are sensitive to basis completeness. The spectra exhibit a systematic redshift and peak sharpening as the basis size increases from DZP to QZTP, indicating that extended diffuse orbitals are crucial for describing high-lying unoccupied states. This trend aligns with observations in other local-orbital implementations, such as SIESTA, where confined basis sets (or those with shorter cutoff radii) are known to artificially shift optical spectra toward higher energies \cite{siesta_tddft_tutorial}.

Our QZTP results show satisfactory agreement with the plane-wave benchmark calculated using the PWDFT module in the DGDFT package \cite{j.jcp.2011.11.032,jcp.1.4931732} by Jia \textit{et al.} \cite{acs.jctc.8b00580} across the entire $0$--$20\ \mathrm{eV}$ range (Fig.~\ref{fig:abs_C14H10}(e)).

\subsubsection{Magic-Size Hexameric Cadmium Selenide Cluster}

We calculated the optical properties of a ``magic-size'' (CdSe)\textsubscript{6} cluster, a prototype for quantum dots \cite{Kasuya2004,jacs.ja003598j}. The cluster was placed in a supercell with a $20\ \textrm{Å}$ vacuum layer to isolate periodic images. We employed the PBE functional and a high-precision numerical atomic orbital basis set: Cd ($4s2p2d1f$) and Se ($3s3p2d1f$), both defined with a $10\ \mathrm{a.u.}$ cutoff radius. The time step was set to $\Delta t = 0.00484\ \mathrm{fs}$ with a total simulation duration of $10000$ steps.
We computed the optical response using the length, velocity, and hybrid gauges. As illustrated in Fig.~\ref{fig:abs_CdSe6}(c), the spectra obtained from these three formulations overlap almost perfectly.

We benchmark our results against the reference data reported by Nadler and Sanz \cite{Nadler2013}, who performed RT-TDDFT calculations using the CP2K code \cite{jcp.5.0007045,j.cpc.2004.12.014} with the PBE functional and a short-range, molecularly optimized double-zeta (m-SR-DZVP) basis set. Our calculated spectrum faithfully reproduces the characteristic three-peak topology observed in the reference within the $1.5$--$4.0\ \mathrm{eV}$ window. The relative oscillator strengths are well-preserved, with the most intense absorption feature appearing in the central region around $3.1\ \mathrm{eV}$.

A systematic redshift of about $0.05\ \mathrm{eV}$ is observed relative to the reference. Consistent with the trend discussed above in Sec.~\ref{sec:Anthracene}, this shift is likely attributed to the higher completeness of our basis set (which includes $f$-polarization functions) compared to the compact m-SR-DZVP basis employed by Nadler and Sanz.

\begin{figure}[!htbp]
    \centering
    \includegraphics[width=0.48\textwidth]{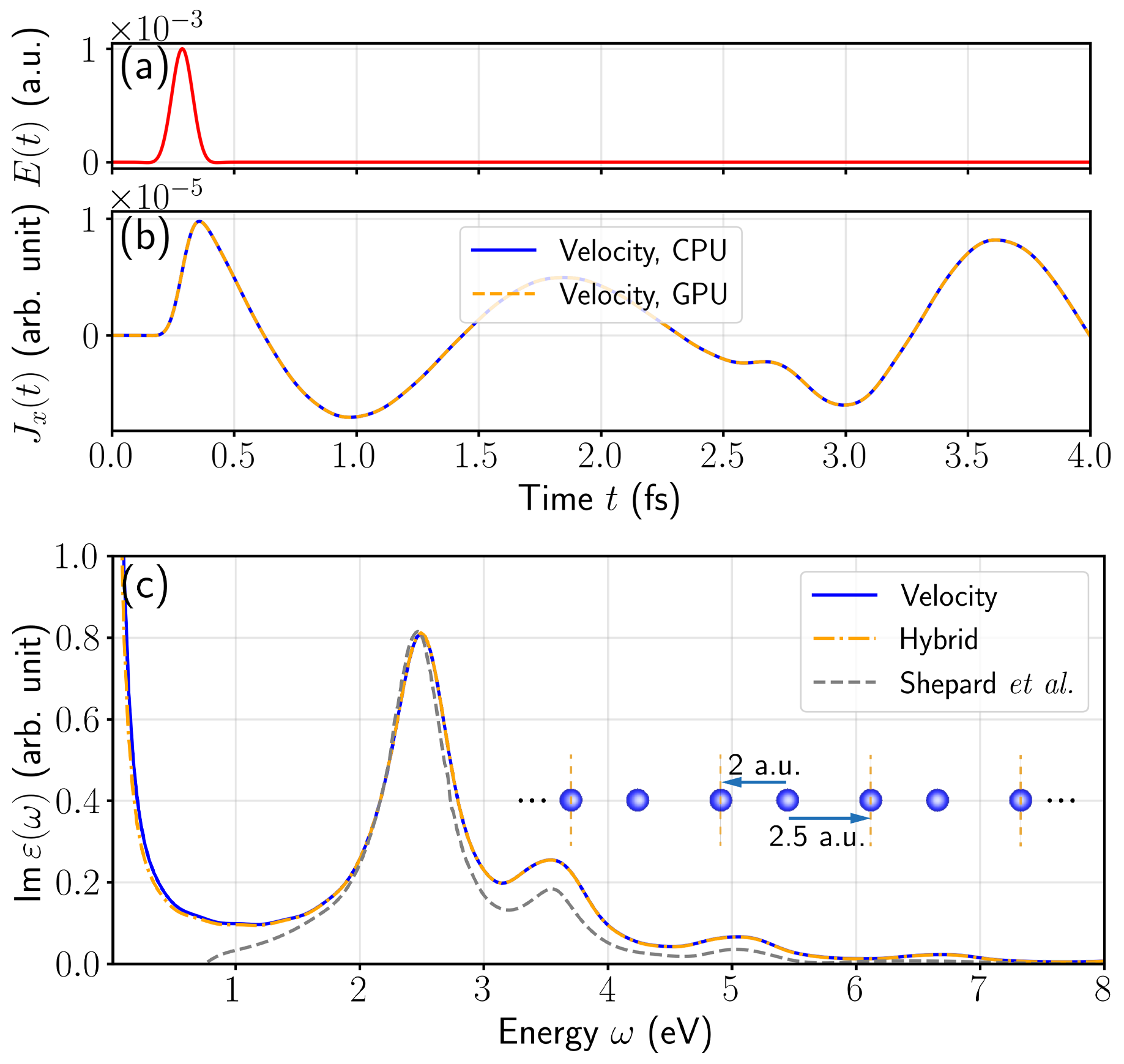}
    \caption{Optical properties of the periodic hydrogen chain. (a) Time-dependent external electric field profile. (b) Time evolution of the macroscopic current density, demonstrating consistency between CPU and GPU calculations. (c) Absorption spectra computed using velocity and hybrid gauges, compared with Qbox benchmark results \cite{jcp.5.0211238}.}
    \label{fig:abs_HChain}
\end{figure}

\subsubsection{One-Dimensional Hydrogen Chain}

To test the description of periodic systems with low dimensionality, we simulated a linear hydrogen chain exhibiting Peierls distortion \cite{Peierls1955,RevModPhys.60.1129}. The unit cell contains 58 hydrogen atoms (29 H\textsubscript{2} units) arranged with alternating bond lengths of $2.0\ \mathrm{a.u.}$ and $2.5\ \mathrm{a.u.}$. A vacuum of $20\times 20\ \mathrm{a.u.}$ was applied in the perpendicular directions ($y$ and $z$) to minimize interactions between periodic images. Calculations were performed at the $\mathit{\Gamma}$ point using the LDA functional with Slater exchange \cite{Bloch1929, Dirac_1930} and the Perdew-Zunger (PZ) correlation \cite{PhysRevB.23.5048}. The basis set consisted of $3s2p$ orbitals with a $10\ \mathrm{a.u.}$ cutoff. The time step was set to $\Delta t = 0.00242\ \mathrm{fs}$ with a total simulation duration of $5000$ steps.

We benchmark our results against the work of Shepard \textit{et al.} \cite{jcp.5.0211238}, as shown in Fig.~\ref{fig:abs_HChain}(c). The reference data were calculated using the plane-wave pseudopotential code Qbox (Qb@ch version) \cite{jcp.5.0057587,jcp.1.4758792}. It is worth mentioning that Shepard \textit{et al.} utilized a gauge transformation to maximally localized Wannier functions (MLWFs) \cite{PhysRevB.56.12847} to apply the electric field in the length gauge, avoiding the periodic boundary issues typically associated with the position operator in extended systems.

As illustrated in Fig.~\ref{fig:abs_HChain}(c), our spectra obtained using both the velocity and hybrid gauges show excellent agreement with the reference in the finite-frequency region. The primary absorption peak around $2.50\ \mathrm{eV}$ and the secondary peak near $3.55\ \mathrm{eV}$ are well-reproduced in terms of both position and intensity. 

A notable difference, however, appears in the $\omega \to 0$ limit. Our results exhibit a divergent behavior, whereas the reference spectrum by Shepard \textit{et al.} correctly vanishes. This divergence is a known numerical artifact (often referred to as a ``fake plasmon'') \cite{PhysRevB.62.7998,Rodrigues_Pela_2021} inherent to the velocity gauge formulation in periodic systems. In contrast, the MLWF-based length gauge formulation employed in the reference naturally avoids this artifact. Apart from this zero-frequency discrepancy, the consistency of the spectral features confirms the accuracy of our implementation for periodic quasi-1D systems.

\begin{figure}[!htbp]
    \centering
    \includegraphics[width=0.48\textwidth]{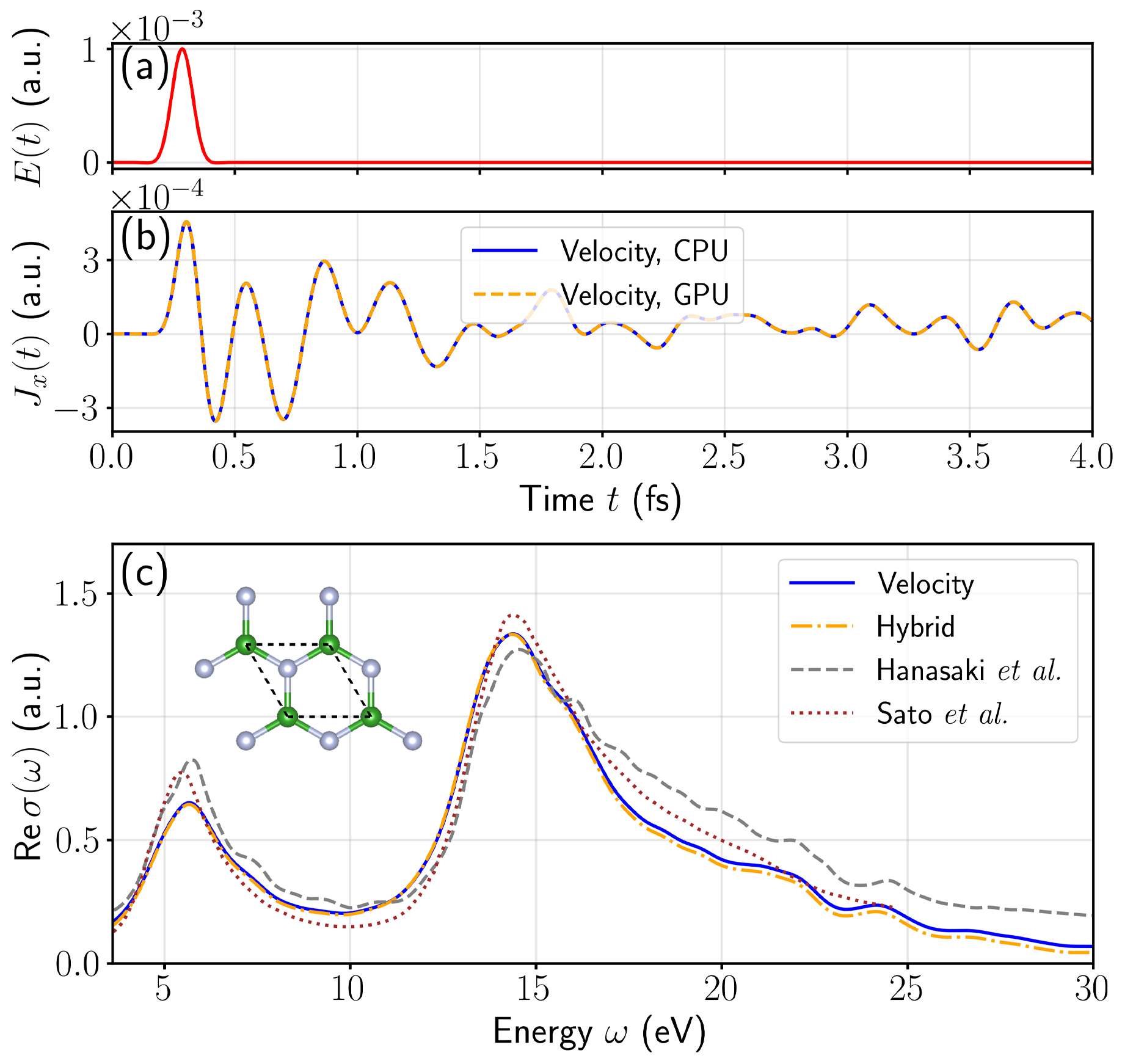}
    \caption{Optical properties of the \textit{h}-BN monolayer. (a) Time-dependent external electric field profile. (b) Time evolution of the macroscopic current density, demonstrating consistency between CPU and GPU calculations. (c) Real part of the optical conductivity, compared with Octopus \cite{app8101777} and PySCF \cite{jcc.27058} benchmark results.}
    \label{fig:abs_hBN}
\end{figure}

\subsubsection{Two-Dimensional Hexagonal Boron Nitride}

For two-dimensional materials, we examined monolayer hexagonal boron nitride (\textit{h}-BN). A vacuum layer of $15\ \textrm{Å}$ was added along the $z$-direction to eliminate spurious interlayer interactions. The Brillouin zone was sampled with a $15\times 15\times 1$ $\bm{k}$-point mesh. We utilized the LDA functional with Slater exchange and PZ correlation, and a rich numerical atomic orbital basis set composed of B ($3s3p2d1f$) and N ($3s3p2d$) with $10\ \mathrm{a.u.}$ cutoffs. The time step was set to $\Delta t = 0.00484\ \mathrm{fs}$ with a total simulation duration of $8000$ steps.

Fig.~\ref{fig:abs_hBN}(c) displays the real part of the optical conductivity $\operatorname{Re}\sigma(\omega)$. The spectrum exhibits a characteristic double-peak structure with a low-energy feature around $5.6\ \mathrm{eV}$ and a prominent high-energy peak near $14.5\ \mathrm{eV}$. To benchmark our results, we compared them with two distinct theoretical approaches: the real-space grid implementation using the Octopus code \cite{Octopus2015} by Sato \textit{et al.} \cite{app8101777} and the Gaussian-basis implementation using the PySCF code \cite{jcp.5.0006074} by Hanasaki \textit{et al.} \cite{jcc.27058}. Note that our macroscopic current is conventionally defined per unit volume, see Eq.~\eqref{eq:current_density}. To properly compare with the reference literature which reports the conductivity in atomic units (a.u.) normalized by the basal area to exclude the artificial vacuum layer, we scaled our results by the $z$-direction vacuum thickness.

Our calculations using the SIAB numerical atomic orbitals show remarkable agreement with both reference methods, despite the fundamental differences in basis set representations. Physically, the lower-energy peak corresponds to interband transitions primarily located around the $\mathit{\Gamma}$ point of the Brillouin zone, while the intense peak at $14.5\ \mathrm{eV}$ is attributed to transitions involving a specific band with anomalous dispersion near the $M$ point \cite{app8101777}.

\begin{figure*}[!htbp]
    \centering
    \includegraphics[width=0.96\textwidth]{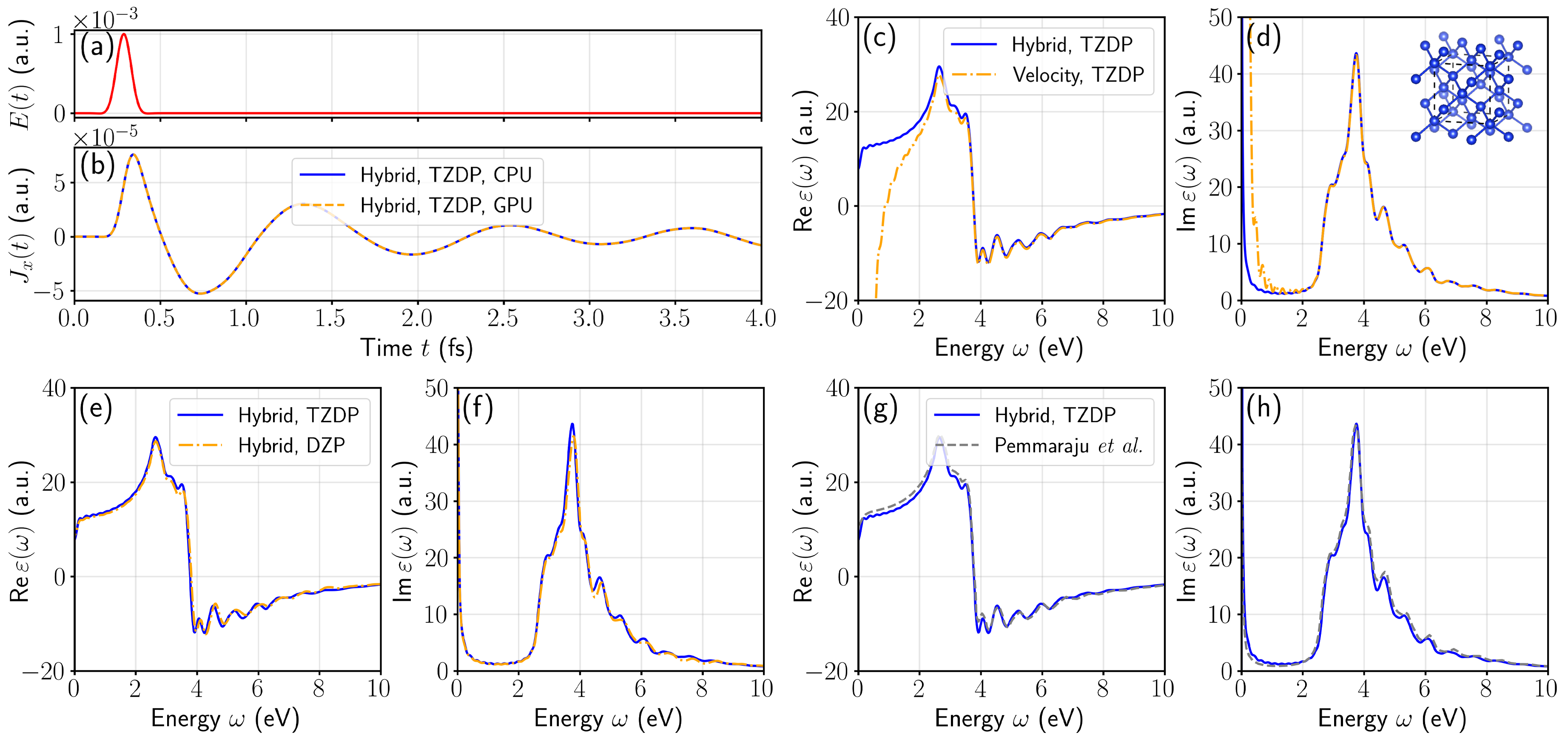}
    \caption{Optical properties of bulk silicon. (a) Time-dependent external electric field profile. (b) Time evolution of the macroscopic current density, demonstrating consistency between CPU and GPU calculations. (c)-(d) Real and imaginary parts of the dielectric function, illustrating the divergence artifact in the velocity gauge. (e)-(f) Basis set convergence test (DZP vs. TZDP). (g)-(h) Comparison of the dielectric function with SIESTA benchmark results \cite{j.cpc.2018.01.013}.}
    \label{fig:abs_Si}
\end{figure*}

\subsubsection{Three-Dimensional Bulk Silicon}\label{sec:bulk_Si}

Finally, we validated the implementation for 3D bulk solids using an 8-atom cubic silicon supercell. The system was sampled using a dense $16\times 16\times 16$ $\bm{k}$-mesh and the PBE functional. Two basis sets were compared: DZP ($2s2p1d$) and TZDP ($3s3p2d$), both with $10\ \mathrm{a.u.}$ cutoffs. The time step was set to $\Delta t = 0.00242\ \mathrm{fs}$ with a total simulation duration of $8000$ steps.

We first examine the numerical behavior of different gauges in the localized basis framework. As illustrated in Figs.~\ref{fig:abs_Si}(c) and (d), the results confirm the theoretical analysis presented in Sec.~\ref{sec:Hybrid_Gauge}. The standard velocity gauge (orange dashed-dotted line) suffers from a severe numerical artifact in the low-frequency limit: the real part of the dielectric function diverges to negative infinity. This behavior stems partly from the lack of position-dependent phase factors in the standard NAO implementation \cite{acs.jctc.5c00111}. In contrast, the hybrid gauge (blue solid line) effectively rectifies this issue, yielding a finite static dielectric constant at zero frequency.

Regarding basis set convergence (Figs.~\ref{fig:abs_Si}(e) and (f)), the difference between DZP and TZDP is minimal compared to the molecular case. Here, the DZP basis contains 13 orbitals per Si atom, whereas the TZDP basis contains 22 orbitals per Si atom. The DZP basis yields a slightly lower intensity at the main absorption peak around $3.7\ \mathrm{eV}$, but the overall spectral shape remains robust.

To verify the validity of our results over the broad spectral range ($0$--$10\ \mathrm{eV}$), we benchmarked our calculations against the work of Pemmaraju \textit{et al.} \cite{j.cpc.2018.01.013}. Their reference data were generated using the SIESTA code \cite{SIESTA_2002} within an LCAO framework. Our results obtained with the TZDP basis faithfully reproduce the key spectral features reported in the reference, where 27 orbitals per Si atom were used, as shown in Figs.~\ref{fig:abs_Si}(g) and (h).

However, it should be noted that standard local or semi-local functionals, such as the LDA/PBE employed here and in above references, are known to underestimate band gaps and fail to capture excitonic effects in bulk semiconductors \cite{RevModPhys.74.601}. Accurate comparison with experimental optical spectra would strictly require advanced methods like the \textit{GW}-BSE formalism \cite{PhysRevLett.81.2312,PhysRevB.62.4927} or time-dependent hybrid functionals \cite{PhysRevB.78.121201,jcp.1.4937379,10.1145/3295500.3356144}. It is worth noting that while ABACUS supports hybrid functionals in the RT-TDDFT framework \cite{ABACUS_Hybrid_functional_2025}, this work restricts itself to the semi-local level. The primary objective here is numerical verification, demonstrating that our GPU-accelerated implementation yields mathematically consistent solutions to the TDKS equations compared to established codes under similar levels of theory.

\subsection{Physical Validation: Total Energy Conservation in Ehrenfest Dynamics}\label{sec:Physical_Validation_Ehrenfest}

While the preceding benchmarks focused exclusively on electron dynamics under fixed atomic positions, practical RT-TDDFT applications often require the simultaneous and stable evolution of ionic degrees of freedom. To validate the accuracy and numerical stability of our nonadiabatic Ehrenfest molecular dynamics implementation, we tested the total energy conservation of the coupled electron-ion system during both adiabatic and highly nonadiabatic processes.

\subsubsection{Vibration of the NaCl Molecule}

We first investigated the total energy conservation during the vibration of the NaCl molecule, representing a nearly adiabatic process. The simulations were initialized from the near-quilibrium bond length of $2.36\ \textrm{Å}$ with a low initial kinetic energy of $2\ \mathrm{eV}$. Because of this low initial kinetic energy, the vibrational dynamics remain predominantly in the adiabatic regime \cite{jcp.1.3700800,jcp.1.2008258}. All simulations were performed using the PBE functional and a basis set of Na ($5s4p3d2f$) and Cl ($3s3p2d$) with $10\ \mathrm{a.u.}$ cutoffs. The time step was set to $\Delta t = 3, 5, 8, 10, 12, 16, 20\ \mathrm{as}$ respectively, with a total simulation time of $T=200\ \mathrm{fs}$, covering the full vibrational period of the molecule.

\begin{figure}[!htbp]
    \centering
    \includegraphics[height=0.25\textwidth]{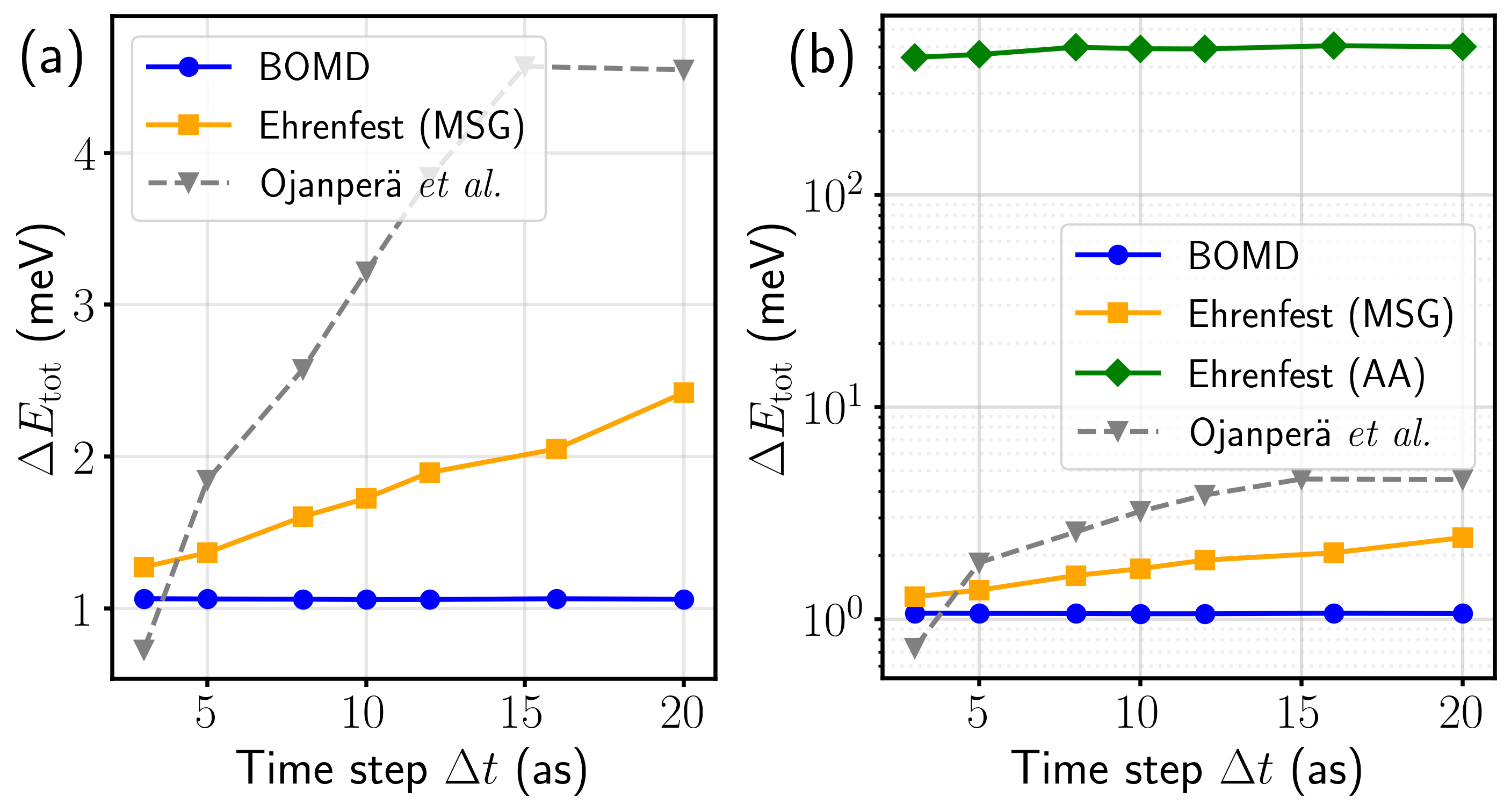}
    \caption{Maximum total energy fluctuation $\Delta E_{\mathrm{tot}}$ of the NaCl molecule during vibrational dynamics as a function of the simulation time step $\Delta t$. (a) The results on a linear scale, comparing BOMD with Ehrenfest dynamics incorporating the moving spatial gauge (MSG), alongside reference data from Ojanperä \textit{et al.} \cite{jcp.1.3700800} using GPAW. (b) The results using the adiabatic approximation (AA) on a logarithmic scale, highlighting the substantial energy drift when the moving basis set derivatives are neglected.}
    \label{fig:energy_fluctuation}
\end{figure}

Define the maximum fluctuation in the total energy within the simulation time $T$ as $\Delta E_{\mathrm{tot}}$:
\begin{equation}
    \Delta E_{\mathrm{tot}} = \max_{t\in [0,T]} |E_{\mathrm{tot}}(t) - E_{\mathrm{tot}}(0)|.
\end{equation}
As illustrated in Fig.~\ref{fig:energy_fluctuation}(a), we compared the energy conservation of our Ehrenfest dynamics implementation incorporating the moving spatial gauge (MSG), which explicitly evaluates the $P$ matrix in Eq.~\eqref{eq:TDKS_Ehrenfest}, against standard Born-Oppenheimer \textit{ab initio} molecular dynamics (BOMD) and reference Ehrenfest data from Ojanperä \textit{et al.} \cite{jcp.1.3700800} using the GPAW code \cite{GPAW_Enkovaara_2010,GPAW_jcp.5.0182685}. For BOMD, the system remains strictly on the ground-state Born-Oppenheimer surface, yielding an exceptionally stable energy baseline with fluctuations of approximately $1\ \mathrm{meV}$. Our Ehrenfest (MSG) approach maintains excellent energy conservation, with $\Delta E_{\mathrm{tot}}$ remaining below $2.5\ \mathrm{meV}$ even at a time step of $20\ \mathrm{as}$. These results demonstrate highly stable numerical performance, exhibiting even lower energy fluctuations compared to the reference data, which drift up to approximately $4.5\ \mathrm{meV}$ under similar conditions.

To emphasize the critical physical role of the moving spatial gauge, we also performed simulations under the adiabatic approximation (AA), where the $P$ matrix in the evolution equation is deliberately neglected as in Eq.~\eqref{eq:TDKS_matrix}. As shown in Fig.~\ref{fig:energy_fluctuation}(b), neglecting the velocity-dependent basis derivatives leads to a catastrophic violation of energy conservation. The maximum energy fluctuation for the AA scheme reaches about $500\ \mathrm{meV}$ (plotted on a logarithmic scale), which is roughly two orders of magnitude higher than that of the MSG scheme. This massive energy drift confirms that the explicit time-dependence of the basis functions introduces a non-trivial moving spatial gauge; without the $P$ matrix correction, the norm of the electronic wavefunctions is not properly preserved during ionic motion, leading to unphysical forces, severe energy accumulation, and numerical instability.

\subsubsection{Torsional Rotation of the \protect\ce{H2C=NH2+} Molecule}

To thoroughly assess the algorithm's performance in the presence of strong nonadiabaticity, we simulated the torsional rotation of the \ce{H2C=NH2+} molecule about its internal axis. The molecule, initially in a planar equilibrium geometry, was assigned a high initial torsional kinetic energy of 10 eV. The simulations were performed using the PBE functional and a basis set of H ($3s2p$), C ($3s2p2d$) and N ($3s3p2d$) with $10\ \mathrm{a.u.}$ cutoffs. The time step was set to $\Delta t = 5\ \mathrm{as}$ with a total simulation time of $T=60\ \mathrm{fs}$.

\begin{figure}[!htbp]
    \centering
    \includegraphics[width=0.48\textwidth]{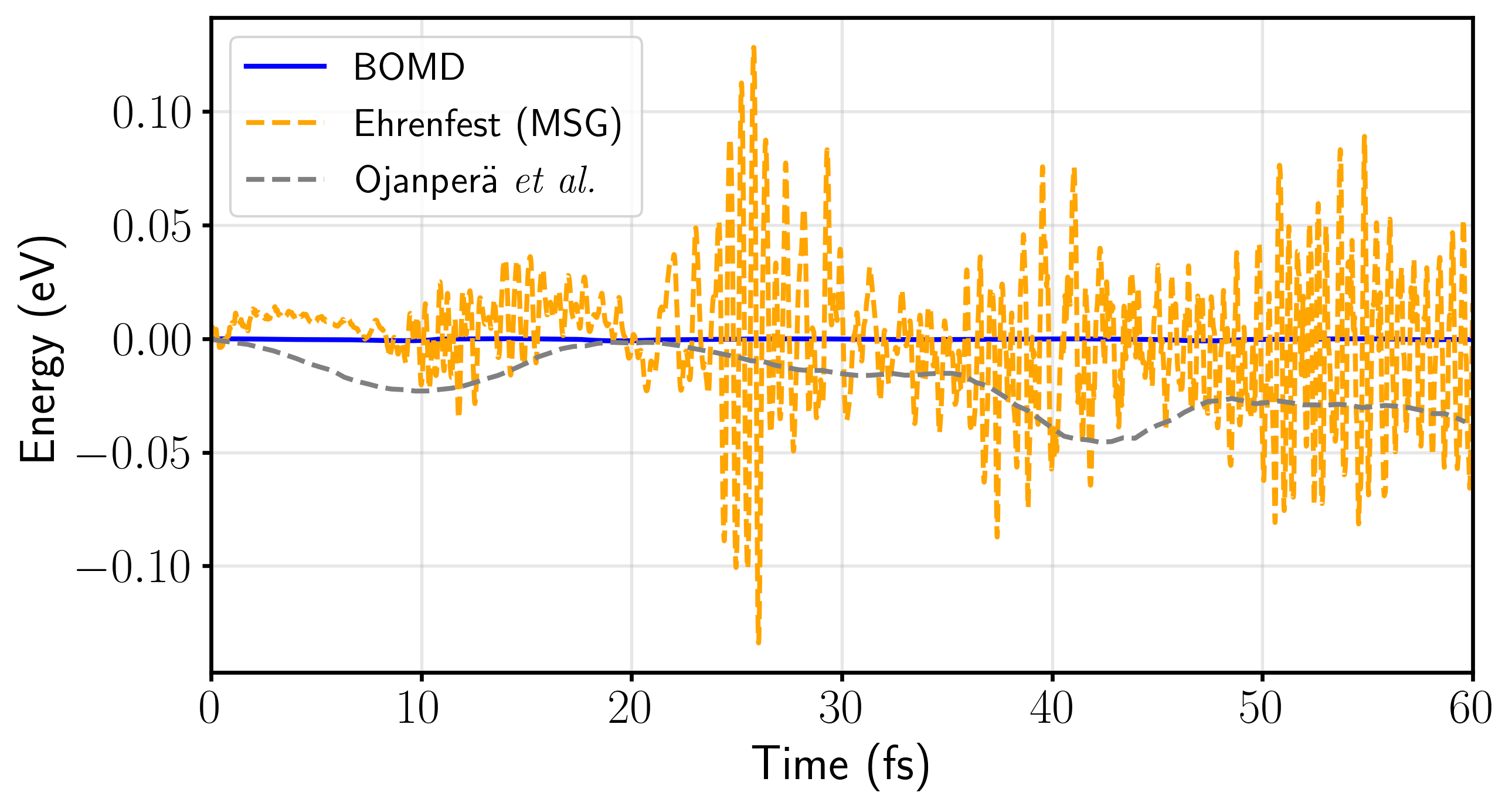}
    \caption{Total energy fluctuation of the \protect\ce{H2C=NH2+} molecule during highly nonadiabatic torsional rotation as a function of time. Results from our Ehrenfest dynamics with the moving spatial gauge (MSG) are compared with BOMD and reference data from Ojanperä \textit{et al.} \cite{jcp.1.3700800}.}
    \label{fig:nonadiabatic_ehrenfest}
\end{figure}

Under these highly energetic conditions, the potential energy surface (PES) generated by Ehrenfest MD would deviate rapidly from the corresponding Born-Oppenheimer PES, indicating a significant onset of nonadiabatic effects and electronic excitations \cite{jcp.1.3700800,jcp.1.2008258}. Accurate energy conservation in such nonadiabatic regimes is notoriously challenging.

Fig.~\ref{fig:nonadiabatic_ehrenfest} presents the total energy fluctuation (shifted to zero at $t=0$) over the simulation period. As expected, BOMD maintains a strictly flat energy baseline. For our Ehrenfest dynamics incorporating the moving spatial gauge (MSG), we observe high-frequency oscillations in the total energy. Despite these rapid fluctuations, the energy is consistently constrained within a stable envelope of approximately $\pm 0.1\ \mathrm{eV}$ around the initial energy, without exhibiting catastrophic long-term divergence.

In comparison, the reference data from Ojanperä \textit{et al.} \cite{jcp.1.3700800} is notably smoother, with an overall energy deviation of less than $0.05\ \mathrm{eV}$. While our implementation exhibits more pronounced oscillatory behavior, the absolute magnitude of the energy fluctuation remains strictly on the same order of magnitude as the reference data. We attribute this more severe fluctuation primarily to the intrinsic mathematical characteristics of the numerical atomic orbitals utilized in our framework. The localized nature and numerical interpolation of NAOs can inherently introduce higher-frequency numerical noise when evaluating explicit basis-set derivatives (the $P$ matrix) under highly energetic, nonadiabatic atomic motions, compared to the PAW method or plane-wave bases. 

Furthermore, we note that the results under the adiabatic approximation (omitting the $P$ matrix) are excluded from Fig.~\ref{fig:nonadiabatic_ehrenfest}, as their energy fluctuation reaches up to $2\ \mathrm{eV}$. This reinforces our earlier conclusion from the NaCl tests: even in the presence of numerical oscillations, incorporating the moving spatial gauge is absolutely indispensable for maintaining the physical validity and stability of nonadiabatic Ehrenfest dynamics.

To sum up, all these benchmarks demonstrate that our GPU-accelerated RT-TDDFT implementation in ABACUS produces quantitatively accurate and physically consistent results across a wide range of material classes and different application scenarios. The seamless integration of CPU and GPU computations, combined with robust gauge handling, basis-set flexibility, and energy-conserving Ehrenfest dynamics, establishes a reliable foundation for large-scale, long-timescale simulations of ultrafast electron dynamics in realistic materials.

\subsection{Performance Analysis}\label{sec:Performance_Analysis}

\subsubsection{System Size Scaling Test}\label{sec:System_Size_Scaling_Test}

To systematically evaluate the computational efficiency and scalability of our heterogeneous RT-TDDFT implementation, we conducted benchmark calculations on bulk silicon supercells with system sizes ranging from 48 to 1200 atoms. All benchmarks were performed on two distinct hardware configurations to ensure a fair comparison between the CPU-only baseline and the GPU-accelerated implementation.

The CPU-only benchmarks were executed on a dual-socket node equipped with two Intel Xeon Gold 6348 CPUs (Ice Lake architecture, 2.60 GHz), providing a total of 56 physical cores. To rigorously evaluate the CPU performance, we optimized the parallelization strategy by exploring different hybrid MPI/OpenMP configurations. Since the efficiency of different modules in ABACUS varies with the parallel decomposition, we tested four distinct ratios while maintaining full utilization of the 56 cores: 1 MPI rank with 56 OpenMP threads ($\mathtt{Np}=1$, $\mathtt{Nt}=56$), 2 MPI ranks with 28 OpenMP threads ($\mathtt{Np}=2$, $\mathtt{Nt}=28$), 4 MPI ranks with 14 OpenMP threads ($\mathtt{Np}=4$, $\mathtt{Nt}=14$), and 8 MPI ranks with 7 OpenMP threads ($\mathtt{Np}=8$, $\mathtt{Nt}=7$). This ensures that our speedup comparisons are made against an optimized CPU baseline rather than a default configuration.

The GPU-accelerated benchmarks were conducted on a heterogeneous node featuring an NVIDIA A800 GPU (80 GB VRAM, PCIe Gen4 interface) hosted by an Intel Xeon Platinum 8358P CPU (Ice Lake architecture, 2.60 GHz, 32 cores). For these calculations, we employed a single MPI process offloading to the GPU, with the host-side OpenMP threads set to 32 to fully utilize the available host CPU resources for non-offloaded tasks.

We compared the performance of two gauge formulations, the velocity gauge and the hybrid gauge, across both hardware platforms. This comparison is particularly critical, as the velocity gauge introduces additional computational complexity due to the position-dependent phase factors in nonlocal pseudopotentials, a specific bottleneck our GPU implementation aims to address. The length gauge was excluded from this scaling test for two reasons: theoretically, the length gauge is incompatible with the periodic boundary conditions of the bulk silicon systems studied here, as stated in Sec.~\ref{sec:Length_Gauge}; computationally, the hybrid gauge shares a nearly identical algorithmic structure and cost profile with the length gauge, effectively serving as a performance proxy for the latter.

\begin{table}[!htbp]
    \centering
    \small
    \caption{Configuration of bulk silicon systems used for the system size scaling tests.}
    \label{tab:system_size}
    \vspace{1ex}
    \begin{tabularx}{0.4\textwidth}{>{\centering\arraybackslash}X
        >{\centering\arraybackslash}X
        >{\centering\arraybackslash}X}
        \toprule
        $N_{\text{atoms}}$ & Supercell & $N_{\text{basis}}$ \\
        \midrule
        48   & $1\times2\times3$ & 1056   \\
        96   & $2\times2\times3$ & 2112   \\
        144  & $2\times3\times3$ & 3168   \\
        192  & $2\times3\times4$ & 4224   \\
        240  & $2\times3\times5$ & 5280   \\
        320  & $2\times4\times5$ & 7040   \\
        384  & $3\times4\times4$ & 8448   \\
        480  & $3\times4\times5$ & 10560  \\
        576  & $3\times4\times6$ & 12672  \\
        640  & $4\times4\times5$ & 14080  \\
        800  & $4\times5\times5$ & 17600  \\
        1000 & $5\times5\times5$ & 22000  \\
        1200 & $5\times5\times6$ & 26400  \\
        \bottomrule
    \end{tabularx}
\end{table}

Regarding the computational parameters, we employed a TZDP numerical atomic orbital basis set for silicon, specifically configured as $3s3p2d$ with a radial cutoff of $8\ \mathrm{a.u.}$. This configuration yields 22 basis functions per atom, resulting in a total basis set size ($N_{\text{basis}}$) of $22 \times N_{\text{atoms}}$. Consequently, the dimensions of the dense Hamiltonian and overlap matrices handled during the propagation are $N_{\text{basis}} \times N_{\text{basis}}$. The energy cutoff (\texttt{ecutwfc}) was set to 100 Ry. Given the large real-space dimensions of the simulated supercells, the Brillouin zone was simply sampled using a single $\mathit{\Gamma}$ point. The specific supercell configurations used in the scaling tests, constructed by expanding an 8-atom cubic unit cell, are detailed in Table \ref{tab:system_size}.

\begin{figure*}[!htbp]
    \centering
    \includegraphics[height=0.27\textwidth]{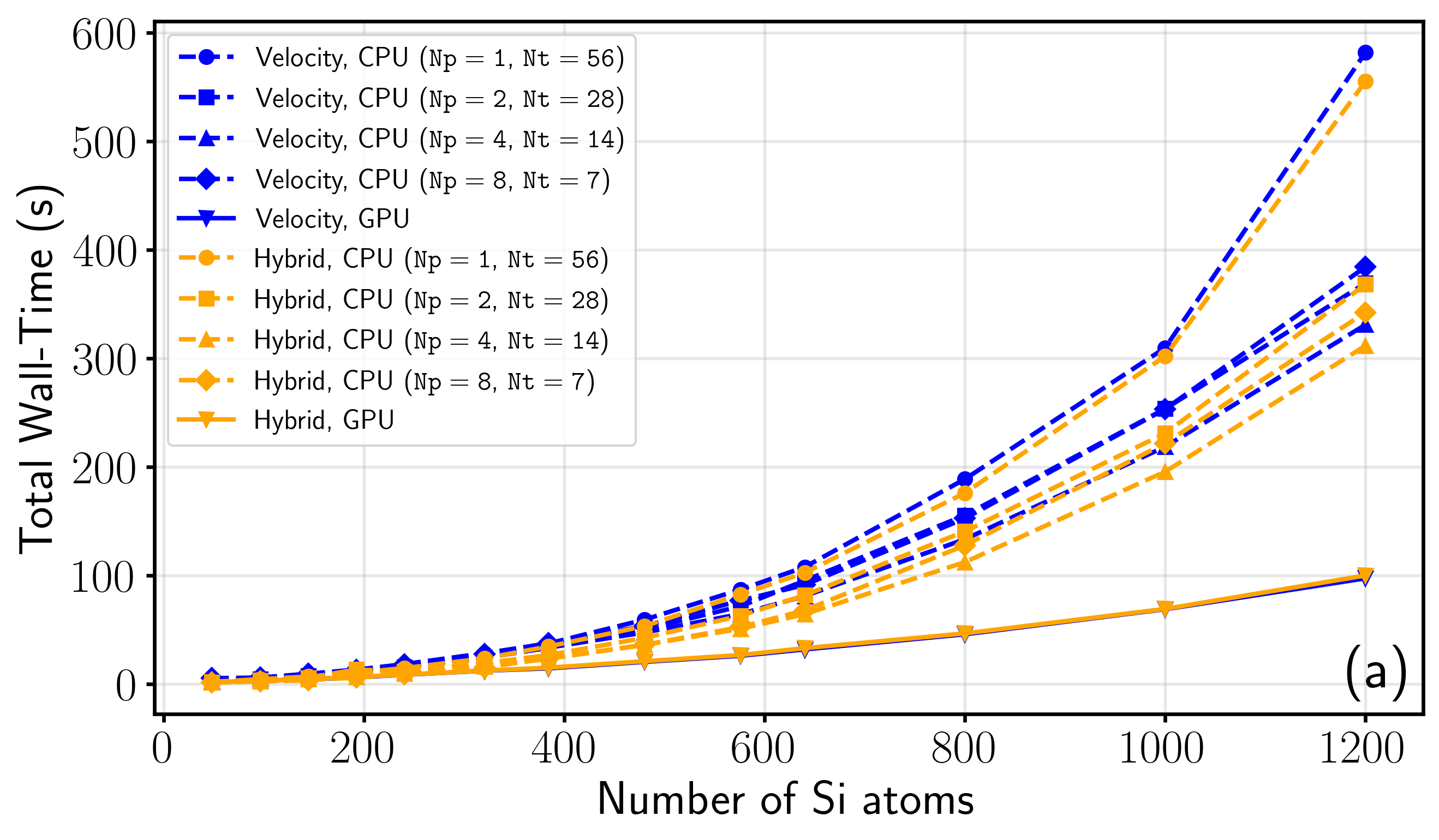}
    \includegraphics[height=0.27\textwidth]{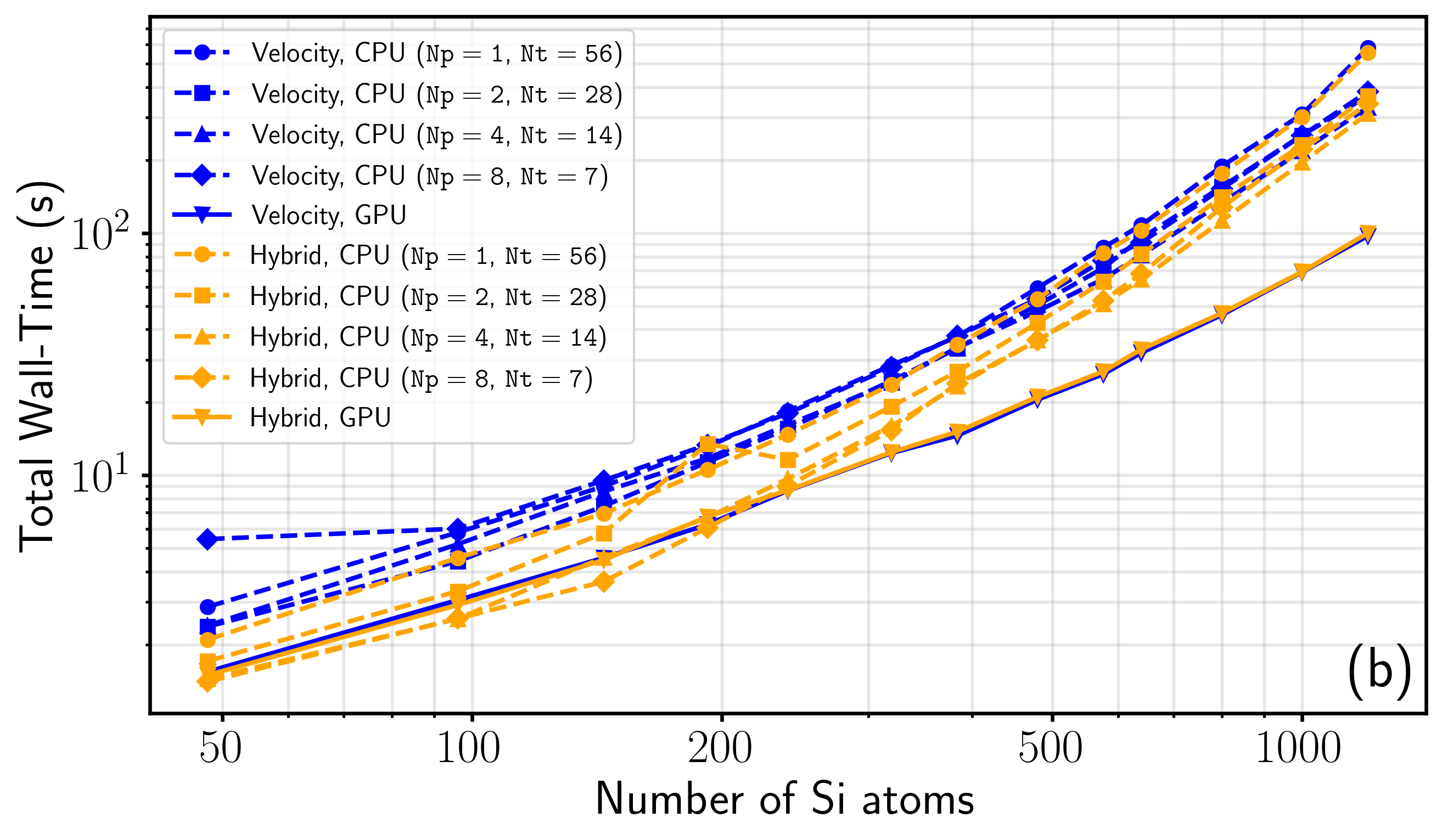}
    \caption{Total wall-clock time normalized by the number of electronic time-propagation steps (\texttt{evolve\_k} calls) for Si systems with varying numbers of atoms. (a) Total wall-clock time on a linear scale. (b) Total wall-clock time on a log-log scale.}
    \label{fig:size_total_time}
\end{figure*}

\textbf{Overall Performance.} The scaling of the total wall-clock time with respect to the system size ($N_{\text{atoms}}$) is presented in Fig.~\ref{fig:size_total_time}. Fig.~\ref{fig:size_total_time}(a) shows the absolute wall-clock time on a linear scale, while Fig.~\ref{fig:size_total_time}(b) provides a logarithmic view to better illustrate the power-law scaling behavior.

Several key trends can be observed from the results. First, regarding the CPU baseline performance, the hybrid MPI/OpenMP strategy plays a decisive role, revealing a non-trivial trade-off between thread synchronization and inter-process communication. As shown by the dashed lines in Fig.~\ref{fig:size_total_time}, simply maximizing the number of MPI ranks does not monotonically improve performance. Instead, we observe an optimal ``sweet spot'' at the intermediate configurations. Specifically, for large systems, the $\mathtt{Np}=4$, $\mathtt{Nt}=14$ configuration (triangle markers) consistently yields the lowest wall-clock time. In contrast, the single-process configuration ($\mathtt{Np}=1$, $\mathtt{Nt}=56$, circle markers) exhibits the poorest performance in the large-size regime. However, aggressively increasing the MPI parallelism to 8 ($\mathtt{Np}=8$, $\mathtt{Nt}=7$, diamond markers) also degrades performance, rendering it slower than both the 2-MPI and 4-MPI cases. For smaller systems, this trend is even more pronounced, where the 8 MPI configuration using velocity gauge becomes the slowest overall. This indicates that while MPI decomposition improves data locality, excessive domain decomposition introduces significant communication overhead that outweighs the computational benefits.

Second, the GPU-accelerated implementation (solid lines) demonstrates a dramatic performance advantage over the CPU baseline, particularly for large systems. For small systems ($N_{\text{atoms}} < 200$), the GPU speedup is modest due to the initial overhead of memory transfers and kernel launches, which takes up a portion of the runtime when the workload is insufficient to saturate the device. However, as the system size increases, the massive parallelism of the GPU is fully exploited. At $N_{\text{atoms}} = 1200$, the GPU implementation completes one time step in approximately 100 seconds, whereas the best CPU configuration requires over 300 seconds for the hybrid gauge and nearly 350 seconds for the velocity gauge. This corresponds to a speedup factor of approximately $3\times$ to $4\times$ relative to a fully utilized 56-core dual-socket CPU node.

Third, a comparison between the velocity gauge (blue lines) and the hybrid gauge (orange lines) reveals the efficacy of our specific optimizations. On the CPU, the velocity gauge is notably slower than the hybrid gauge. This discrepancy arises from the computationally expensive numerical integration of phase-modulated nonlocal projectors on spherical grids, as discussed in Sec.~\ref{sec:Spherical_Grid_Integration}. In contrast, on the GPU, the performance curves for the velocity gauge and hybrid gauge are nearly indistinguishable (solid blue vs. solid orange lines). This convergence indicates that our specialized GPU kernel for spherical grid integration effectively eliminates the overhead associated with the velocity gauge (as we are about to demonstrate subsequently), rendering the choice of gauge neutral with respect to computational cost.

Finally, the logarithmic plot in Fig.~\ref{fig:size_total_time}(b) confirms that the overall scaling of the GPU implementation remains favorable. The slope of the GPU curves is comparable to or slightly better than that of the CPU curves, suggesting that the offloading strategy does not introduce unfavorable scaling bottlenecks within the tested range.

\begin{figure*}[!htbp]
    \centering
    \includegraphics[height=0.27\textwidth]{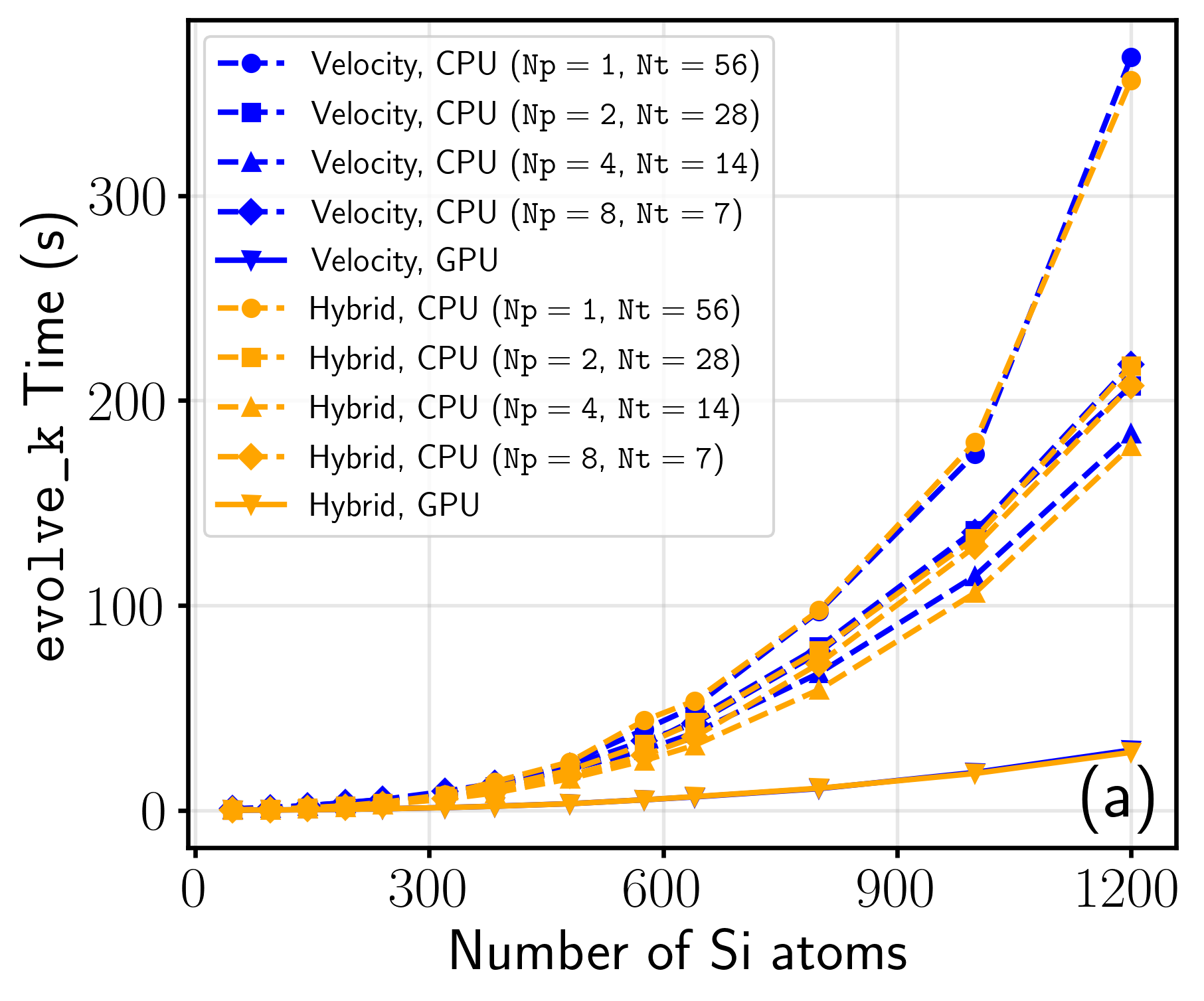}
    \includegraphics[height=0.27\textwidth]{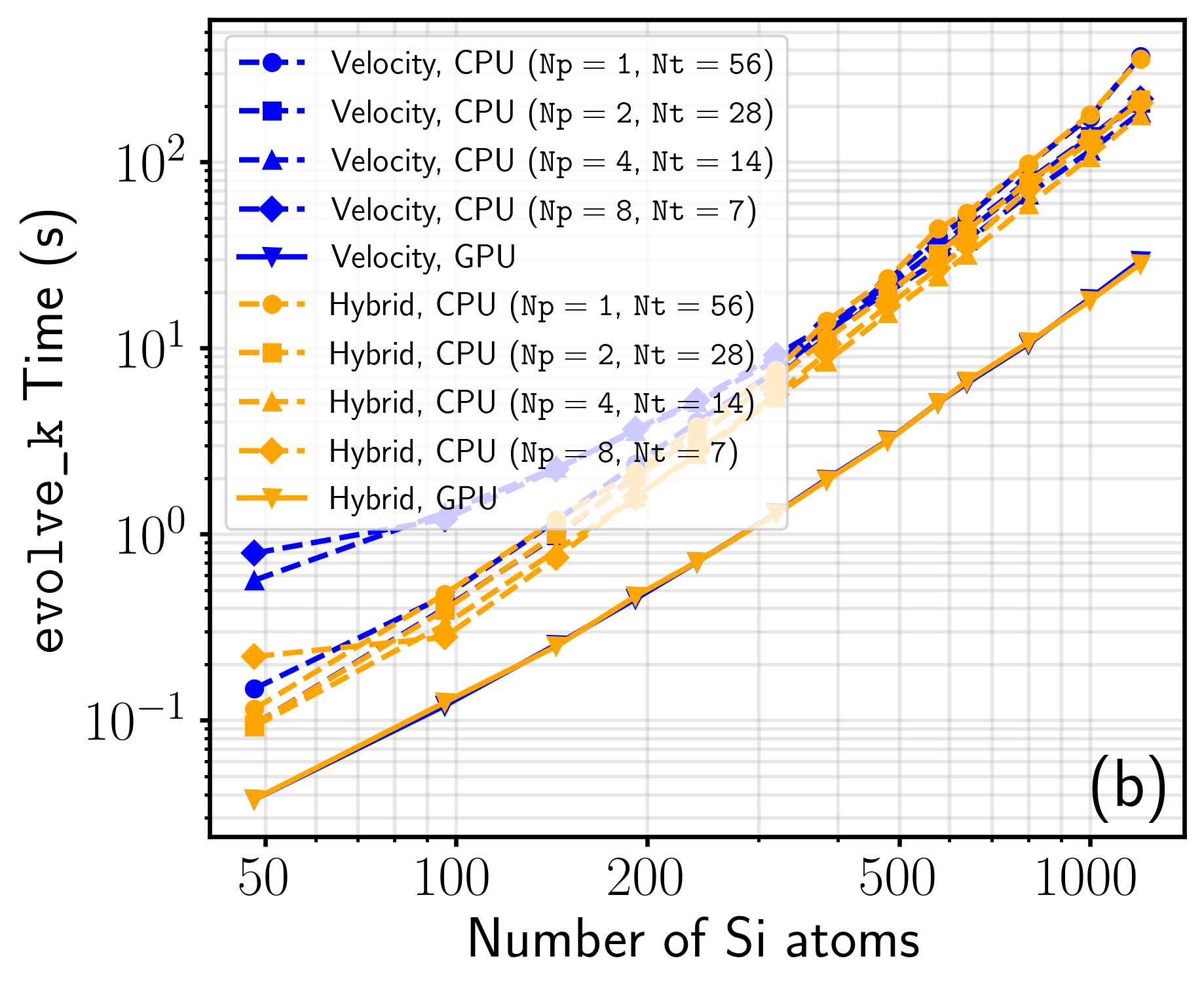}
    \includegraphics[height=0.27\textwidth]{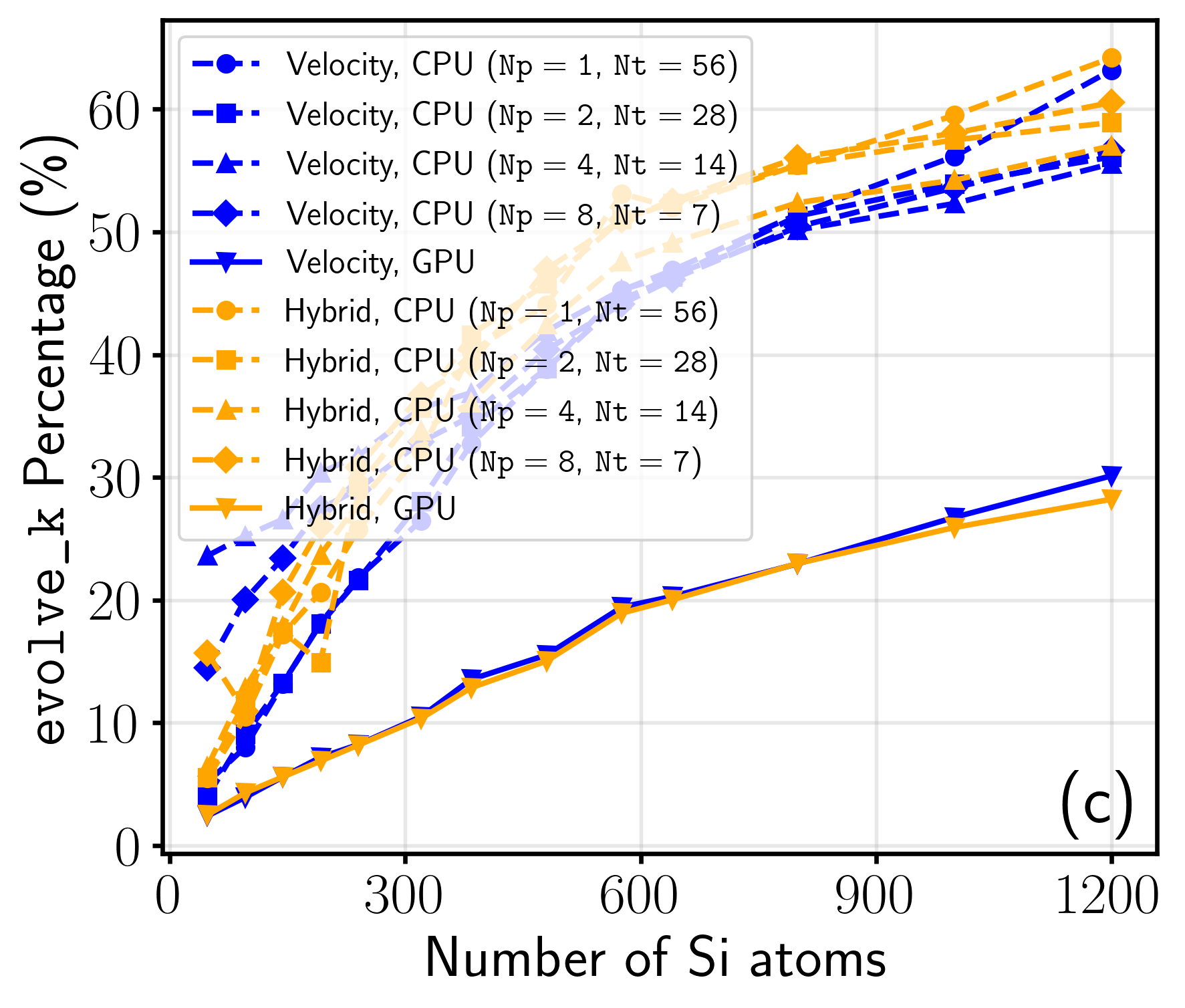}
    \caption{Performance analysis of the wavefunction propagation routine (\texttt{evolve\_k}). The wall-clock time is normalized by the number of electronic time-propagation steps (\texttt{evolve\_k} calls). (a) Wall-clock time on a linear scale. (b) Wall-clock time on a log-log scale. (c) The percentage of the \texttt{evolve\_k} routine relative to the total wall-clock time.}
    \label{fig:size_evolve_k}
\end{figure*}

\textbf{Wavefunction Propagation.} The propagation of the wavefunction, handled by the \texttt{evolve\_k} routine, relies heavily on dense linear algebra operations. Specifically, for the Crank-Nicolson propagator used throughout this work, the algorithm requires constructing matrices and solving linear systems. In our single-GPU implementation (as distinguished from the multi-GPU implementation discussed later), we propagate the wavefunction by solving the linear equation Eq.~\eqref{eq:Crank_Nicolson}. To avoid the computationally expensive explicit matrix inversion, this system is solved via LU factorization (\texttt{getrf}) of the operator on the left-hand side, followed by a forward and backward substitution (\texttt{getrs}) applied to the right-hand side, as described in Sec.~\ref{sec:Linear_Algebra_Operators}. These operations formally scale as $\mathcal{O}(N^3)$ with respect to the basis set size $N$, making them the theoretical bottleneck for large-scale simulations. The linear algebra routines are offloaded to the GPU using the cuBLAS and cuSOLVER libraries. However, porting the implementation to other heterogeneous computing platforms is straightforward due to the unified linear algebra interface encapsulated in our design, as detailed in Sec.~\ref{sec:Linear_Algebra_Operators}. This abstraction implies that no modifications are required at the physical algorithm level; enabling new hardware support simply involves adding the corresponding linear algebra library bindings within the abstraction layer.

The performance scaling of this routine is detailed in Fig.~\ref{fig:size_evolve_k}. Fig.~\ref{fig:size_evolve_k}(a) and (b) reveal a clear difference in how different hardware architectures handle these cubic-scaling workloads. On the CPU, the performance is extremely sensitive to the parallel decomposition strategy. Notably, the single-MPI configuration (blue/orange circles) exhibits a drastic increase in runtime for large systems, reaching approximately 360 seconds at $N_{\text{atoms}}=1200$. In contrast, configurations with higher MPI counts (especially the $\mathtt{Np}=4$, $\mathtt{Nt}=14$ configuration) effectively mitigate this issue, reducing the runtime to approximately 180--220 seconds by distributing the matrix blocks more effectively using the 2D block-cyclic distribution in the ScaLAPACK library \cite{ScaLAPACK_1996}.

However, the GPU implementation (solid lines) significantly shifts the performance bottleneck. By offloading these dense linear algebra kernels to GPU, the computation time is reduced by an order of magnitude. At $N_{\text{atoms}}=1200$, the GPU completes the \texttt{evolve\_k} step in roughly 30 seconds. This represents a speedup of approximately $6\times$ to $7\times$ compared to the best CPU configuration, and over $12\times$ compared to the single-MPI CPU baseline.

The impact of this acceleration on the overall simulation is best illustrated by the percentage breakdown in Fig.~\ref{fig:size_evolve_k}(c). On the CPU (dashed lines), the cost of wavefunction propagation grows disproportionately with system size. For the largest system, \texttt{evolve\_k} consumes over 60\% of the total runtime in the 1 MPI case and remains above 55\% even for optimized MPI configurations. This confirms that on standard CPU nodes, time propagation is the dominant computational bottleneck. Conversely, on the GPU (solid lines), the contribution of \texttt{evolve\_k} is significantly suppressed. Even at 1200 atoms, it accounts for only about 30\% of the total time. This shift demonstrates that our heterogeneous implementation successfully alleviates the primary bottleneck of RT-TDDFT, allowing the simulation cost to be more evenly distributed among other tasks such as grid integration.

Additionally, unlike the total time analysis, the performance difference between velocity gauge and hybrid gauge within \texttt{evolve\_k} is negligible on both platforms. This is expected, as the gauge choice primarily affects the complexity of the Hamiltonian construction, while the propagation step itself operates on the already-assembled Hamiltonian and overlap matrices, whose dense structure remains invariant under gauge transformation.

\begin{figure*}[!htbp]
    \centering
    \includegraphics[height=0.27\textwidth]{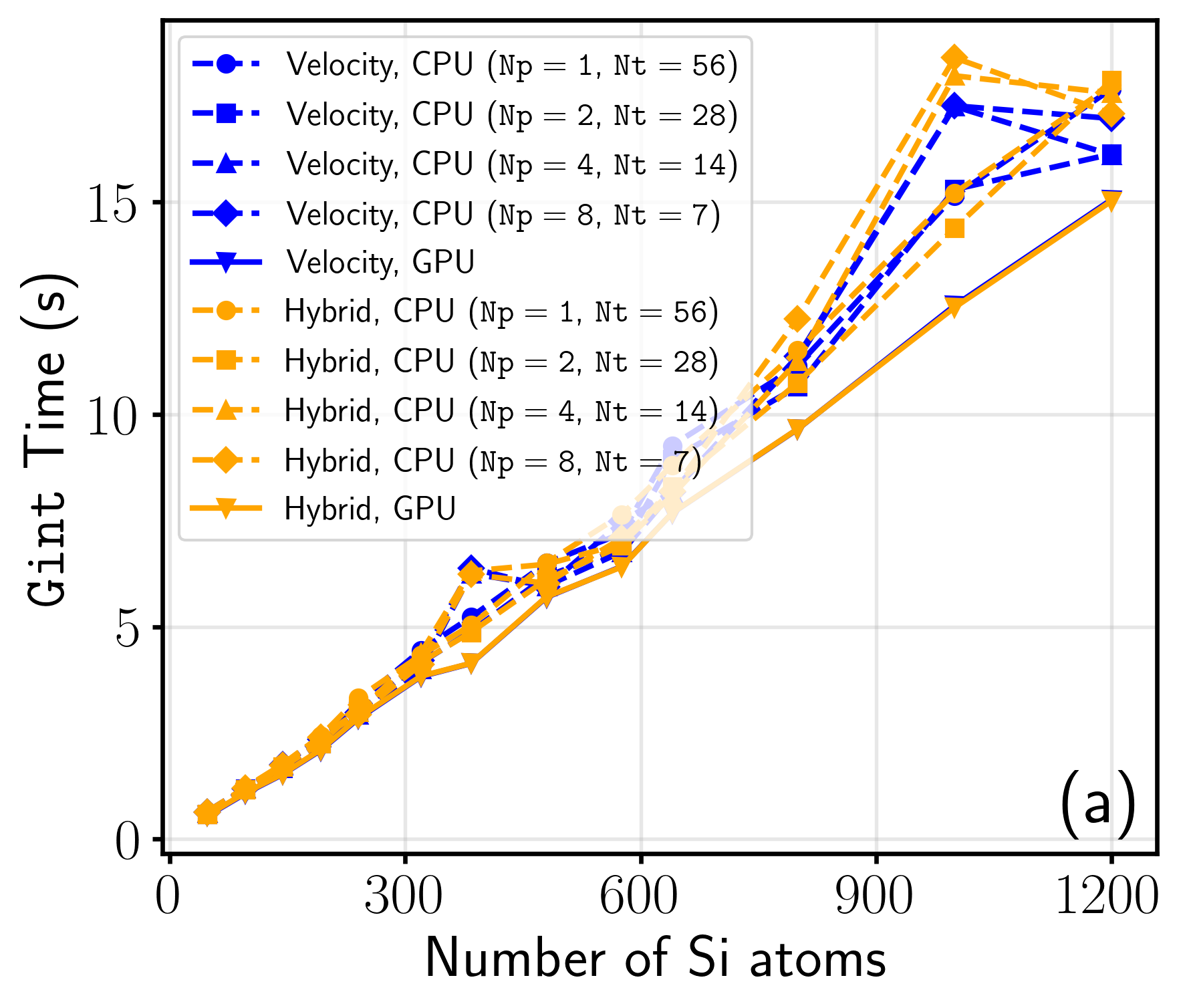}
    \includegraphics[height=0.27\textwidth]{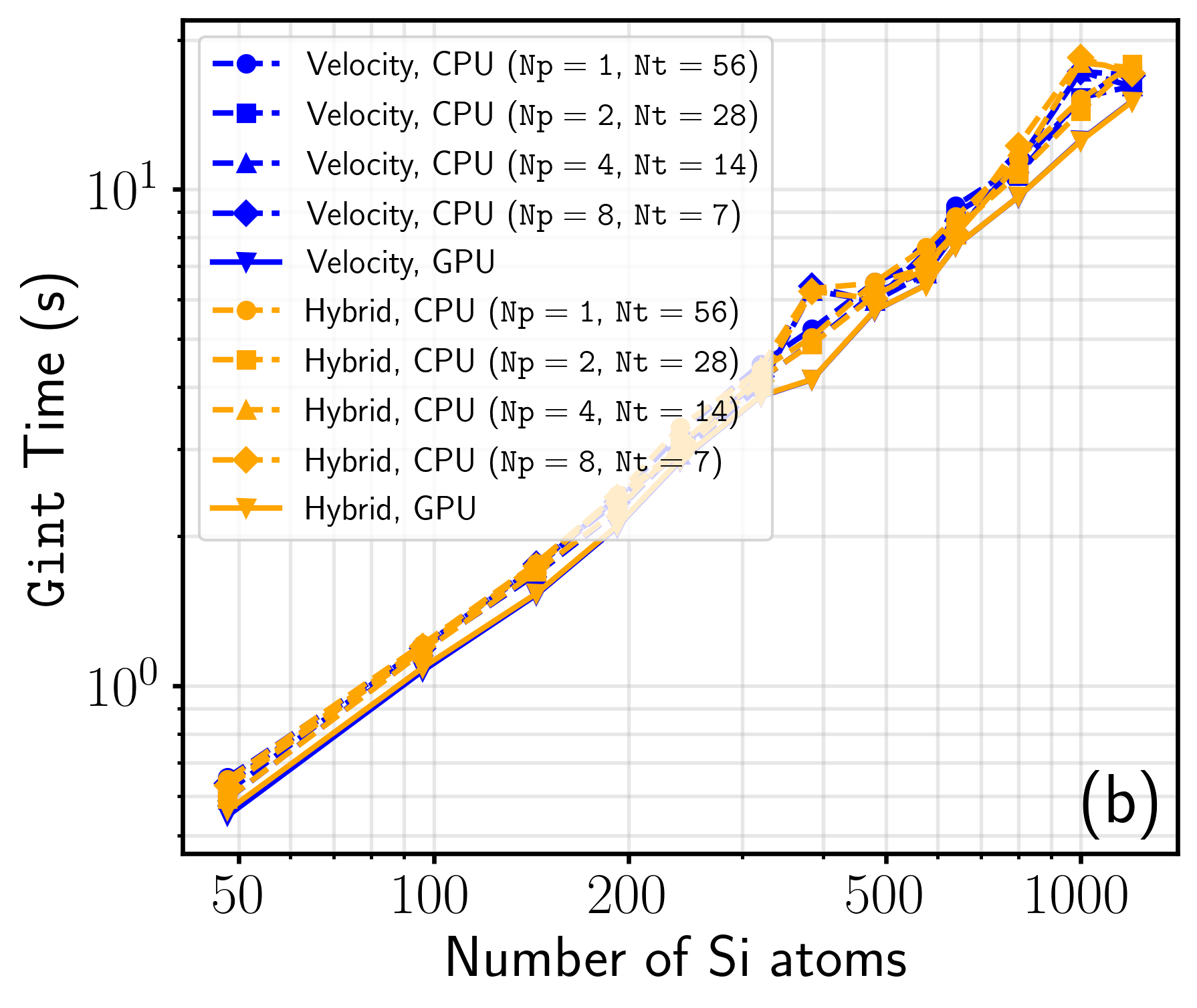}
    \includegraphics[height=0.27\textwidth]{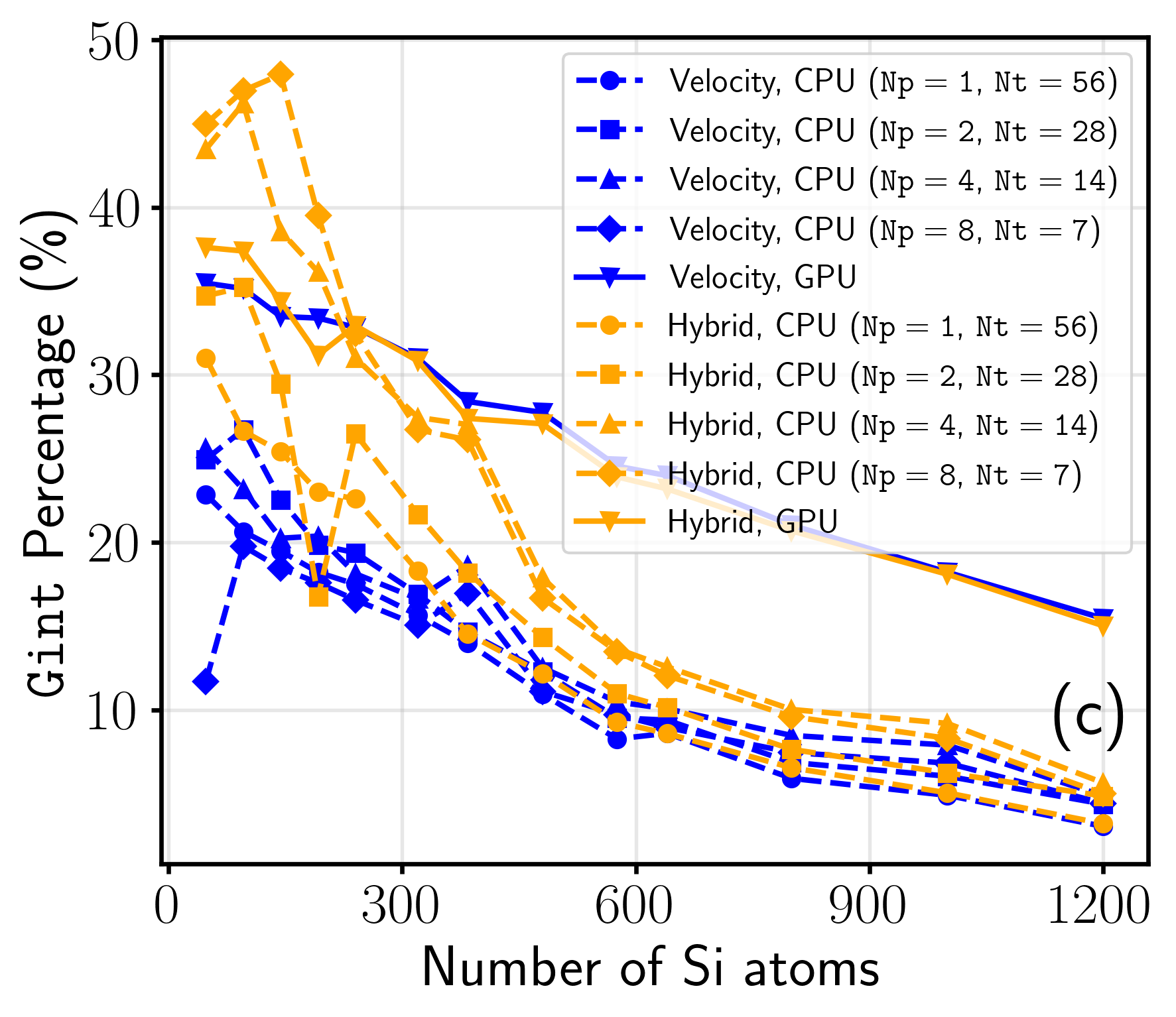}
    \caption{Performance analysis of the uniform grid integration module (\texttt{Gint}). The wall-clock time is normalized by the number of electronic time-propagation steps (\texttt{evolve\_k} calls). (a) Wall-clock time on a linear scale. (b) Wall-clock time on a log-log scale. (c) The percentage of the \texttt{Gint} module relative to the total wall-clock time.}
    \label{fig:size_Gint}
\end{figure*}

\textbf{Uniform Grid Integration.} The \texttt{Gint} module is responsible for computing physical quantities defined on the uniform real-space mesh, as stated in Sec.~\ref{sec:Uniform_Grid_Integration}. In the context of RT-TDDFT, the total time for this module is the sum of three primary components: the calculation of the charge density (\texttt{cal\_gint\_rho}), the local potential term (\texttt{cal\_gint\_vl}), and the corresponding forces (\texttt{cal\_gint\_fvl}). As established in our previous work \cite{ABACUS_GPU}, these operations are parallelized on GPUs using a batched strategy that groups atomic orbitals and grid points to exploit data locality, transforming irregular memory accesses into efficient matrix operations.

The performance scaling of the \texttt{Gint} module is presented in Fig.~\ref{fig:size_Gint}. Unlike the cubic scaling observed in wavefunction propagation, the grid integration exhibits a strict linear scaling, $\mathcal{O}(N)$, with respect to the system size. This linear trend is clearly visible in the log-log plot of Fig.~\ref{fig:size_Gint}(b), where the slope is approximately unity for all configurations.

In terms of absolute wall-clock time (Fig.~\ref{fig:size_Gint}(a)), the performance gap between the CPU and GPU implementations is much narrower than that of the propagation step. For the largest system ($N_{\text{atoms}}=1200$), the CPU configurations complete the integration in approximately 17 seconds per step, while the GPU implementation finishes in about 15 seconds. The fact that a single A800 GPU can marginally outperform a full dual-socket 56-core CPU node on this task is notable. Grid integration is typically memory-bandwidth bound due to the large volume of grid data and the need for atomic accumulation of results. The GPU implementation effectively leverages the high-bandwidth memory of the A800 to maintain high throughput, preventing this step from becoming a bottleneck.

An interesting feature of the heterogeneous acceleration is revealed in the percentage analysis (Fig.~\ref{fig:size_Gint}(c)). Here, we observe an inversion of the trends compared to the wavefunction propagation. For the CPU runs (dashed lines), the percentage of time spent on \texttt{Gint} drops precipitously as the system size increases, falling below 5\% for the 1200-atom system. This is not because \texttt{Gint} becomes faster, but because the cubic-scaling \texttt{evolve\_k} routine grows so rapidly that it dominates the computational cost compared to the linear-scaling grid operations.

Conversely, for the GPU runs (solid lines), the \texttt{Gint} module accounts for a significantly larger portion of the total runtime, approximately 15\% to 30\% for large systems. This behavior is consistent with Amdahl's law \cite{Amdahl1967}. Since the GPU has accelerated the dominant bottleneck (\texttt{evolve\_k}) by an order of magnitude, the remaining linear-scaling components, such as grid integration, naturally become more prominent in the relative breakdown. This result is highly favorable: it confirms that we have successfully suppressed the primary computational hotspot to the point where the formerly secondary $\mathcal{O}(N)$ operations now constitute a substantial part of the remaining, much reduced, total execution time.

\begin{figure*}[!htbp]
    \centering
    \includegraphics[height=0.27\textwidth]{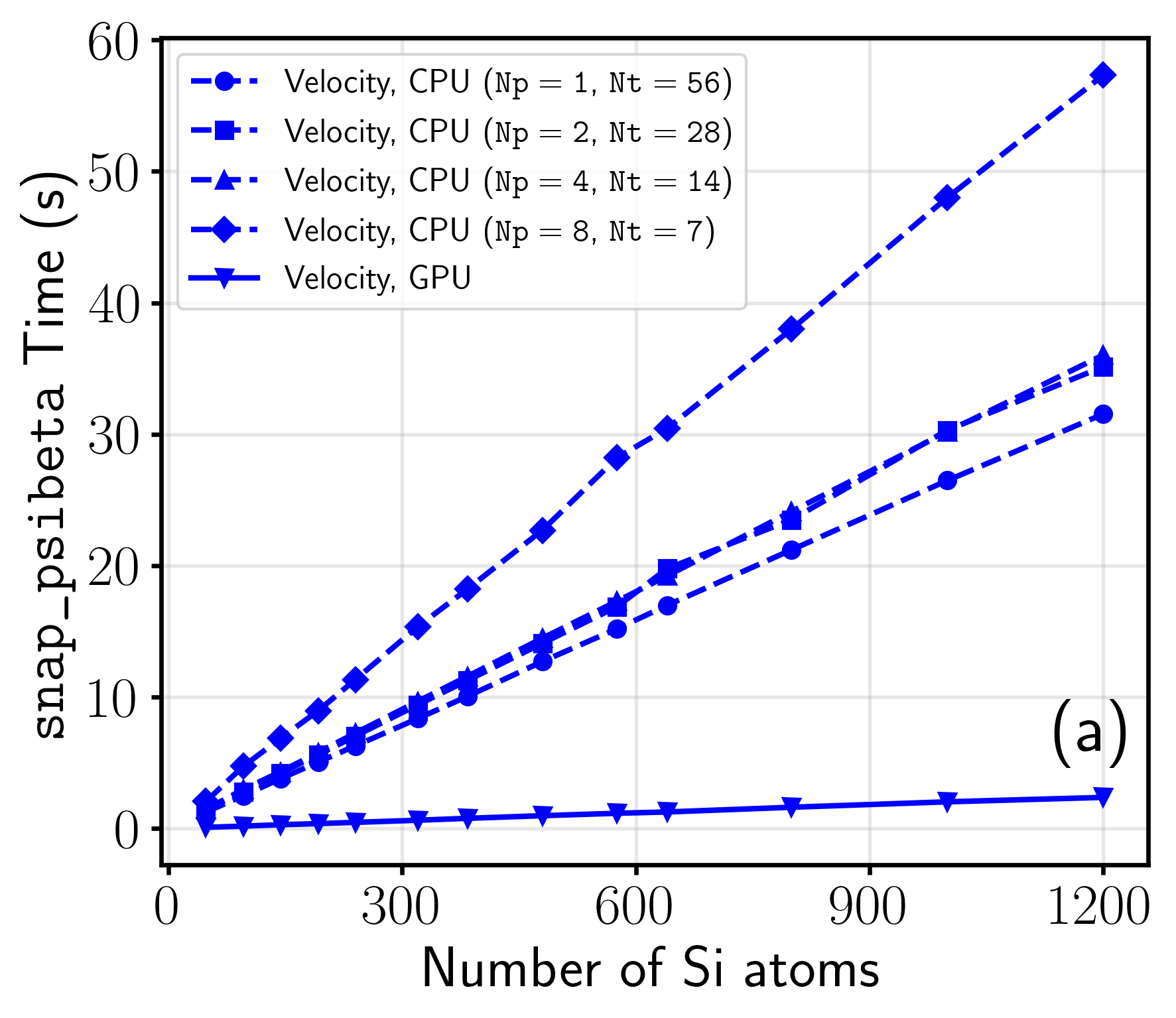}
    \includegraphics[height=0.27\textwidth]{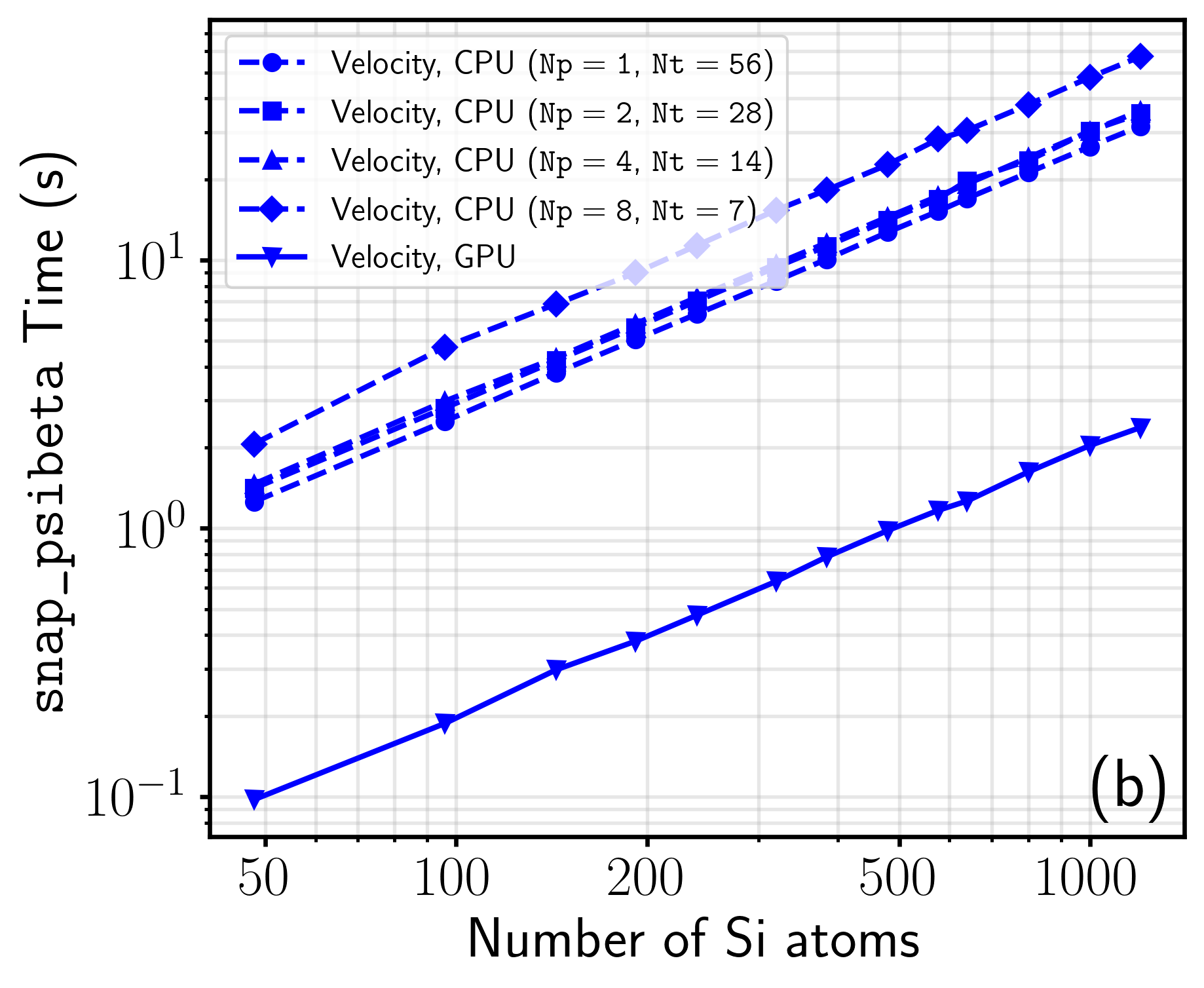}
    \includegraphics[height=0.27\textwidth]{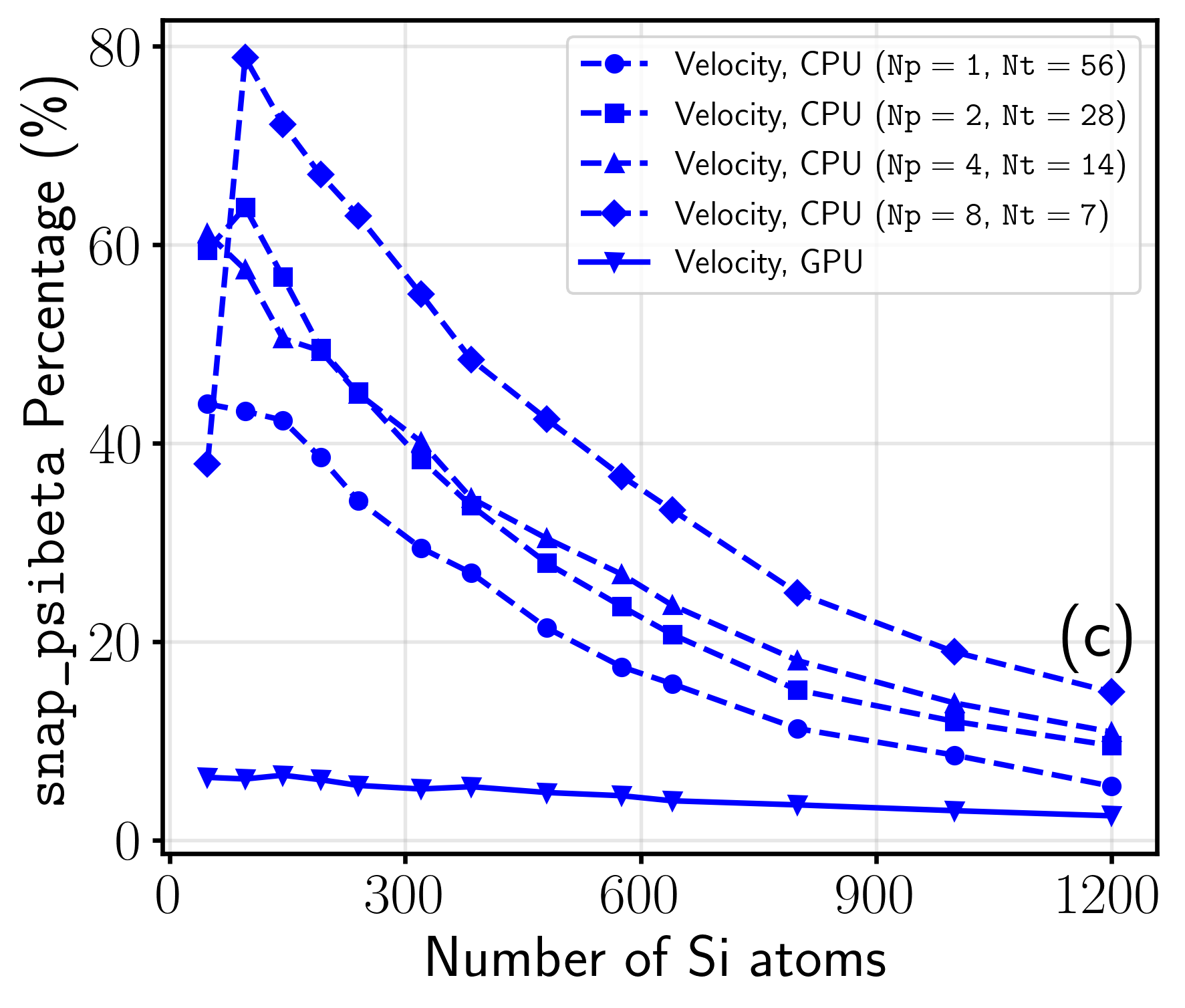}
    \caption{Performance analysis of the spherical grid integration module (\texttt{snap\_psibeta}), which is specific to the velocity gauge implementation. The wall-clock time is normalized by the number of electronic time-propagation steps (\texttt{evolve\_k} calls). (a) Wall-clock time on a linear scale. (b) Wall-clock time on a log-log scale. (c) The percentage of the \texttt{snap\_psibeta} module relative to the total wall-clock time.}
    \label{fig:size_snap_psibeta}
\end{figure*}

\textbf{Spherical Grid Integration.} A critical bottleneck identified in Sec.~\ref{sec:Velocity_Gauge} and Sec.~\ref{sec:Spherical_Grid_Integration} is the evaluation of nonlocal pseudopotential projector overlaps in the velocity gauge, denoted as \texttt{snap\_psibeta}. This step is unique to the velocity gauge formulation, as it requires integrating the position-dependent phase factor $\mathrm{e}^{-\mathrm{i}\bm{A}(t)\cdot\bm{r}}$ against the atomic projectors. Since this phase factor breaks the two-center integral symmetry used in standard LCAO methods, these integrals must be computed numerically on atom-centered spherical grids at every time step.

The performance scaling of this module is shown in Fig.~\ref{fig:size_snap_psibeta}. As expected, the computational cost scales linearly ($\mathcal{O}(N)$) with the number of atoms, as shown in the log-log plot (Fig.~\ref{fig:size_snap_psibeta}(b)). However, the prefactor of this linear scaling differs drastically between CPU and GPU architectures.

On the CPU (dashed lines in Fig.~\ref{fig:size_snap_psibeta}(a)), the single-MPI configuration ($\mathtt{Np}=1$, $\mathtt{Nt}=56$, circle markers) achieves the best performance, completing the calculation for the 1200-atom system in approximately 31 seconds. In contrast, increasing the number of MPI ranks degrades performance, with the $\mathtt{Np}=8$, $\mathtt{Nt}=7$ configuration (diamond markers) being the slowest at approximately 57 seconds. This behavior can be attributed to the parallelization strategy implemented in the CPU code. As described in our methodology in Sec.~\ref{sec:Spherical_Grid_Integration}, this module relies on OpenMP to parallelize the loop over neighboring atoms for each central atom. Since the number of neighbors can vary, the single-process configuration utilizing 56 OpenMP threads with dynamic scheduling achieves superior load balancing compared to the static domain decomposition enforced by multiple MPI ranks. Nevertheless, even in the best-case CPU scenario, this operation remains expensive, consuming nearly double the time of the uniform grid integration.

In sharp contrast, the GPU implementation (solid line) dramatically reduces this overhead. By mapping the irregular neighbor-loop structure to a batched kernel, the GPU completes the task for the 1200-atom system in less than 2.5 seconds. This represents a speedup of over $12\times$ compared to the fastest CPU configuration and nearly $23\times$ compared to the slowest.

The strategic impact of this acceleration is evident in Fig.~\ref{fig:size_snap_psibeta}(c). On the CPU, the \texttt{snap\_psibeta} module is a major contributor to the total runtime, accounting for 40\%--80\% of the cost in small-to-medium systems where matrix operation is not yet dominant. However, on the GPU, the contribution of this module is suppressed to below 5\% across all system sizes. This result confirms that our heterogeneous implementation effectively eliminates the ``gauge penalty'', allowing researchers to choose the appropriate gauge for their physical system without being constrained by computational cost.

\begin{figure*}[!htbp]
    \centering
    \includegraphics[height=0.27\textwidth]{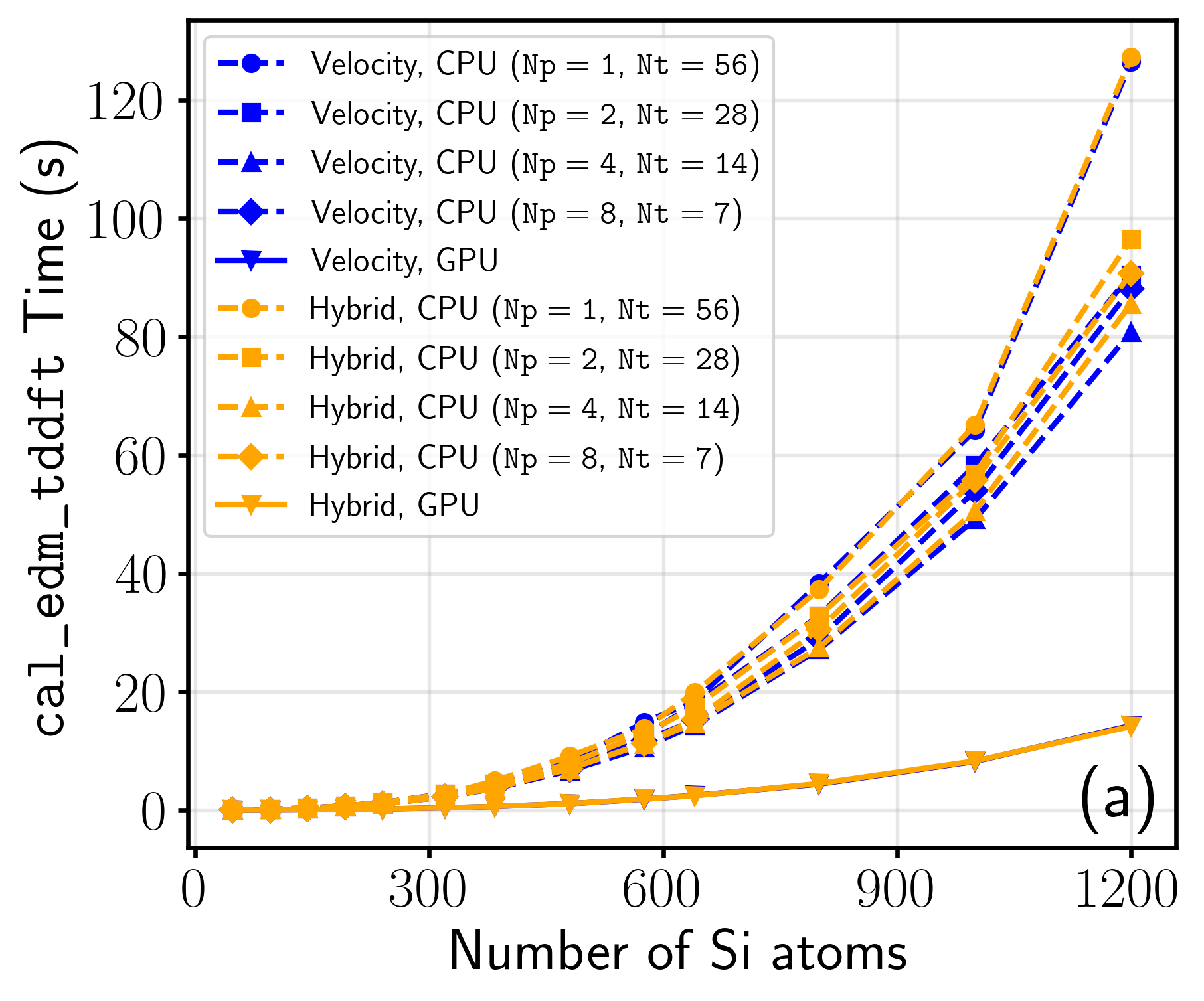}
    \includegraphics[height=0.27\textwidth]{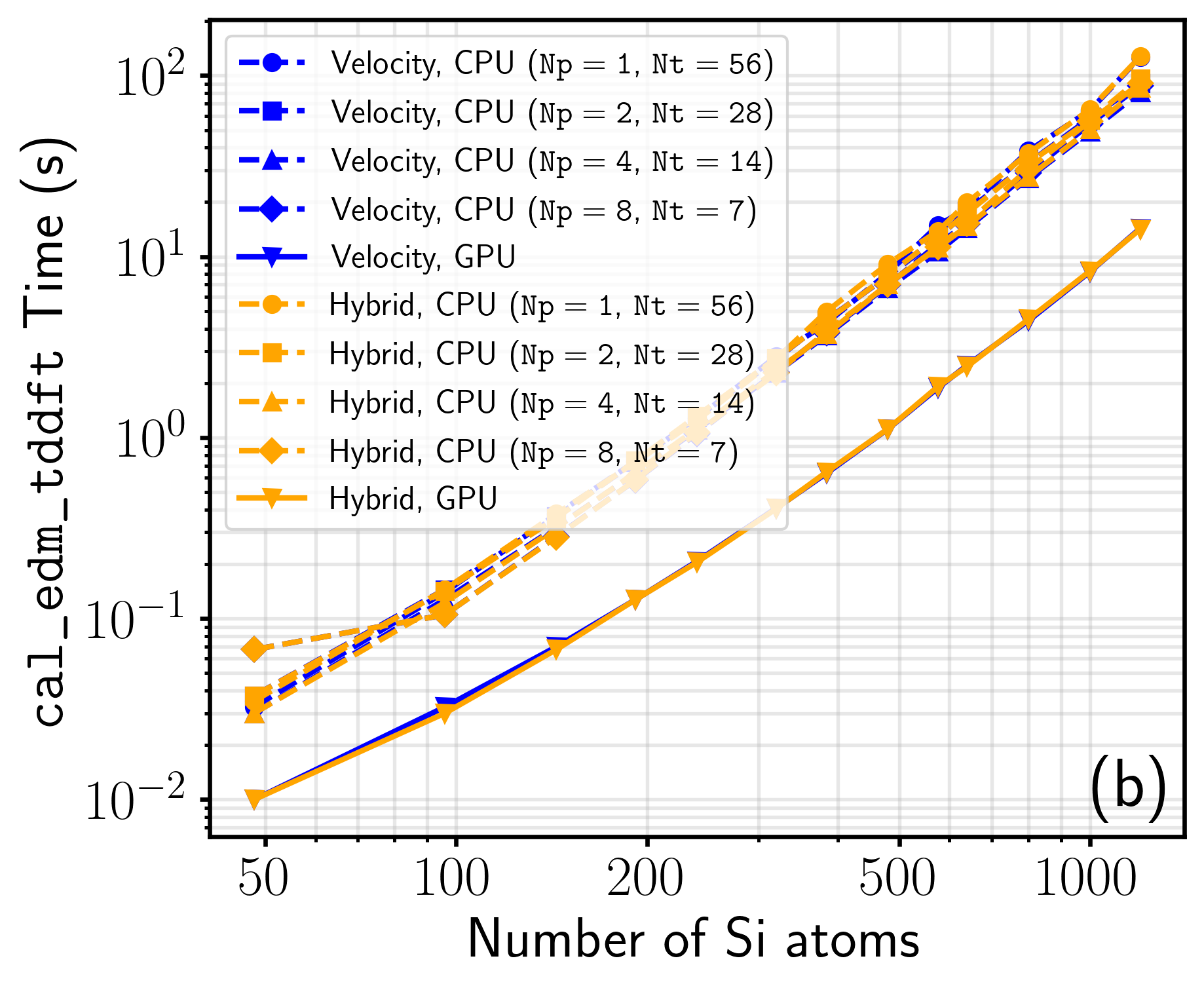}
    \includegraphics[height=0.27\textwidth]{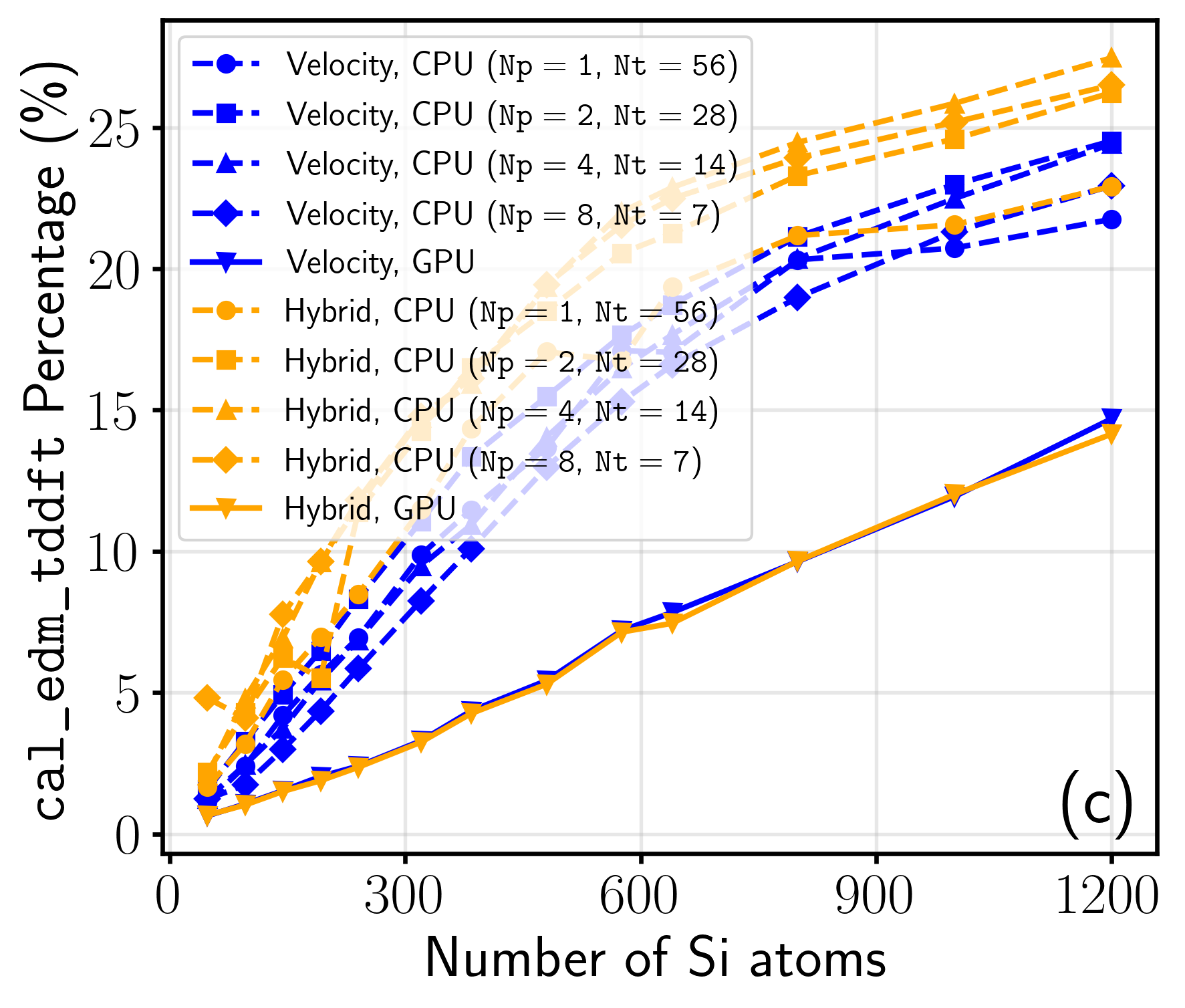}
    \caption{Performance analysis of the energy-density matrix calculation module (\texttt{cal\_edm\_tddft}). The wall-clock time is normalized by the number of electronic time-propagation steps (\texttt{evolve\_k} calls). (a) Wall-clock time on a linear scale. (b) Wall-clock time on a log-log scale. (c) The percentage of the \texttt{cal\_edm\_tddft} module relative to the total wall-clock time.}
    \label{fig:size_cal_edm}
\end{figure*}

\textbf{Energy-Density Matrix.} Finally, we examine the calculation of the EDM, a prerequisite for evaluating the nonorthogonality force in Ehrenfest dynamics. As defined in Eq.~\eqref{eq:EDM_definition}, this step involves explicitly computing the inverse of the overlap matrix and performing a sequence of dense matrix multiplications. Consequently, this module shares the same cubic scaling behavior ($\mathcal{O}(N^3)$) and computational characteristics as the wavefunction propagation.

The performance trends shown in Fig.~\ref{fig:size_cal_edm} closely mirror those of the \texttt{evolve\_k} routine. On the CPU, the execution time is heavily influenced by the MPI process count, with the $\mathtt{Np}=4$, $\mathtt{Nt}=14$ configuration again proving optimal for minimizing ScaLAPACK communication overheads while maintaining sufficient thread parallelism.

The GPU acceleration for this module is equally effective. By leveraging the cuSOLVER and cuBLAS libraries to handle the matrix inversion and multiplication on the device, the computation time is reduced by approximately an order of magnitude compared to the CPU baseline. For the 1200-atom system, the GPU implementation completes the EDM calculation in 14 seconds, whereas the CPU implementation requires significantly longer, ranging from 80 seconds (best case) to 130 seconds (worst case) depending on the MPI configuration, corresponding to speedups of approximately $6\times$ to $9\times$.

The percentage analysis in Fig.~\ref{fig:size_cal_edm}(c) confirms that while the EDM calculation can consume a noticeable fraction of the runtime on CPUs (up to about 25\%), it remains a relatively smaller contributor on the GPU (less than 15\%).

\begin{figure}[!htbp]
    \centering
    \includegraphics[height=0.24\textwidth]{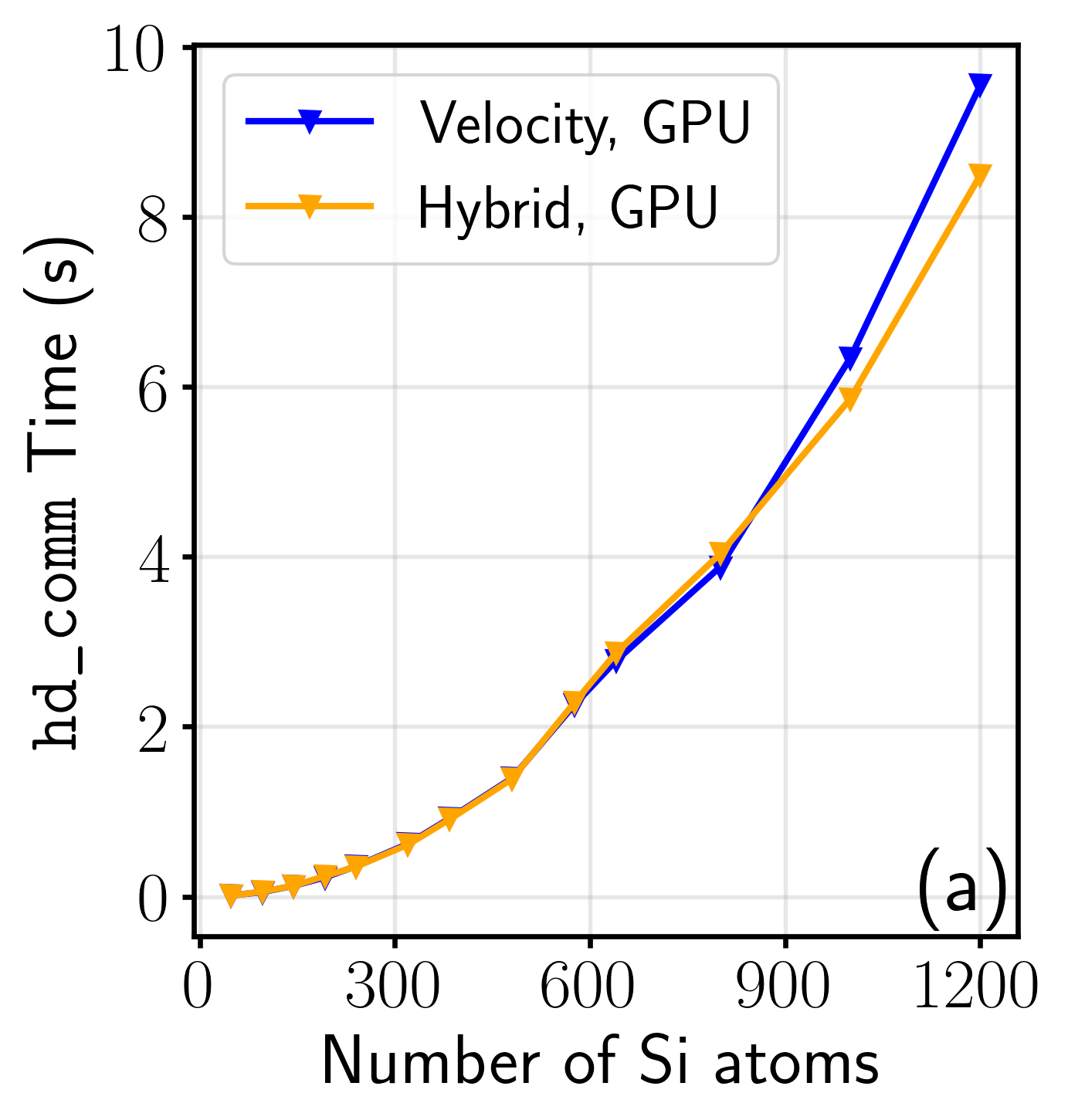}
    \includegraphics[height=0.24\textwidth]{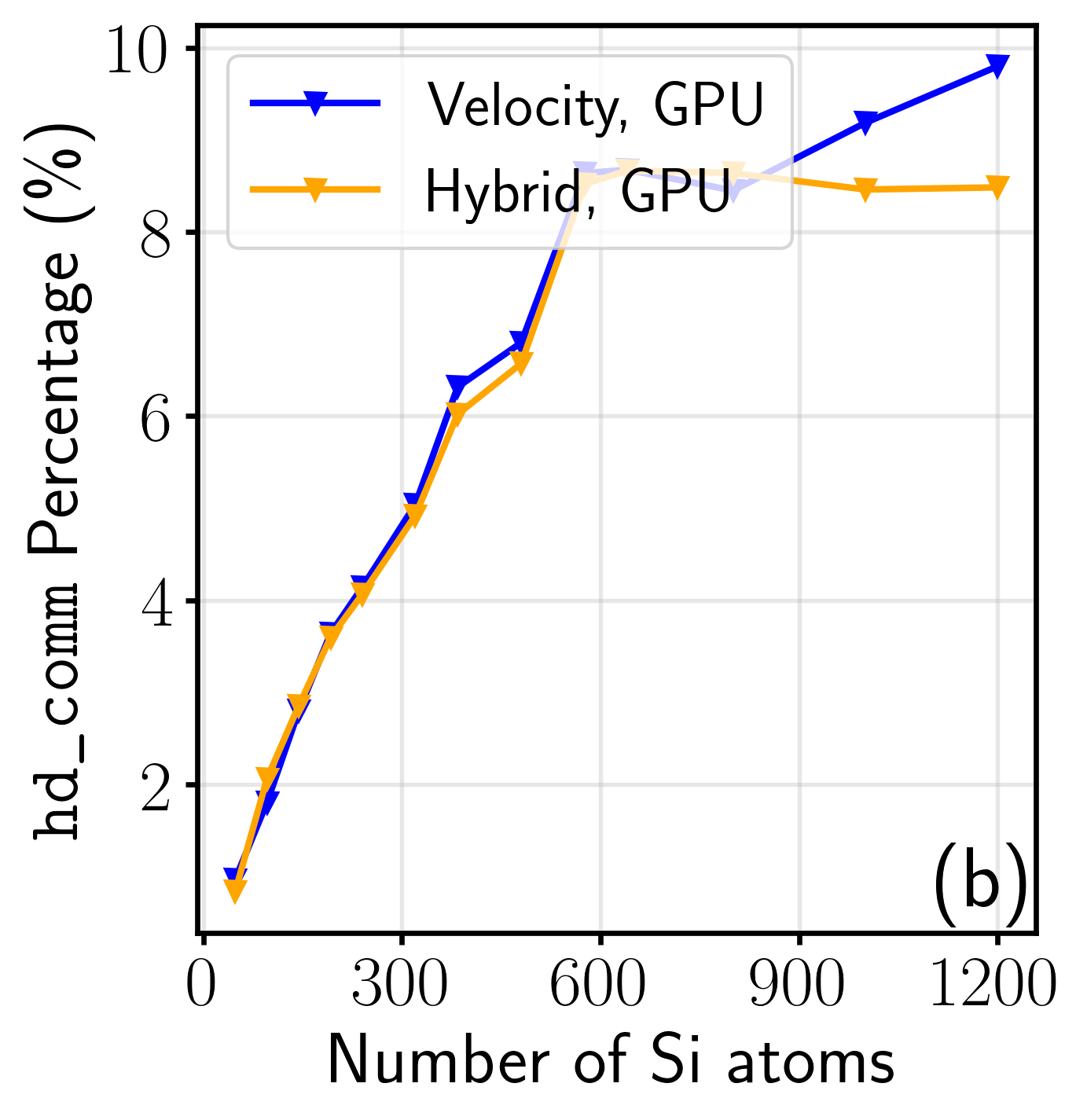}
    \caption{Analysis of the host-device (HD) communication overhead. The wall-clock time is normalized by the number of electronic time-propagation steps (\texttt{evolve\_k} calls). (a) Total HD communication time on a linear scale. (b) The percentage of HD communication time relative to the total wall-clock time. Note that these communication costs are already embedded within the timing statistics of the computational modules (e.g., \texttt{evolve\_k} and \texttt{cal\_edm\_tddft}) and do not represent an independent additive component.}
    \label{fig:hd_comm}
\end{figure}

\textbf{Communication Overhead.} Additionally, we briefly comment on the data transfer overhead between the host CPU and the GPU device, a common concern in heterogeneous high-performance computing. Fig.~\ref{fig:hd_comm} quantifies the total time spent on PCIe data transfers. While the absolute communication time grows with system size (reaching approximately 9 seconds for the largest system), its proportion relative to the total runtime remains moderate, peaking at around 9\%.

It is important to note that the majority of this communication overhead originates from the transfer of dense matrices required for the wavefunction propagation and EDM calculation. However, these linear algebra operations are fundamentally compute-bound. As the system size $N$ increases, the computational complexity for matrix operations scales as $\mathcal{O}(N^3)$, whereas the data transfer volume scales quadratically as $\mathcal{O}(N^2)$. Consequently, for large systems, the computational load grows significantly faster than the communication cost, naturally masking the latency of data movement.

For the grid integration modules, the communication overhead is further mitigated through algorithmic optimization. As detailed in \cite{ABACUS_GPU}, we employ a CUDA stream-based pipeline strategy that overlaps data transfer with kernel execution. By asynchronously transferring batch $i+1$ while the GPU computes batch $i$, the effective communication time perceived by the application is minimized. This combination of high arithmetic intensity in matrix operations and latency hiding in grid operations ensures that the PCIe bandwidth does not become a limiting bottleneck in our implementation.

\subsubsection{Strong Scaling Test}

Following the assessment of algorithmic scaling with system size, we evaluated the strong scaling performance of our heterogeneous implementation. In this regime, the problem size is held fixed while the number of computational resources (GPUs) is increased, providing insights into the parallel efficiency of the code.

The strong scaling benchmarks were conducted on a high-frequency heterogeneous computing cluster distinct from the nodes used in the previous section. Each computing node is equipped with four NVIDIA Tesla V100-SXM2 GPUs (32 GB VRAM per GPU) interconnected via NVLink. The host processor is an AMD Ryzen 9 9950X3D (16 cores, 32 threads) overclocked to 5.75 GHz. This specific configuration, featuring a high-frequency consumer-grade CPU, allows us to minimize the serial bottlenecks on the host side.

\begin{table}[!htbp]
    \centering
    \small
    \caption{Configuration of bulk silicon systems used for strong scaling tests.}
    \label{tab:strong_scaling_systems}
    \vspace{1ex}
    \begin{tabularx}{0.48\textwidth}{
        >{\centering\arraybackslash}X
        >{\centering\arraybackslash}X
        >{\centering\arraybackslash}X
        >{\centering\arraybackslash}X
    }
    \toprule
    Test Type & $N_{\text{atoms}}$ & Supercell & GPU Count \\
    \midrule
    \multirow{2}{*}{Intra-node} & 240  & $2\times3\times5$ & 1--4 \\
                                & 640  & $4\times4\times5$ & 1--4 \\
    \midrule
    \multirow{2}{*}{Inter-node} & 1000 & $5\times5\times5$ & 16--40 \\
                                & 1728 & $6\times6\times6$ & 16--40 \\
    \bottomrule
    \end{tabularx}
\end{table}

To cover different workload granularities, we designed two sets of scaling tests. \textit{Intra-node scaling}: performed on a single node using 1 to 4 GPUs. We selected two medium-sized silicon supercells containing 240 and 640 atoms, respectively. These sizes are representative of typical production runs where the workload per GPU is relatively high. \textit{Inter-node scaling}: performed across 4 to 10 computing nodes, corresponding to a range of 16 to 40 GPUs. For these large-scale tests, we employed larger supercells containing 1000 and 1728 atoms to ensure sufficient computational load to saturate the devices.

The configuration for these systems is summarized in Table \ref{tab:strong_scaling_systems}. Consistent with the system size scaling tests, we utilized the TZDP basis set (22 basis functions per atom) and a 100 Ry energy cutoff. For all multi-GPU calculations, we adopted a one-MPI-process-per-GPU strategy.

\begin{figure*}[!htbp]
    \centering
    \includegraphics[height=0.27\textwidth]{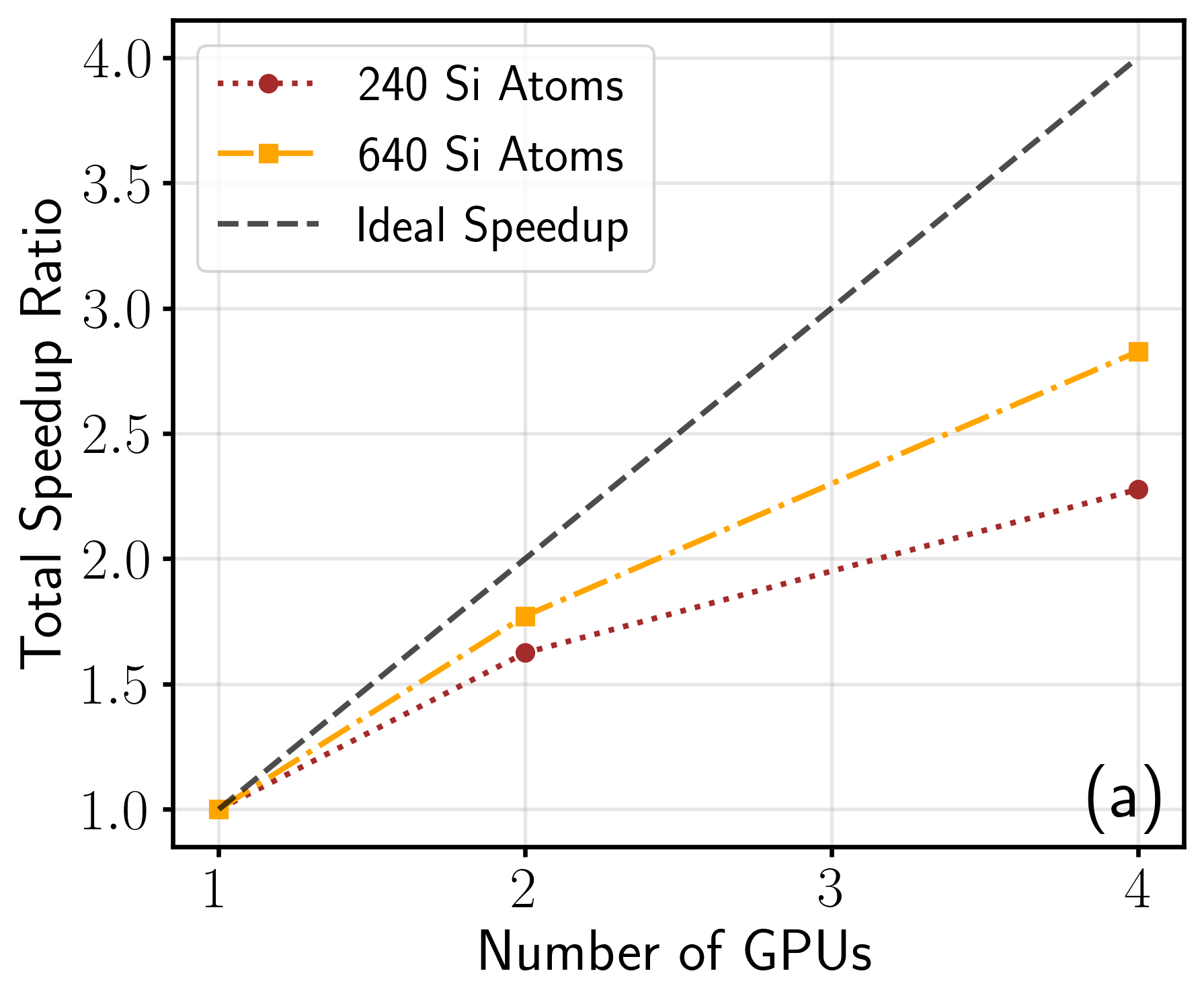}
    \includegraphics[height=0.27\textwidth]{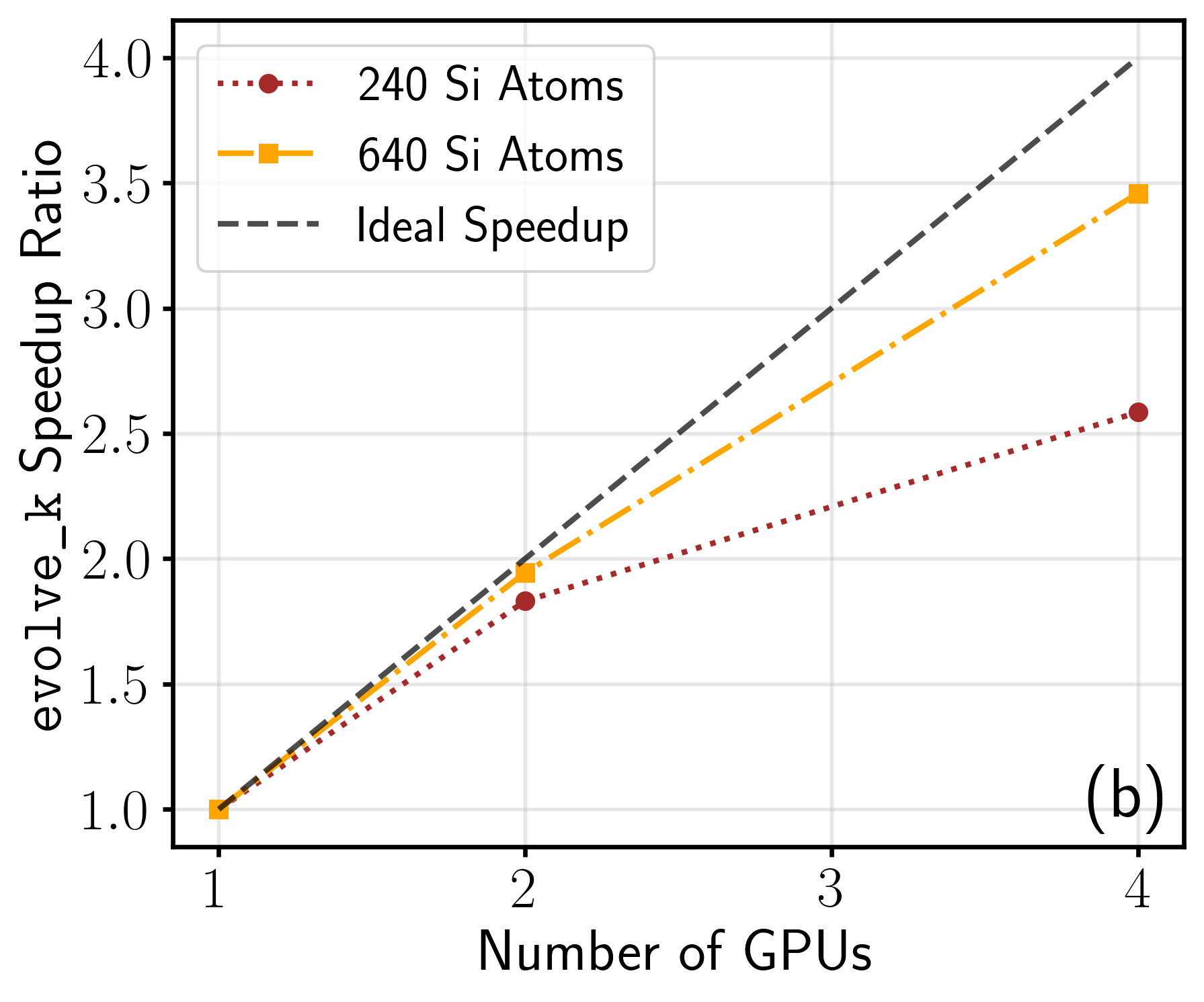}
    \includegraphics[height=0.27\textwidth]{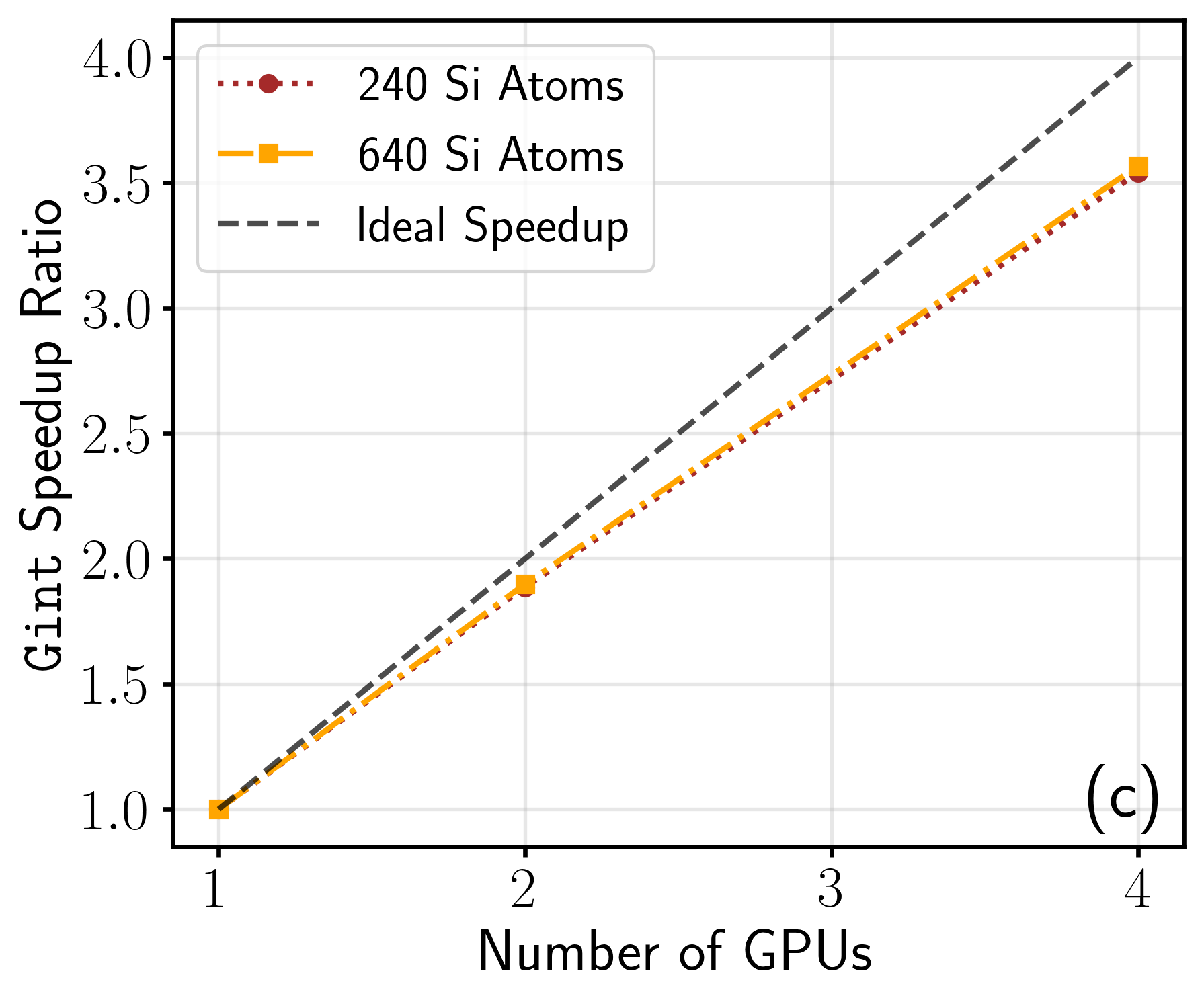}
    \caption{Intra-node strong scaling performance on a single computing node. The speedup ratios are calculated relative to the single-GPU baseline for two systems: 240 Si atoms (red dotted lines) and 640 Si atoms (orange dash-dotted lines). The ideal linear speedup is indicated by the grey dashed line. (a) Speedup of the total wall-clock time. (b) Speedup of the wavefunction propagation module (\texttt{evolve\_k}). (c) Speedup of the uniform grid integration module (\texttt{Gint}).}
    \label{fig:strong_scaling_one_node}
\end{figure*}

\textbf{Multi-GPU Implementation Strategy.} It is important to note that the multi-GPU acceleration capability for RT-TDDFT described herein is currently available as an experimental feature in the \texttt{develop} branch of ABACUS. While this functionality is not included in the stable version v3.9.0.24, it has been successfully integrated into the codebase and is scheduled for official inclusion starting from version v3.11.0-beta.1 as a preliminary multi-GPU support feature.

Before discussing the strong scaling results, we address a critical algorithmic adaptation required for the multi-GPU implementation using the NVIDIA cuBLASMp and cuSOLVERMp libraries. While the single-GPU implementation utilizes LU factorization (\texttt{getrf}) and substitution (\texttt{getrs}) to solve the Crank-Nicolson propagator equation, directly porting this workflow to the distributed multi-GPU backend encounters a specific library limitation.

According to the current cuSOLVERMp documentation \cite{cuSOLVERMp_docs}, the distributed linear solver routines based on LU factorization (\texttt{cusolverMpGetrs}) and Cholesky factorization (\texttt{cusolverMpPotrs}) support only a single right-hand side vector ($\mathtt{NRHS}=1$). In the context of RT-TDDFT, we must solve for the propagated wavefunction coefficients for all electronic bands simultaneously, where the number of bands ($N_{\text{band}}$) typically ranges from hundreds to thousands. Relying on the current \texttt{cusolverMpGetrs} implementation would necessitate a serial loop over $N_{\text{band}}$, effectively stripping the solver of parallelism across the bands and severely degrading computational efficiency for large systems.

To overcome the single right-hand side limitation of the LU-based or Cholesky-based solver in cuSOLVERMp, we adopted an alternative QR-based strategy to solve the Crank-Nicolson equation $AU=B$. The solution procedure enables fully batched parallel solving through three distinct steps: (1) \textit{QR Factorization}, where the denominator matrix $A$ is decomposed into unitary $Q$ and upper triangular $R$ matrices using \texttt{cusolverMpGeqrf}; (2) \textit{Orthogonal Transformation}, where the numerator matrix $B$ is updated by applying the conjugate transpose of $Q$ (i.e., $Y = Q^{\mathrm{H}} B$) via \texttt{cusolverMpOrmqr}; and (3) \textit{Triangular Solve}, where the final propagator $U$ is obtained by solving the triangular system $RU = Y$ using \texttt{cublasMpTrsm}. Although this QR-based workflow incurs a higher theoretical operation count than LU factorization, it unlocks massive parallelism across large $N_{\text{band}}$ or $N_{\text{basis}}$, preventing the linear algebra solver from becoming a serial bottleneck in multi-GPU simulations.

From a theoretical complexity perspective, the QR-based approach involves approximately $4N^3/3$ FLOPs for the factorization, which is roughly twice the cost of LU factorization ($2N^3/3$). However, in the distributed multi-GPU regime, this algorithmic overhead is justified and far outweighed by the ability to solve for all $N_{\text{band}}$ right-hand sides in a single parallel kernel launch. This strategy ensures that the massive compute capability of the multi-GPU cluster is fully utilized, preventing the linear algebra solver from becoming a serial bottleneck.

\textbf{Intra-node Scaling.} The strong scaling performance within a single node is presented in Fig.~\ref{fig:strong_scaling_one_node}. We focus our analysis on the two most computationally intensive components identified in the previous section: the wavefunction propagation (\texttt{evolve\_k}) and the uniform grid integration (\texttt{Gint}). These are the only modules currently targeted for multi-GPU acceleration, as they dominate the runtime for large-scale simulations.

The total speedup, shown in Fig.~\ref{fig:strong_scaling_one_node}(a), reveals a clear dependency on the problem size. For the smaller system (240 atoms), the speedup saturates early, reaching approximately $2.3\times$ on 4 GPUs. In contrast, the larger system (640 atoms) exhibits better scalability, achieving a speedup of $2.8\times$. This behavior highlights the importance of device saturation; larger systems provide sufficient arithmetic intensity to amortize the fixed overheads associated with kernel launches and inter-device communication.

A deeper insight is provided by decomposing the performance into specific modules. Fig.~\ref{fig:strong_scaling_one_node}(b) illustrates the scaling of the \texttt{evolve\_k} routine, which relies on the distributed matrix solver. For the 640-atom system, this module demonstrates near-ideal scaling behavior up to 2 GPUs and maintains a high parallel efficiency of 87\% at 4 GPUs ($3.5\times$ speedup). This confirms that our strategy effectively parallelizes the dense linear algebra workload despite the theoretical increase in operation count. However, for the 240-atom system, the efficiency drops to 65\% at 4 GPUs ($2.6\times$ speedup). This drop is expected for distributed dense linear algebra: as the local matrix block size shrinks, the ratio of communication (scaling as $\mathcal{O}(N^2)$) to computation (scaling as $\mathcal{O}(N^3)$) increases, making it more difficult to achieve ideal scaling.

On the other hand, the grid integration module (\texttt{Gint}) exhibits excellent scalability regardless of system size, as shown in Fig.~\ref{fig:strong_scaling_one_node}(c). Both the 240-atom and 640-atom systems achieve speedups of around $3.5\times$ on 4 GPUs, closely tracking the ideal linear scaling curve.

\begin{figure*}[!htbp]
    \centering
    \includegraphics[height=0.27\textwidth]{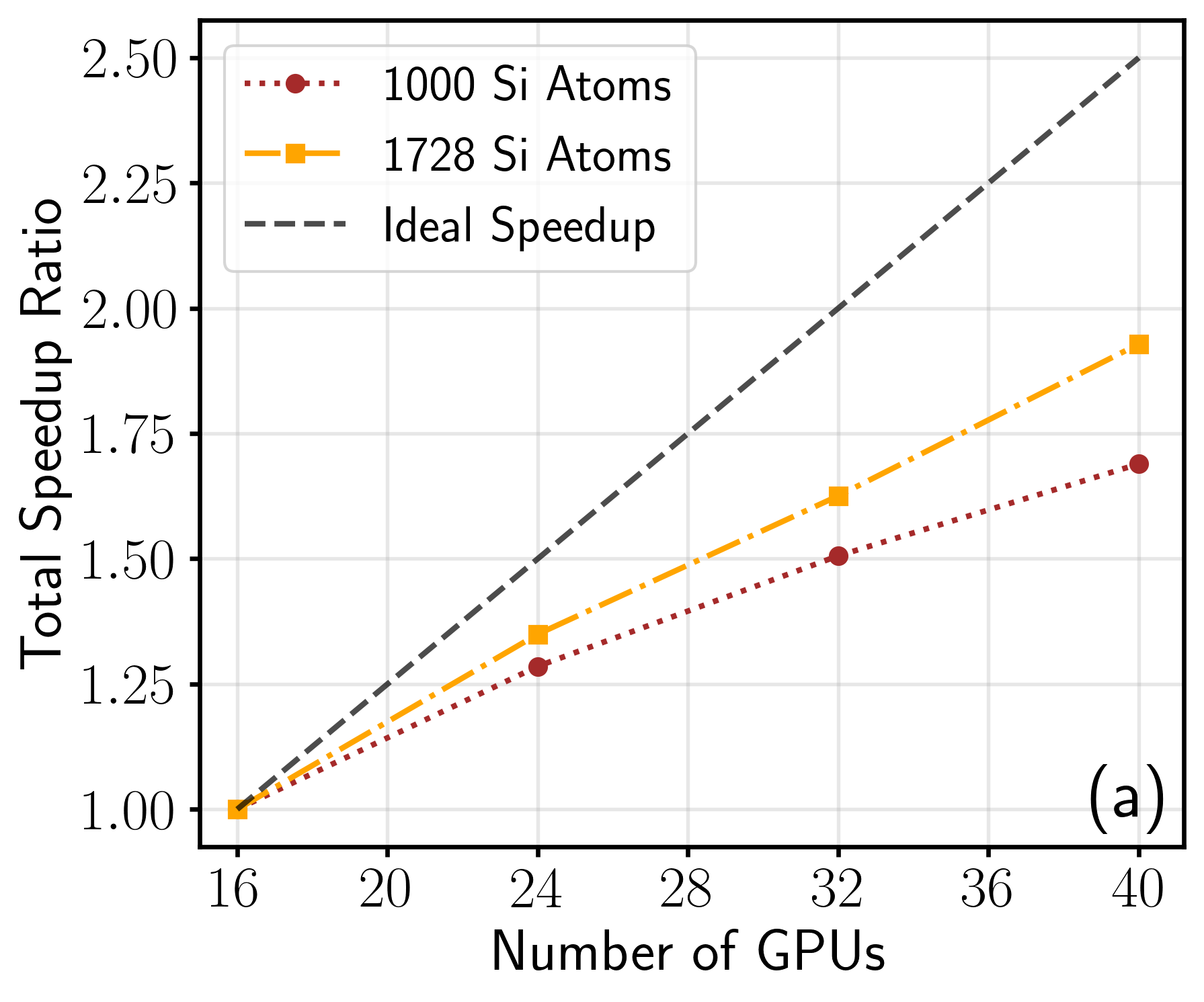}
    \includegraphics[height=0.27\textwidth]{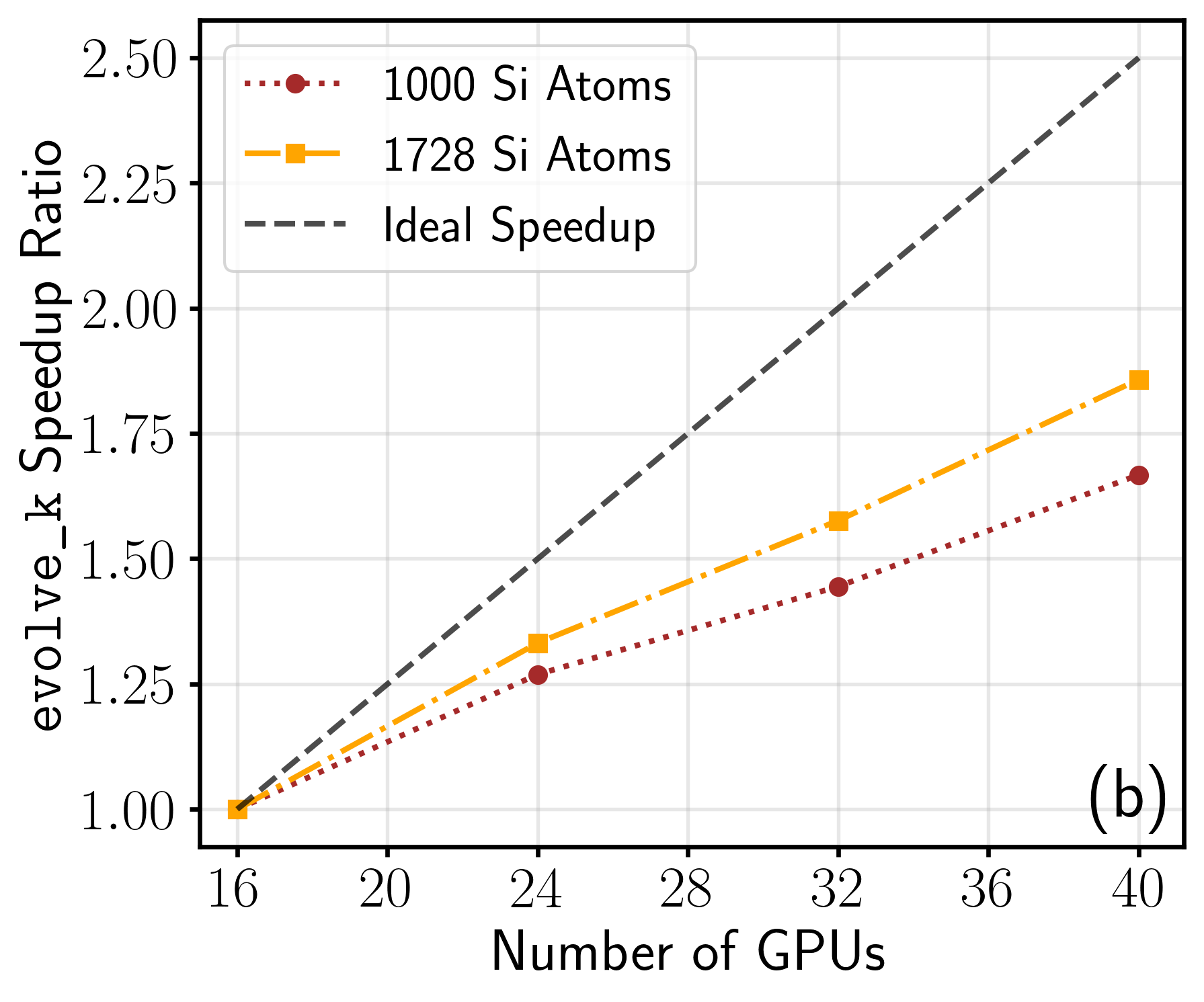}
    \includegraphics[height=0.27\textwidth]{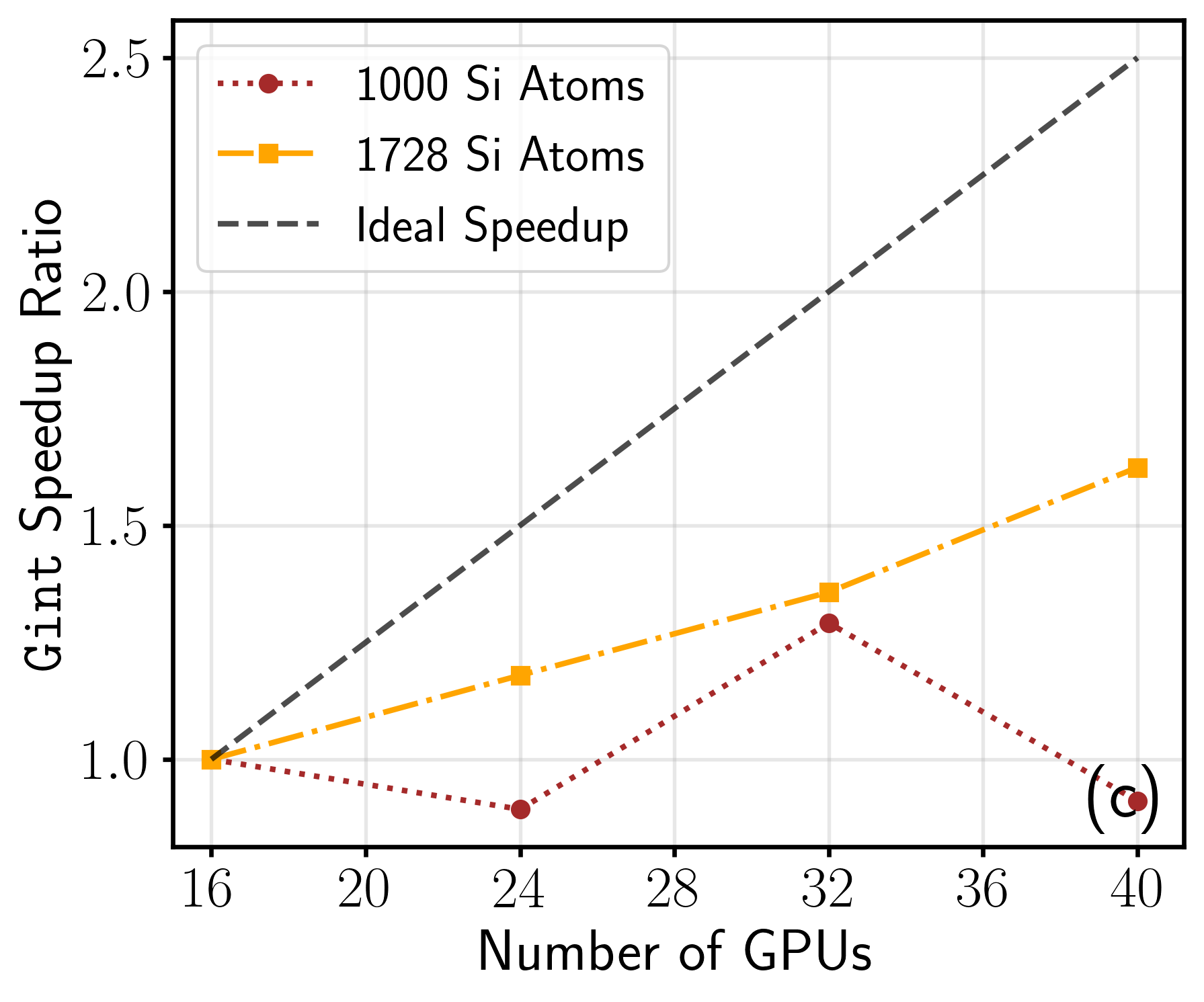}
    \caption{Inter-node strong scaling performance across multiple computing nodes (4 to 10 nodes, corresponding to 16 to 40 GPUs). The speedup ratios are calculated relative to the 4-node (16-GPU) baseline for two large-scale systems: 1000 Si atoms (red dotted lines) and 1728 Si atoms (orange dash-dotted lines). The ideal linear speedup is indicated by the grey dashed line. (a) Speedup of the total wall-clock time. (b) Speedup of the wavefunction propagation module (\texttt{evolve\_k}). (c) Speedup of the uniform grid integration module (\texttt{Gint}).}
    \label{fig:strong_scaling_multi_node}
\end{figure*}

\textbf{Inter-node Scaling.} Finally, we evaluate the scalability of our implementation across multiple computing nodes, pushing the parallelization to 40 GPUs. The results for the 1000-atom and 1728-atom systems are presented in Fig.~\ref{fig:strong_scaling_multi_node}. Note that speedups are normalized against the 16-GPU (4-node) baseline.

The total performance scaling, shown in Fig.~\ref{fig:strong_scaling_multi_node}(a), demonstrates the capability of our code to leverage cluster-level resources. For the largest system (1728 atoms), we achieve a total speedup of approximately $1.9\times$ when scaling from 16 to 40 GPUs (1), corresponding to a parallel efficiency of roughly 76\%. While the 1000-atom system shows slightly lower efficiency ($1.7\times$ speedup, 68\% efficiency), this is expected as the workload per GPU decreases.

The performance is primarily driven by the wavefunction propagation \texttt{evolve\_k}, as shown in Fig.~\ref{fig:strong_scaling_multi_node}(b). Despite the high communication overhead inherent in distributed dense linear algebra, the matrix solver maintains respectable scalability. For the 1728-atom case, the speedup reaches $1.86\times$ on 40 GPUs. This confirms that for sufficiently large matrices, the $\mathcal{O}(N^3)$ computational intensity effectively amortizes the inter-node communication latency.

In contrast, the \texttt{Gint} module in Fig.~\ref{fig:strong_scaling_multi_node}(c) exhibits significant saturation. The speedup for the 1728-atom system tops out at $1.6\times$ on 40 GPUs, diverging from the ideal linear curve. This behavior is a direct consequence of the algorithm's $\mathcal{O}(N)$ linear scaling. As the number of GPUs increases, the local grid volume per GPU shrinks rapidly, causing the surface-to-volume ratio, and thus the halo exchange communication overhead, to dominate. However, this sub-linear scaling is acceptable in the context of the overall simulation. As established in the system size scaling analysis in Sec.~\ref{sec:System_Size_Scaling_Test}, the grid integration consumes a progressively smaller fraction of the total runtime as the system size grows. Consequently, the saturation of \texttt{Gint} has a limited impact on the total time, which remains dominated by the scaling behavior of the cubic-scaling linear algebra components.

\section{Conclusion and Outlook}\label{sec:Conclusion}

In this work, we have presented a comprehensive heterogeneous implementation of RT-TDDFT within the ABACUS software package. By establishing a hardware-agnostic computing framework centered on the unified data abstraction \texttt{Tensor}, we have successfully decoupled high-level physical algorithms from low-level device management. This architecture not only facilitates the efficient execution of complex electronic dynamics simulations on modern GPU platforms but also ensures the long-term maintainability and portability of the codebase. A distinct contribution of this study is the resolution of the computational cost associated with the velocity gauge in simulations using numerical atomic orbitals. Through the development of a specialized GPU-accelerated spherical grid integration scheme, we have effectively eliminated the prohibitive cost of evaluating position-dependent phase factors.

Our implementation has been validated across a diverse range of systems, from isolated molecules to extended bulk solids. The results demonstrate excellent physical consistency with established benchmarks and reveal significant performance gains. On a single NVIDIA GPU, our code achieves a wall-clock speedup of $3\times$--$4\times$ compared to a fully utilized 56-core CPU node, with the critical spherical integration kernel accelerating by over $12\times$. Furthermore, the multi-GPU implementation, powered by a distributed linear solver strategy, exhibits robust strong scaling up to 40 GPUs with a parallel efficiency of approximately 76\%.

Looking forward, several challenges and opportunities remain. A primary objective for future development concerns the portability of multi-device acceleration. While our current \texttt{Tensor} framework successfully abstracts memory management and single-device linear algebra, enabling the code to run on diverse architectures such as NVIDIA GPUs, AMD GPUs, and Hygon DCUs, the distributed multi-GPU capability currently relies on the cuBLASMp and cuSOLVERMp libraries, restricting high-performance cluster-level simulations to the NVIDIA ecosystem. This limitation arises from a disparity in the software maturity of emerging accelerator platforms: while basic linear algebra libraries (BLAS/LAPACK) are generally available across vendors, robust distributed math libraries equivalent to ScaLAPACK are often less mature or lack standardized interfaces on non-NVIDIA hardware. Bridging this gap will require sustained collaboration across the HPC community to collectively advance open, portable, and high-performance software stacks.

Complementing these architectural efforts, we also envision the integration of mixed-precision arithmetic as a promising direction. While the current default implementation employs complex double-precision calculations to ensure strict numerical stability, the strategic adoption of single or mixed precision in specific computational kernels, leveraging the massive throughput of Tensor Cores on modern GPUs, could offer further performance breakthroughs without compromising physical accuracy.

Beyond architectural and performance optimizations, an important physical extension for future development is the implementation of strictly momentum-conserving forces in Ehrenfest dynamics. While our current formulation utilizes an energy-conserving force that correctly incorporates the moving spatial gauge to yield stable trajectories, rigorous momentum balance requires making the total quantum-classical action stationary. This variational approach introduces additional velocity-dependent force terms related to the spatial derivatives of the time-dependent basis functions. Implementing these exact corrections within the numerical atomic orbital framework is mathematically non-trivial, but it will further elevate the physical fidelity of simulations, particularly for high-energy ion-target collisions and highly directional nonadiabatic processes.

In summary, the modular design of this framework provides a basis for adapting to evolving hardware and algorithmic trends. We hope that the high-performance RT-TDDFT implementation presented here, with its unique combination of localized basis set efficiency and hardware-accelerated computing, will serve as a valuable resource for the community. It offers a practical and efficient means to investigate ultrafast electron dynamics in large-scale complex materials, contributing to the broader effort of understanding non-equilibrium phenomena from first principles.

\section*{Declaration of Interests}

The authors declare that they have no known competing financial interests or personal relationships that could have appeared to influence the work reported in this paper.

\section*{Acknowledgments}

We thank Nuo Chen, Xiaoyang Zhang, and Haozhi Han for their valuable discussions and technical assistance regarding the heterogeneous implementation in ABACUS. We are particularly grateful to Nuo Chen for his comprehensive documentation of the heterogeneous framework, which significantly facilitated our development process. We also acknowledge Dr. Qianrui Liu for providing the GPU implementation of stochastic DFT as a valuable reference for this work. T.B. thanks Haiyu Li from Peking University for constructive comments on the language and phrasing of the manuscript.

This work was supported by the National Key R\&D Program of China under Grant No.\ 2025YFB3003603 and NSFC Excellence Research Group Program No.\ 12588301. This work was partially supported by PKU Kunpeng \& Ascend Center of Excellence. The numerical simulations were performed on the High-Performance Computing Platform of CAPT. We also thank the Open Source Supercomputing Center of S-A-I for providing computational resources.

\bibliographystyle{elsarticle-num}
\bibliography{ref.bib}

\end{document}